\documentclass[useAMS,usenatbib]{mnras}
\usepackage[utf8]{inputenc}
\usepackage{graphicx}
\usepackage{amsfonts}
\usepackage{natbib}
\usepackage{amssymb, amsmath, amsbsy}
\usepackage[dvips]{color}
\usepackage{setspace}
\usepackage[center]{caption}

\usepackage{pythontex}
\definecolor{green}{RGB}{0, 128, 0}
\definecolor{darkgreen}{RGB}{0, 100, 0}
\definecolor{plum}{RGB}{221, 160, 221}
\definecolor{brown}{RGB}{165, 42, 42}
\definecolor{steelblue}{RGB}{70, 130, 180}
\definecolor{orange}{RGB}{255, 165, }
\definecolor{teal}{RGB}{0, 128, 128}
\definecolor{darkred}{RGB}{139, 0, 0}

\defcitealias{muratov15}{M15}
\defcitealias{tumlinson17}{Tumlinson, Peeples, \& Werk (2017)}

\newcommand{\kms}{\mathrm{\,km\,s^{-1}}} 
\newcommand{\vcir}{v_\mathrm{c}} 
\newcommand{\fig}{Figure \ref} 
\newcommand{\eqn}{Equation \ref}
\newcommand{\sect}{Section \ref} 
\newcommand{\msun}{M_\odot} 
\newcommand{\ezw}{\textit{ezw} } 
\newcommand{\tab}{Table \ref} 
\newcommand{\logMstar}{\log(M_*)} 
\newcommand{\sigfof}{\sigma_\mathrm{FoF}} 
\newcommand{\siggal}{\sigma_\mathrm{gal}} 
\newcommand{\mgal}{M_\mathrm{gal}} 
\newcommand{\mvir}{M_\mathrm{vir}} 
\newcommand{\logmvir}{\log(M_\mathrm{vir})} 
\newcommand{\mh}{M_\mathrm{h}} 
\newcommand{\trec}{t_\mathrm{rec}} 
\newcommand{\frec}{f_\mathrm{rec}} 
\newcommand{\so}{\textsc{so}\ } 
\newcommand{\skid}{\textsc{skid} } 
\newcommand{\fL}{f_\mathrm{L}} 
\newcommand{\rvir}{R_\mathrm{vir}} 
\newcommand{\fgas}{f_\mathrm{gas}} 
\newcommand{\mgas}{M_\mathrm{gas}} 

\newcommand{\figpath}{./figures}

\title[Wind Scalings in Cosmological Simulations]{The Impact of Wind Scalings on Stellar Growth and the Baryon Cycle in Cosmological Simulations}
\author[S. Huang et al.]{Shuiyao
Huang$^{1}$\thanks{E-mail:shuiyao@astro.umass.edu},
Neal Katz$^{1}$, Romeel Dav\'e$^{2,3,4}$, Benjamin D. Oppenheimer$^{5}$, 
\newauthor
David H. Weinberg$^{6}$, Mark Fardal$^{1,7}$, Juna A. Kollmeier$^{8}$ and Molly S. Peeples$^{7,9}$
\\$^1$ Astronomy Department, University of Massachusetts, Amherst, MA 01003,
USA
\\$^2$ Institute for Astronomy, Royal Observatory, Univ. of Edinburgh, Edinburgh EH9 3HJ, UK
\\$^3$ University of the Western Cape, Bellville, Cape Town 7535, South Africa
\\$^4$ South African Astronomical Observatories, Observatory, Cape Town 7925,
South Africa
\\$^5$ CASA, Department of Astrophysical and Planetary Sciences, University of
Colorado, Boulder, CO 80309, USA
\\$^6$ Astronomy Department and CCAPP, Ohio State University, Columbus, OH
43210, USA
\\$^7$ Space Telescope Science Institute, Baltimore, MD 21218
\\$^8$ Observatories of the Carnegie Institution of Washington, 813 Santa Barbara Street, Pasadena, CA, 91101, USA
\\$^{9}$ Johns Hopkins University, Department of Physics \& Astronomy
}

\begin{document}

\date{Accepted 0000 October 00. Received 0000 October 00; in original form 0000
October 00}

\pagerange{\pageref{firstpage}--\pageref{lastpage}} \pubyear{0000}

\maketitle

\label{firstpage}

\begin{abstract}
Many phenomenologically successful cosmological simulations employ
kinetic winds to model galactic outflows. Yet
systematic studies of how variations in kinetic wind scalings might alter
observable galaxy properties are rare. Here we employ \textsc{gadget-3}
simulations to study how the baryon cycle, stellar mass function, and
other galaxy and CGM predictions vary as a function of the assumed outflow
speed and the scaling of the mass loading factor with velocity dispersion.
We design our fiducial model to reproduce the measured wind properties at 25\%
of the virial radius from the Feedback In Realistic Environments (FIRE)
simulations. We find that a strong dependence of $\eta \sim \sigma^5$
in low mass haloes with $\sigma < 106\kms$ is required to match the faint end of the stellar mass functions at
$z > 1$. In addition, faster
winds significantly reduce wind recycling and heat more halo gas.
Both effects result in less stellar mass growth in massive haloes and impact
high ionization absorption in halo gas. We cannot simultaneously
match the stellar content at $z=2$ and $z=0$ within a single model,
suggesting that an additional feedback source such as AGN might be
required in massive galaxies at lower redshifts, but
the amount needed depends strongly on assumptions regarding the outflow
properties. We run a 50 $\mathrm{Mpc/h}$, $2\times576^3$
simulation with our fiducial parameters and show that it matches a
range of star-forming galaxy properties at $z\sim0-2$.

\end{abstract}
\section{INTRODUCTION}


Galactic scale outflows (galactic winds) driven by star formation processes
have been recognised as a critical ingredient in galaxy 
evolution. Galactic winds are observed ubiquitously among star forming galaxies
in both the local and
distant Universe, and their properties are often found to correlate with the
properties of the central galaxy such as the star formation rate and the
circular velocity \citep{rupke05, martin05, heckman16}. The short-lived,
massive stars formed in star forming galaxies release a considerable amount of
energy and momentum during their short lifetimes through radiation, stellar
winds, and supernova (SNe) explosions. Collectively, these effects could
efficiently drive the large scale outflow of dense, metal-enriched gas from the
interstellar medium (ISM) to large distances from the galaxy, making a strong
impact on galaxy growth and also on the properties of the circumgalactic medium
(CGM). Galactic winds have been implemented as a sub-grid model in cosmological
simulations, in which they play a critical role in explaining the suppressed
star formation in dwarf galaxies and the metal content in the CGM
\citep[e.g.][]{oppenheimer08, oppenheimer12, ford13}.


However, implementing galactic winds in cosmological simulations remains a
challenge
because of our limited knowledge of the wind driving mechanism, and the limited
resolution of large volume simulations. Self-consistently generating galactic
winds by explicitly modelling the key wind driving mechanisms is still a
challenging problem that is under active study \citep{zhang18}. More 
importantly,
the physical processes that are critical to driving winds occur on scales that
are so small that they remain
unresolvable in even the highest resolution zoom-in simulations of
today \citep[e.g.][]{scannapieco15, schneider17, fire2}. As a consequence, modern cosmological simulations 
adopt a variety of
sub-grid prescriptions that describe how to launch galactic winds from
simulated galaxies
\citep{springel03, stinson06, oppenheimer06, agertz13, schaye15, fire2, pillepich18a}.
This diversity of numerical recipes for
galactic winds leads to many different predictions from these simulations
\citep{scannapieco12, sembolini16b, sadoun16, valentini17}.

The \textit{kinetic feedback} models \citep{springel03, oppenheimer06}, like 
those that we employ, rely on
scaling relations that connect the macroscopic properties
of galactic winds, such as the wind velocity $v_w$ and the mass loading factor $\eta$, defined
as the ratio between the outflow rate ($\dot{M}_w$) and the star formation rate
(SFR) to the resolved properties of their host 
galaxies such as the halo mass
$\mh$, or some characteristic velocity (e.g., the velocity dispersion
$\sigma$). Though the properties of galactic winds and the physical mechanisms
that generate them are still poorly understood, there have been many
constraints on these scaling relations from observations, analytic calculations,
and simulations \citep{rupke05, murray05, murray11}.

The fiducial wind prescription 
that we have used in many of our previous papers
\citep[e.g.][]{oppenheimer06, dave13, ford16}
was motivated by the analytic momentum-driven and energy-driven wind models
developed by \citet{murray05}. In the momentum-driven scenario, the outflow is
driven in a momentum-conserving manner by the radiation pressure from massive
stars and SNe acting on the dust particles that is coupled to the cool gas. The
momentum flux overwhelms the gravitational potential of the dark matter halo
in early phases and accelerates the cool gas from within the star forming
region to an asymptotic velocity at the virial radius of the dark matter halo.
Assuming an isothermal potential and ignoring hydrodynamic forces,
\citet{murray05} derived the
evolution of the wind speed as a function of radius as:
\begin{equation}
v_w(r) = 2\sigma_\mathrm{1D}\sqrt{(\fL - 1)\ln \left(\frac{r}{R_0}\right)}
\label{eqn:vmurray}
\end{equation}
where $\sigma_\mathrm{1D}$ is the one dimensional velocity dispersion measured
for an isothermal sphere, $f_\mathrm{L} = L/L_\mathrm{M}$ is the ratio between
the luminosity of the galaxy and the critical Eddington luminosity, and $R_0$ is
the radius from which the wind is launched. They also derived the scalings between $\eta$ and $\sigma$ as
$\eta\propto \sigma^{-1}$ based on conservation of momentum. In our more recent simulations we
actually assume that $\eta\propto \sigma^{-2}$ for small galaxies,
which is the scaling one would expect for energy driven winds by supernova.

Even if this is not the correct physics behind real galactic winds,
this modified momentum-driven model predicts scaling relations between global
quantities
such as mass loading, wind velocity and the stellar mass that, when included
in cosmological hydrodynamic simulations, are broadly consistent with many
observational constraints \citep{oppenheimer06, oppenheimer08, dave10, oppenheimer10, dave11a, dave11b, dave13, ford16}. However, implementing the wind model
into our simulations is more complicated than suggested by the above equations. 
Instead of launching a wind from any radius
$R_0$ as in equation \ref{eqn:vmurray}, we eject wind particles with an
initial velocity $v_w$ from star forming regions that inhabit the centre of the
galactic potential and let them propagate
out under the combined gravitational and hydrodynamical forces (we ignore
hydrodynamic interactions for a short period
after wind launch; see below for details).  Furthermore, the
dynamical evolution of wind particles in our simulation is very different from
the analytical solution of \citet{murray05} for several reasons. First, the
gravitational potentials in our simulated haloes are steeper than the isothermal
sphere assumed in \citet{murray05}, especially in the central region where
baryonic matter dominates. Second, our simulations do not explicitly include
radiation pressure, which in their calculation accelerates the outflow all the
way out
to the virial radius; Third, wind particles in our simulation are further
slowed down by hydrodynamic interactions with the gas in the CGM or the
intergalactic medium (IGM). Finally, these interactions are probably not
accurately evolved owing to resolution and other numerical issues.


Recent zoom-in simulations of individual galaxies provide further insights into
the scaling relations between the launched winds and their host galaxies
\citep{muratov15, christensen16}. Capable of resolving GMC scales and the
turbulent nature of the ISM, these simulations drive winds by explicitly
modelling physical processes that depend on the local ISM properties and analyse
how the wind behaviours depend on the global properties of their host galaxies,
therefore better bridging the gap between the governing physics on small scales
and the impact of the winds in the broader context of galaxy formation
(but still not resolving all the scales critical for driving winds \citep[e.g.][]{schneider17}).

Using a series of simulations that span four decades in halo mass up to
$10^{12}M_\odot$ and covering a redshift range from $z=0$ to $z=4$, the
Feedback In Realistic Environments (FIRE) project \citep[][hereafter \citetalias{muratov15}]{muratov15} derives
how mass loading factors and wind speeds depend on the circular velocity, the
halo mass, and the stellar mass of the host galaxies. They report faster wind
speeds in massive haloes than in our previous simulations using the fiducial
wind model described above (and in more detail below). They also report a 
stronger scaling between the mass loading factor and the circular velocity 
with $\eta \propto \vcir^{-3.3}$, steeper than the energy-driven wind scaling,
$\eta \propto \vcir^{-2}$, which we assume in our simulations for low mass 
galaxies. 
\citet{christensen16} simulate and analyse over twenty spiral and dwarf
galaxies covering halo masses from $10^{9.5}M_\odot$ to $10^{12}M_\odot$.
Despite using a very different feedback model, they obtain a similar scaling
for the mass loading factor, $\eta \propto \vcir^{-2.2}$.

One key issue is that \citetalias{muratov15} report their results at
$R_{25}$, one quarter of the virial radius, while by necessity we impose our
wind scalings at wind launch,
which occurs inside star forming regions within the galaxy at much smaller 
radii. Clearly, it makes more sense to talk about galactic wind properties
outside the galaxy and $R_{25}$ is a reasonable radius to choose. As we discuss
below, \citetalias{muratov15} motivated us to look at our wind scaling
properties at $R_{25}$, and we find that they are very different from those at launch. 

Motivated by this recognition, in this paper we revisit the basic assumptions
made
in our sub-grid wind model. In particular, we re-calibrate our prescriptions
for launching winds from galaxies using the scaling relations found in
the FIRE simulations as constraints. We will examine how the new prescription,
now capable of qualitatively reproducing the wind behaviours seen in the FIRE
simulations, will affect some of the basic predictions of our cosmological
simulations, such as the galactic stellar mass functions (GSMFs) and the
galactic mass-metallicity relation (MZR) at various redshifts. Furthermore, we
also experimented with a range of wind parameters, all allowed by current
observational and theoretical constraints on galactic winds, to examine the
robustness of these predictions to small changes in the wind implementation and
were surprised to find that these basic observational quantities were actually
very sensitive to small changes in the wind scalings, changes that are much
smaller than the differences between wind models employed by different 
simulation groups \citep[e.g.][]{agertz13, schaye15, mufasa, pillepich18a}.

Recent cosmological simulations also often adopt a 
sub-grid AGN feedback model and
show that it is crucial to reproduce the number density of massive galaxies and
the fraction of red galaxies at low redshifts to match observations.
The simulations presented in this paper, like our past published work, do not
include any such sub-grid model for AGN feedback, or any other
mechanism that has been proposed in the literature to specifically quench star
formation in massive galaxies \citep[e.g.][]{crain15, mufasa, weinberger18}. Simulations without AGN
feedback tend to produce too many blue massive galaxies, indicating that the
stellar feedback alone is insufficient to keep these galaxies quenched.
However, the implementation of AGN feedback often has little effect at higher
redshift or in small galaxies, where stellar feedback dominates galaxy growth.
The implementation of any AGN feedback model that includes free parameters that
are tuned to match observations will thus inevitably be affected by the stellar
feedback models. Therefore, in this paper we will not focus on reproducing the
observed Universe in every detail, but we
will rather explore how sensitively galaxy evolution depends on the star 
formation driven wind model. This knowledge will also help characterise the
limits of what stellar feedback alone can accomplish and thus provide further
constraints on any additional required feedback mechanism.

The paper is organised as follows. \sect{sec:method} reviews our original
sub-grid model for launching winds in our simulations and also introduces the
new wind algorithm. \sect{sec:simulations} describes our cosmological
simulations and introduces the test simulations that we use in this paper.
\sect{sec:sensitivity} studies how sensitively our simulations depend on the
parameters of the wind algorithm by comparing results from the test
simulations. \sect{sec:discussion} studies in detail how stellar mass grows 
within galaxies in our simulations by focusing on their accretion and merger
histories, and how it is affected by certain wind parameters. It also discusses
the challenge of matching observations relying only on stellar driven winds and
the requirements for any additional feedback mechanism.
\sect{sec:fiducial_simulation} presents results from our high resolution
simulation using the new wind algorithm with a fiducial choice of wind
parameters and compares them to results from earlier published
versions of our cosmological simulations. \sect{sec:summary} summarises our
results.

%
%

\section{The Kinetic Feedback Model}
\label{sec:method}
\subsection{Our published wind model}
\label{sec:ezw}

We based our original sub-grid wind model \citep[\textit{ezw,}][]{dave13,
ford16} on the analytic formulation of energy-driven and momentum-driven
winds from \citet{murray05}. Here we summarise our numerical algorithm for
including it in our simulations. For any SPH particle $i$ in a galaxy that is
above a critical density threshold $\rho_{SF}$ for star formation,
we determine whether or not to
turn it from a normal SPH particle into a wind particle according to a
probability $p_i$ that scales with the local star formation rate:
$$p_i \propto \eta \times \mathrm{SFR}_i$$
We choose the critical density threshold as $\rho_{SF} = 0.13\ \mu m_\mathrm{H}$
\citep{springel03}, where $\mu$ is the mean 
atomic weight and $m_\mathrm{H}$ is the mass of
the hydrogen atom. Once an SPH particle becomes a wind particle, we add a
velocity boost of $v_w$ to the particle in the direction of
$\mathbf{v_i}\times\mathbf{a_i}$, where $\mathbf{v_i}$ and $\mathbf{a_i}$ are
the velocity and acceleration of the particle, respectively, before launch, as
outflows are often seen perpendicular to the disc where the resistance from
the cold dense ISM is minimised. All hydrodynamical interactions relating to the
newly created wind particle are ignored for an interval of $t_\mathrm{delay} = 20\ \mathrm{kpc}/v_w$ or
until the particle has reached a density threshold of $\rho_{th}=0.1\rho_{SF}$.
This decoupling from
hydrodynamical forces ensures that wind particles are able to escape the disc
where the resolution is insufficient to correctly model the hydrodynamical
interactions \citep{dallavecchia08}. The two free parameters, $\eta$ and
$v_w$, are crucial to the wind model, whose values are constrained from the
analytical scalings that correlate them with the
galaxy velocity dispersion $\sigma$. For our preferred published model
\citep{dave13}, which we refer to as the $ezw$ model, those scalings are:

\begin{equation}
  v_{w; ezw} = 4.29\sigma\sqrt{\fL - 1} + 2.9\sigma
  \label{eqn:vwind_ezw}
\end{equation}

\begin{equation}
\eta=\left\{
\begin{split}
  &\frac{150\kms}{\sigma}\frac{75\kms}{\sigma} &(\sigma \leqslant 75\kms)\\
  &\frac{150\kms}{\sigma} &(\sigma \geqslant 75\kms)
\end{split}
,\right.
\label{eqn:eta_ezw}
 \end{equation}
where $\fL$ depends on the metallicity of the SPH particle as constrained by
observations \citep{rupke05}, and we adjust the normalisation factor 
$\sigma_0=150$km/s to match the enrichment level of the high-redshift
intergalactic medium \citep{oppenheimer08}.

This original wind velocity formula (\eqn{eqn:vwind_ezw}) rescales the launch wind 
velocity by adding $2.9\sigma$ in an attempt to get the correct asymptotic
velocity at the virial radius, to account for the dynamical evolution of the winds inside the
halo. However, as we will show in \sect{sec:v25vc}, this does not work very well.

The formula for $\eta$ introduces a characteristic velocity $\sigma_{ezw} = 75\
\kms$ that separates the momentum-driven wind scaling regime from the 
energy-driven one. The momentum-driven mass-loading factor, which scales with
$\sigma^{-1}$, applies to relatively large systems where outflows could be driven
primarily by the momentum flux from young stars and supernovae while the
thermal energy from SNe would be dissipated too quickly to become dynamically
important. However, in dwarf galaxies with $\sigma$ below this limit, we assume
that energy feedback from supernovae starts to dominate, based on analytical
and numerical models by \citet{murray10} and \citet{hopkins12}. In this energy
conserving regime, we assume $\eta \propto \sigma^{-2}$. Whether or not the
physical models behind these scaling relations are correct, \citet{dave13}
show that this hybrid scaling leads to better agreement with both the low mass 
stellar mass and HI mass functions at $z=0$.

We determine the velocity dispersion $\sigma$ of the host halo on-the-fly. We
identify galaxies using a friends-of-friends (FoF) algorithm that binds star
forming particles to their closest neighbours. We estimate the velocity
dispersion using the total mass of the galaxy $\mgal$:
\begin{equation}
\sigfof =
200\left(\frac{\mgal}{5\times10^{12}M_\odot}\frac{H(z)}{H_0}\right)^{1/3} \kms
\label{eqn:sigma}
\end{equation}
where $\mgal$ is the total mass of the FoF group, and $H(z)$ and $H_0$ are the
Hubble constants at redshift $z$ and 0, respectively. $\mgal$ includes dark
matter, gas, and stars and we choose the FoF linking length to be smaller
than one that would make $\mgal$ equal to the virial mass (see \S\ref{sec:fof}
for details).  In principle, we could
measure the velocity dispersion $\sigfof$ for each galaxy
directly. However, uncertainties arise owing to poor resolution particularly in
the inner regions of each galaxy \citep{oppenheimer06}. Moreover, the numerical
noise would in some cases yield an unrealistic $\sigfof$ that would lead to
unphysical results. Finally, satellite galaxies often have their $\sigfof$
overestimated owing to contamination by particles that actually belong to
the central galaxy but that are impossible to separate.
Therefore, we use the above empirical relation given that
any error that arises from using this relation is sub-dominant to the
uncertainties that come from our assumptions about the wind model itself.



\subsection{A New Algorithm for Launching Winds}
The wind speed formula above (\eqn{eqn:vwind_ezw}) derives from the analytic
calculations of \citet{murray05} (Equation \ref{eqn:vmurray} in this paper)
for radiation driven winds. However, as discussed in the introduction, the
actual propagation of wind particles in SPH simulations is very different from
that assumed in this analytic model. In the simulations, a wind particle is initially
decoupled from the hydrodynamics until it reaches the critical density
$\rho_\mathrm{th}$, typically several kiloparsecs from the galactic centre.
After that, the particle can interact hydrodynamically with the surrounding
gas and will slow down and heat. How this occurs depends on the numerical 
details of the hydrodynamic solver, since the interaction is poorly resolved.
In dwarf galaxies, the winds are typically much faster
than the escape velocity and are able to escape their host haloes, but most
winds in massive galaxies only travel to a certain distance from the galaxy and
eventually fall back within a recycling timescale $\trec$
\citep{oppenheimer10}. We will show in section \ref{sec:sensitivity} that both
the distances to which wind particles travel and their recycling timescales
are highly sensitive to the initial wind speed and numerical resolution. This
leads to large uncertainties in the evolution of galaxies and their CGM
properties because the behaviour of the ejected wind particles has a crucial
impact on the gas supply for star formation inside galaxies, and the density,
temperature, and metal distribution in the surrounding halo gas.

Here we present an improved algorithm to determine how winds are ejected from
their host galaxies. In this new method, we keep the velocity of a wind particle
relative to its host galaxy constant until it reaches the density threshold
$\rho_\mathrm{th}$ where the particle recouples hydrodynamically to the
other gas. We choose the density threshold to be $0.1\
\rho_\mathrm{SF}$, where $\rho_\mathrm{SF}$ is the physical density threshold
above which star formation occurs in the simulation. Therefore, before 
recoupling, the wind particle effectively also decouples from gravity so
that its kinetic evolution remains temporarily independent of the central
potential dominated by baryons. The wind particle
still contributes to the overall gravitational field as it leaves the disc, 
preventing galaxies with strong outflows from having large artificial
dynamical disturbances in the disc.

We also adopt a new formula for the initial wind speed that is parameterized
differently from before. As we will see in \sect{sec:v25vc}, this results in 
$v_w\propto \vcir$ at $R_{25}$ as found in \citetalias{muratov15} and
\citet{murray05}. We keep the tangential velocity relative to the galaxy fixed
so that the wind particle conserves its angular momentum as it is launched. We
determine the radial component of the velocity by
\begin{equation}
  v_w = \alpha_v\sigma\sqrt{f_\mathrm{L}}\left(\frac{\sigma}{50\ \mathrm{km\
s^{-1}}}\right)^{\beta_v}
  \label{eqn:vwind}
\end{equation}
where $\alpha_v$ and $\beta_v$ are free parameters that will be discussed
later, and $\fL$ is the metallicity dependent ratio between the galaxy
luminosity and the Eddington luminosity. We adopt the \citet{oppenheimer06}
formula for $\fL$:
\begin{equation}
  f_\mathrm{L} = f_{\mathrm{L};\odot}\times 10^{-0.0029(\log Z_\mathrm{gal} + 9
)^{2.5} + 0.417694}
\end{equation}
where $f_{\mathrm{L};\odot}$ varies randomly between 1.05 and 2. Here we 
now use the mass 
weighted average metallicity $Z_\mathrm{gal}$ of the host galaxy to compute 
$\fL$, instead of directly using the metallicity of the wind particle as in 
their paper and in our past work.
This is more appropriate since it is the global properties of
the galaxy that will determine the wind properties at $R_{25}$ and the
metallicities may have large variances inside a galaxy. In most galaxies, $\fL$ is only slightly above 1. Note that the
$\sqrt{\fL-1}$ term in the original \ezw velocity formula (\eqn{eqn:vwind_ezw})
becomes $\sqrt{\fL}$ after adding in the kinetic energy lost to the
gravitational field as the wind particle climbs up the potential, so that our
winds can match the asymptotic velocity predicted by \eqn{eqn:vmurray}. In the
original formula, this correction is included as the second term $2.9\sigma$,
which is normalised at a radius of $R_\mathrm{esc} = 0.1\rvir$. 
Note that this is different from the $R_0$ that appears in the analytic formula
(\eqn{eqn:vmurray}). \citet{oppenheimer06} chose $R_0 = 0.01\rvir$ to obtain
the constant factor 4.29 in the first term of \eqn{eqn:vwind_ezw}. In our new
formula, we use the same normalisation radius $R_\mathrm{esc} = R_0$ for the two
terms. We incorporate the freedom of choosing $R_0$ into the parameter
$\alpha_v$.

The parameters $\alpha_v$ and $\beta_v$ determine the overall wind speed and
its scaling with the depth of the halo potential. Since $\sigma$ scales with
$\mgal^{1/3}$, the wind speed scales with the halo mass by a power
law index $(1 + \beta_v)/3$. The parameter $\beta_v$, therefore, characterises
how much momentum the wind particles need to overcome the central,
baryon dominated gravitational potential and reflects how the central potential
varies with halo mass. The parameter $\alpha_v$ reflects the uncertainties in
choosing the normalisation radius $R_0$ in \eqn{eqn:vmurray} and in measuring
the $\sigma$ from simulations. We calibrate our parameters to make our wind scalings at
$R_{25}$ consistent with the results from \citetalias{muratov15} (see \sect{sec:v25vc} for details).
Note that this wind speed should not be directly compared with
observations because this wind speed formula only applies to
winds that are close to the disc ($R_0$), where they are launched, while observationally
the location of the winds is usually unknown.

In addition to the wind speed, we also explore different choices for the mass
loading factor scalings. Instead of the original formula (\eqn{eqn:eta_ezw}),
we now parameterise $\eta$ as:

\begin{equation}
\eta=\left\{
\begin{split}
  &\alpha_\eta\left(\frac{150\kms}{\sigma}\right)\left(\frac{\sigma_{ezw}}{\sigma}\right)^{\beta_\eta}(1 +z)^{1.3} &(\sigma \leqslant \sigma_{ezw})\\
  &\alpha_\eta\left(\frac{150\kms}{\sigma}\right)(1 +z)^{1.3} &(\sigma \geqslant \sigma_{ezw})
\end{split}
\right.
\label{eqn:eta_w}
\end{equation}
where $\alpha_\eta$ is a normalisation factor and $\beta_\eta$ is the power law
index. 

Therefore, the mass loading factor $\eta$ still follows a momentum-driven wind
scaling $\eta \propto \sigma^{-1}$ in massive systems above a certain threshold
$\sigma_{ezw}$. Below that threshold, $\eta \propto \sigma^{-(1+\beta_\eta)}$.
The original energy-driven scaling in small galaxies corresponds to $\beta_\eta
= 1$. However, we will show in later sections that to match the
observed number densities of small galaxies at high redshifts requires a higher
$\beta_\eta$ value. We also adopt a redshift dependent factor $(1 +
z)^{1.3}$ motivated by the results from the FIRE simulations
\citepalias{muratov15}. To avoid unphysically large $\eta$ at high redshifts, we only allow $\eta$ to change with $z$ at $z < 4$ and use a constant factor of $5^{1.3}$ at $z > 4$.

\subsection{The FoF Group Finder}
\label{sec:fof}
Both the mass loading factor and the wind speed in our wind algorithm depend
on the properties of the galaxy from which the winds are launched.
To identify galaxies and their host haloes on the fly, we use a FoF group finder with a linking length of
0.04 times the mean interparticle spacing, and multiply the resulting mass by a
constant factor determined via a
calibration against the results from using \textsc{so}, a more accurate but more
computationally costly spherical
overdensity halo finder \citep{keres05,huang19}. Above the mass
resolution limit, the on-the-fly FoF group finder typically underestimates the
total mass (including both baryons and dark matter) by a factor of 2 to 3 with a scatter of $\sim$ 0.1 dex. The discrepancies
between the estimated halo masses are nearly scale-independent, introducing only
a small additional factor of $\sigma^{0.05}$ to the wind scalings
(\eqn{eqn:vwind} and \eqn{eqn:eta_w}).

Furthermore, we now identify the
group centre and velocity centroid by including all cold baryons (including
star forming gas and stars) within the FoF group instead of just using stars as
in our previous work. This is especially important for dwarf galaxies that
have only begun assembling, which can be almost devoid of stars.



\subsection{Wind energy}

In this section we calculate the total kinetic energy flux of the star
formation driven winds according to our new wind speed formula 
(\eqn{eqn:vwind}).  For
a total mass $M_*$ of stars formed, the total kinetic energy added to the winds
that are generated as a consequence is:
\begin{equation}
E_\mathrm{kin}(M_*) = \frac{1}{2}M_w\bar{v}_w^2
\end{equation}
where the amount of winds launched over a given time $M_w$ relates to the
amount of star formation by $M_w \equiv \eta M_*$, and $\bar{v}_w$ is the
average wind speed. Combining this equation with the definition of $\eta$ and
$v_w$, results in:
\begin{equation}
E_\mathrm{kin} = 0.7\bar{f_\mathrm{L}}\alpha_v^2\sigma^2\left(
\frac{106}{\sigma} \right)^{\beta_\eta-2\beta_v+1}(1+z)^{1.3}M_*
\end{equation}
where $\bar{\fL} \sim 1.0$ is the average value for the luminosity factor that
appears in \eqn{eqn:vmurray} and \eqn{eqn:vwind}. Therefore, the wind energy
generated per solar mass of star formation is
\begin{equation}
  \begin{split}
    E_\mathrm{kin}(\msun) = & 1.6\times10^{47}\\
    & \times \alpha_v^2  (2.1)^{2\beta_v} \left( \frac{106}{\sigma}\right)^{\beta_\eta-2\beta_v-1}\\
    & \times(1+z)^{1.3}\ \mathrm{ergs}
  \end{split}
\end{equation}
Using the parameters for our fiducial simulation: $\alpha_v = 4.0$, $\beta_v =
0.6$ and $\beta_{\eta} = 4.0$, the wind energy for massive galaxies with
$\sigma > 106\kms$ is:
\begin{equation}
  E_\mathrm{kin}(\msun) = 6.3\times10^{48}\left( \frac{106}{\sigma}
\right)^{-2.2}(1+z)^{1.3}\ \mathrm{ergs}
\end{equation}
One can compare this value to the amount of energy available from
Type II SNe. According to a Chabrier IMF, the average number of type II SNe per
one solar mass is $\eta_\mathrm{SN} = 8.3\times10^{-3}$. Assuming each
supernova produces $E_\mathrm{SN} = 10^{51}\ \mathrm{ergs}$ of energy, the
total energy produced by supernova per solar mass of stars formed is 
$\epsilon_\mathrm{SN} = 8.3\times10^{48}\
\mathrm{ergs}$. In our new wind model this equals the wind energy from a galaxy
at $z = 0$ with $\sigfof \sim 120\kms$. After taking into account the factor of
$\sim 3^{1/3}$ underestimate of the real $\sigma$ by the on-the-fly FoF finder
and using $\vcir = \sqrt{2}\sigma$, this corresponds to a circular velocity of
$\vcir \sim 240\kms$. For lower mass galaxies the wind energy is a fraction
($< 1$) of
the energy available from supernovae, but for massive galaxies it is larger.

Hence, in addition to \eqn{eqn:vwind}, we adopt an upper limit for the
wind speed
that requires the total kinetic energy of winds to be less than 5 times the
total available energy from type II SNe. Of course, in these more massive 
systems
one expects the winds to be dominated by photon momentum if one takes the model
seriously.  Even so we think it prudent to limit the total wind
energy. The calculation above shows that the upper
limit only matters at high redshift or in the most massive systems.

%
%

\section{Simulations}
\label{sec:simulations}
We implemented the new wind algorithm into our SPH code based on
\textsc{gadget-3} (see \citet{springel05} for reference). The
code includes several recent numerical improvements in the SPH technique
\citep{huang19}. To summarise, we use the pressure-entropy formulation
\citep{hopkins13} of SPH to integrate the fluid equations and a quintic spline
kernel to measure fluid quantities over 128 neighbouring particles. We also use
the \citet{cd10} viscosity algorithm and artificial conduction as in
\citet{read12} to capture shocks more accurately and to reduce numerical noise.
Both the artificial viscosity and the conduction are turned on
only in converging flows with $\nabla \cdot \mathbf{v} < 0$ to minimise
unwanted numerical dissipation. We also include the Hubble flow while
calculating the velocity divergence. Our fiducial code leads to considerable
improvements in resolving the instabilities at fluid interfaces in subsonic
flows and produces consistent results with other state-of-art hydrodynamic
codes in various numerical tests \citep{sembolini16a, sembolini16b, huang19}.

In the current version we also add metal line cooling including
photo-ionization effects for 11 elements as in \citet{wiersma09}, and we 
recalculate cooling rates according to an updated ionizing background
\citep{haardt12}. The star formation processes are modelled as in
\citet{springel03}, which includes a subgrid model for the multiphase ISM in
dense regions with $n_\mathrm{H} > 0.13\ \mathrm{cm^{-3}}$, and a star
formation recipe that is scaled to match the Kennicutt-Schmidt relation. In 
this paper we will distinguish SPH particles as star-forming or
non star-forming based on whether or not their densities are higher than this
density threshold. We specifically trace the enrichment of four metal species C,
O, Si, Fe that are produced from type II SNe, type Ia SNe and AGB stars as in
\citet{oppenheimer08}. These processes also generate energy that we put in the simulations with sub-grid models. 
However, the input energy from these feedback processes only have sub-dominant effects to galaxy formation compared to the wind feedback \citep{oppenheimer08}.

We evolve the fiducial simulation (RefHres) for this paper in a comoving
periodic box with $L = 50\ h^{-1}$ Mpc on each side that initially contains
$2\times576^3$ gas and dark matter particles. The initial mass of each gas
particle and dark matter matter are $m_\mathrm{gas}=1.1\times10^7M_\odot$ and
$m_\mathrm{dark}=6.6 \times 10^7M_\odot$, respectively. 
Our Plummer equivalent gravitational softening of 1.2 kpc determines our
spatial resolution.  In addition, we run
several simulations with the same initial conditions in $L = 50\ h^{-1}$ Mpc
boxes but at a lower resolution with $2\times288^3$ particles (with
eight times worse mass resolution and two times worse spatial resolution).
Most of these
simulations use the same wind algorithm as in the fiducial simulation with only
differences in the wind parameters. We use these simulations to test the
numerical convergence of the wind algorithm and also to determine the
sensitivity of the simulations to the wind parameters.

We choose the $50\ h^{-1}$ Mpc size to balance between a
decent numerical resolution and a representative volume, given the computational resources that are feasible.
Furthermore, many of our previous results are based on simulations of similar volume and resolution.
We are therefore able to verify the robustness of these previous results with the volume chosen.
Zoom-in simulations \citep[e.g.][]{sadoun16, valentini17} provide a wide range of observables and diagnostics different from those probed by cosmological simulations
and are therefore complementary to this present study.

In \tab{tab:simulations} we summarise and classify all the simulations into
three categories. The
first category simulations differ from the fiducial simulation only in
terms of the mass loading factor. The second category simulations
differ only in terms of wind speed. In \sect{sec:sensitivity}, we will focus on
comparing simulations of each category to demonstrate the sensitivity to the
wind parameters. The rest of the simulations differ in both the mass loading
factor and the wind speed or use different wind algorithms such as our original
hybrid \ezw wind model. In addition, the ezwDESPH simulation is the only
simulation that uses the traditional SPH technique and, therefore, is the
simulation that is closest to the original numerical model used in our 
previously published
simulations \citep[e.g.][]{dave13, ford16}. In \sect{sec:fiducial_simulation},
we will focus on results from the fiducial simulation and compare them to
observations as well as the the original \ezw model to show how much the new
fiducial wind algorithm changes some of our basic results from our previous
simulations.

\begin{table*}
\begin{minipage}{160mm}
  \begin{center}
\caption{Simulations and their wind parameters}
\begin{tabular}{@{}lcccccll} \hline
\label{tab:simulations}
Simulation & $\alpha_\eta$ & $\beta_\eta$ & $\sigma_{ezw}$ \footnote{The first
three parameters, $\alpha_\eta$, $\beta_\eta$ and $\sigma_{ezw}$ determine the
mass loading factor according to \eqn{eqn:eta_w}. The \textit{ezw} wind model
uses a slightly different formula (\eqn{eqn:eta_ezw}) where these parameters
have a similar effect.} & $\alpha_v$ & $\beta_v$ \footnote{The next two
parameters, $\alpha_v$ and $\beta_v$ determines the initial wind speed
according to \eqn{eqn:vwind}. The wind speed in the \textit{ezw} model is
formulated in a quite different way so that the parameters do not apply.} &
Colour\footnote{We use a consistent colour scheme for the entire paper to
distinguish simulations from each other. This column indicates the unique colour
that is used to represent the corresponding simulation.} & Remark\\
\hline
RefHres & 0.1 & 4 & 106 & 4.0 & 0.6 & black & Fiducial wind model, high resolution with $N_{gas}=576^3$\\
Ref & 0.1 & 4 & 106 & 3.5 & 0.6 & \textcolor{magenta}{magenta} & Fiducial wind model, $N_{gas}=288^3$ \\
ezw & 4.29 & 1 & 75 & - & 0.0 & \textcolor{darkred}{darkred} & The \ezw model from \citet{dave13}\\
ezwFast & 4.29 & 1 & 75 & 3.5 & 0.6 & \textcolor{teal}{teal} & \ezw mass loading, but wind speeds of Ref\\
ezwDESPH & 4.29 & 1 & 75 & - &
0.0 & \textcolor{blue}{blue} & The \ezw wind model, traditional SPH methods\\ 
StrongFB & 0.2 & 4 & 106 & 3.5 & 0.6 & \textcolor{red}{red} & $\alpha_\eta = 0.2$ instead of the fiducial value 0.1\\
WeakFB & 0.05 & 4 & 106 & 3.5 & 0.6 & \textcolor{orange}{orange} & $\alpha_\eta = 0.05$ instead of the fiducial value 0.1\\
Sigma75 & 0.4 & 4 & 75 & 3.5 & 0.6 & \textcolor{brown}{brown} & $\sigma_{ezw} = 75\kms$ instead of the fiducial $106\kms$\\
Ref$\sigma$4 & 0.1 & 3 & 106 & 3.5 & 0.6 & \textcolor{steelblue}{steelblue} & $\eta$ scales with $\sigma^{-4}$, not $\eta \propto \sigma^{-5}$ as in Ref\\
Ref$\sigma$3 & 0.1 & 2 & 106 & 3.5 & 0.6 & \textcolor{plum}{plum} & $\eta$ scales with $\sigma^{-3}$ not $\eta \propto \sigma^{-5}$ as in Ref\\
RefSlow & 0.1 & 4 & 106 & 3.0 & 0.3 & \textcolor{green}{green} & Same as the Ref simulation but slower winds\\
\hline \end{tabular}
\end{center}
\end{minipage}
\end{table*}

\begin{figure} 
\centering
\includegraphics[width=0.95\columnwidth]{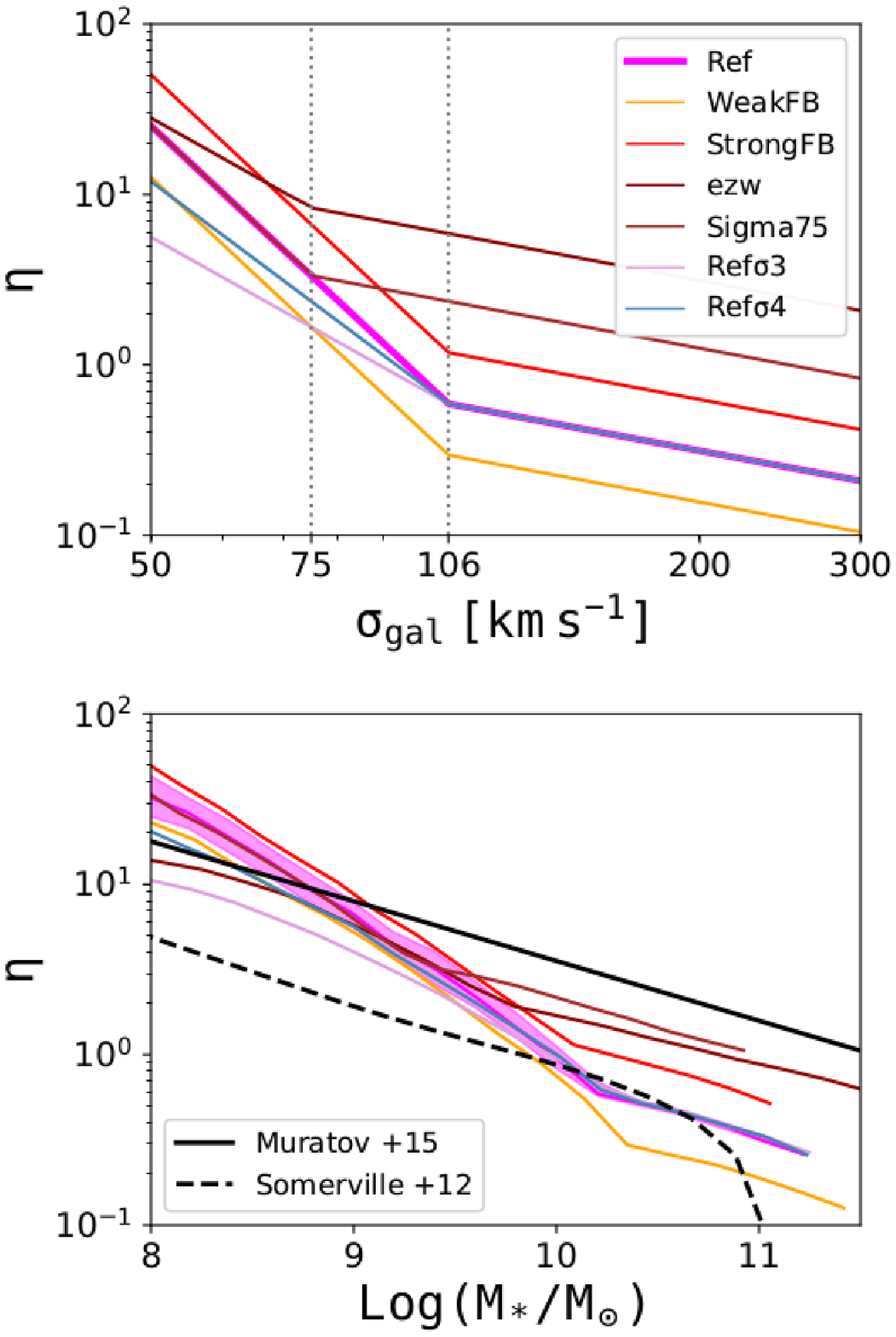}
\caption{\textit{Upper panels}: The mass loading factor $\eta$, as a function
of the halo velocity dispersion $\siggal$ at $z=2$. In the simulations,
We calculate $\siggal$ from the mass of each halo identified by the
on-the-fly FoF group finder. The scalings follow \eqn{eqn:eta_ezw} for the \ezw
winds and \eqn{eqn:eta_w} for the new wind algorithm. \textit{Lower panels}:
$\eta$ as a function of stellar mass $M_*$. The shaded area traces the median
and includes 68\% of galaxies within each mass bin for the Ref simulation.
Different simulations are
colour coded according to \tab{tab:simulations}. We also show the analytic 
formula
from \citetalias{muratov15} (black solid line) and \citet{somerville12} (black
dashed line) for comparison.}
\label{fig:massloadings}
\end{figure}

To illustrate how differences in the mass loading factors affect the
simulations, we show in the upper panel of \fig{fig:massloadings} the input
scaling laws (\eqn{eqn:eta_w}) between the mass loading factor $\eta$ and the
velocity dispersion $\sigma$ measured from the on-the-fly FoF group finder. In the
lower panel we show the actual mass loading factor of individual galaxies in
each simulation as a function of their stellar mass at $z=2$. For comparison,
we also show the empirical fit from the FIRE simulations \citepalias{muratov15}, and
a formula used in the \citet{somerville12} semi-analytic model\footnote{Since
\citet{somerville12} parametrises $\eta$ as a function of halo mass, we used
the SMHM relation at $z=2$ from \citet{behroozi13} to obtain the halo mass from
the stellar mass for any galaxy.}. When making these comparisons remember,
however, that the $\eta$ referred to in the simulations is at wind launch inside
the galaxy while the $\eta$ in the FIRE simulations are measured well outside
the galaxy ($R_{25}$).


\subsection{Wind speed at $R_{25}$}
\label{sec:v25vc}
A major update to our fiducial simulation from our original \textit{ezw} model
\citep{dave13} is the re-adjustment of the initial wind velocity so that
we approximately have $v_w\propto \sigma$ outside the galaxy as opposed to
at wind launch, in better correspondence with the \citet{murray05} model. In this section we will
characterise how this modification changes the behaviour of winds as they
propagate into the halo after they have been launched. We will compare
the wind behaviour to that predicted from the FIRE simulations, which follow the
evolution of wind particles with more detailed physics and at higher resolution.

Using these zoom-in simulations, \citetalias{muratov15} derived an empirical
relation between the wind speed at $R_{25} \equiv 0.25\ R_{vir}$ and the
circular velocity $\vcir$ of the host halo from which the winds are launched.
They find that the median wind speed could be fit with
\begin{equation}
\label{eqn:m15_v25_50}
  v_{w,50} = 0.85 \vcir^{1.1}
\end{equation}
and that the upper 95th percentile wind speed is fit with
\begin{equation}
\label{eqn:m15_v25_95}
  v_{w,95} = 1.9 \vcir^{1.1}
\end{equation}
They obtain these relations from their data at high and medium redshifts but do
not find apparent redshift evolution of these relations. This is very close
to the $\vcir^{1.0}$ in the \citet{murray05} model
and our scaling at wind launch.

\begin{figure*} 
\centering
\includegraphics[width=1.90\columnwidth]{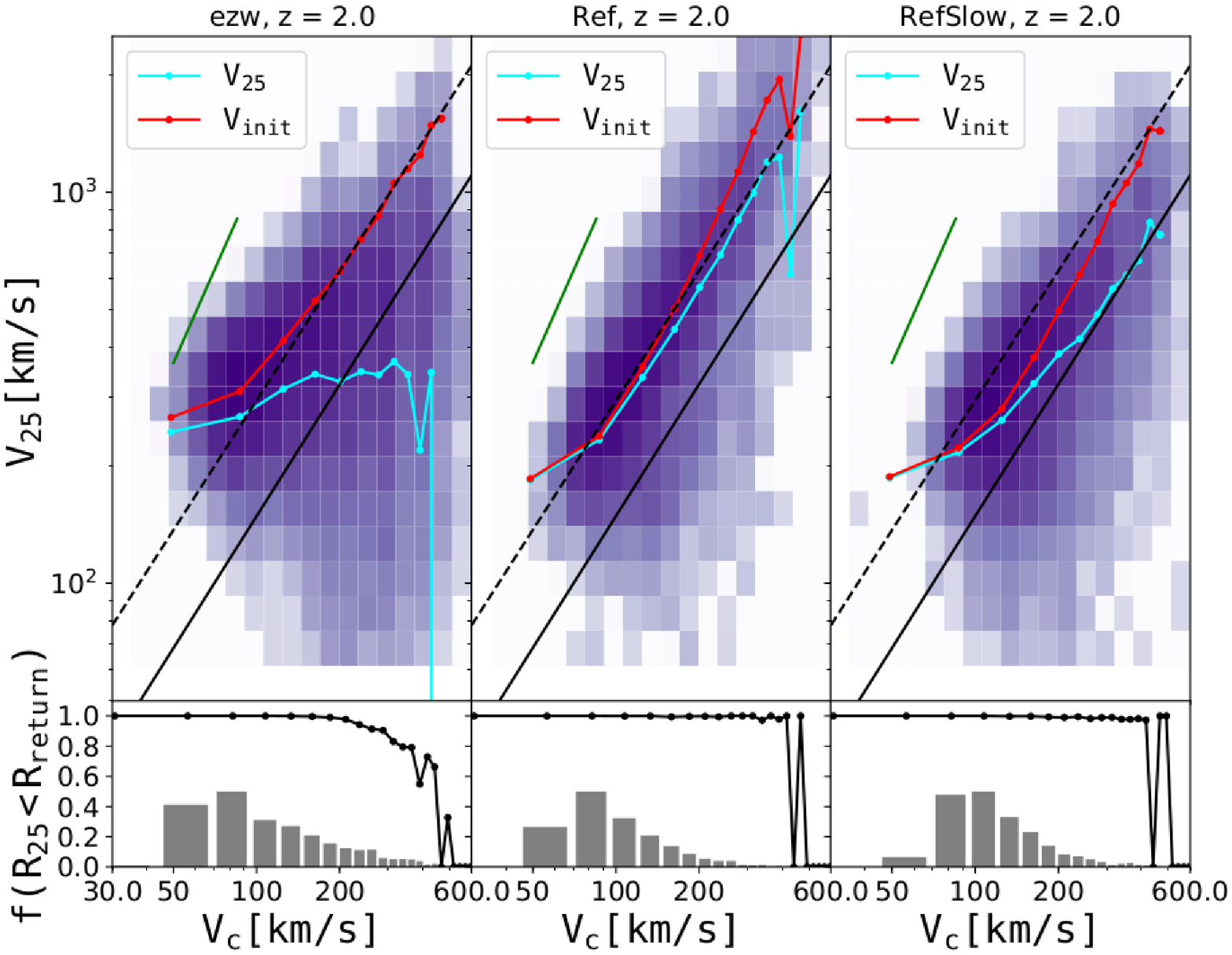}
\caption{
\textit{Upper panels}: The relation between the velocity of wind
particles and the circular velocity $\vcir$ of their host galaxy. These wind
particles are launched within a small redshift window at $z = 2$. In each
panel, the red line shows the running medians of the initial launch velocities
and the cyan line shows $v_{25}$ - the velocities of the wind particles when
they reach 0.25 the virial radius ($R_{25}$). The green segment indicates the 
$v_w \propto \sigma^{1.6}$ scaling imposed at launch for the fiducial simulation.
The colour map shows the distribution
of wind particles according to their $v_{25}$ and $\vcir$. The colour scale
indicates the number of wind particles in each cell. We also include the
empirical fit from the FIRE simulations \citepalias{muratov15}. The black solid
and dashed lines in each panel correspond to their 50th and 95th percentiles,
respectively. \textit{Lower panels}: In each panel, the black line shows the
fraction of wind particles that reach $R_{25}$ in their
host halo. The grey histogram shows the unweighted distribution of $\vcir$ for
all wind particles. The three panels from \textit{left} to \textit{right} are
plotted for the ezw, Ref and RefSlow simulations,
respectively. Agreement between the cyan and black solid
lines thus represents agreement between our wind launch prescription and the
median FIRE results, but agreement between the cyan solid and black dashed lines
may be a better measure for reasons described in the text.
The new wind algorithm in our fiducial simulation is able to
reproduce the trend seen in the FIRE simulations. However, winds in the
original \ezw simulations are in general too slow, particularly in massive
galaxies, where only a small fraction of wind particles is able to reach
$R_{25}$ before falling back.}
\label{fig:v25vc}
\end{figure*}

However, we want to compare with our winds not at wind launch but at $R_{25}$,
where they are measured in \citetalias{muratov15}.
To measure the wind speed at $R_{25}$ in our simulations at a given redshift,
we track the evolution of all wind particles that are ejected within a small
redshift bin centred at that redshift. We identify the host haloes of all
these wind particles from the halo catalogue generated with the post-processing
halo finder \textsc{so} \citep{keres05}.  Then for each wind particle we
define the wind velocity at $R_{25}$,$v_{25}$, as the radial velocity of
the particle when it first crosses the $R_{25}$ of its host halo.

In the left panel of \fig{fig:v25vc}, we compare the speed of wind particles in
our original \ezw model (ezw) to the empirical relations derived from
the FIRE simulations \citepalias{muratov15} measured at $R_{25}$.
The \ezw winds (cyan line) are much slower than in \citetalias{muratov15} in massive galaxies. 
In fact, most wind particles launched in galaxies above a certain $\vcir$ in the ezw
simulation do not have sufficient initial momentum to ever reach $R_{25}$ at
all. This is more clearly illustrated in the bottom panels, which show as solid
lines the fraction of wind particles that reach $R_{25}$. In
the massive galaxies in the ezw simulation, the winds fall back onto their
host galaxies within a very short timescale and, therefore, play little role in
regulating the star formation of their host galaxies. 
Even though the winds were launched with $v_w\propto \vcir$ (red line)
their velocities are almost independent of $\rvir$ at $R_{25}$ (cyan line).
It was this realisation that originally
motivated us to reexamine our wind model.

In the middle panel of \fig{fig:v25vc}, we make the same plot for 
our fiducial simulation (Ref).
Now the distribution of wind particles from our simulation roughly agrees with
the empirical scaling relations from the FIRE simulations \citepalias{muratov15}.
In detail the median velocities of
our winds (cyan lines) are higher than their medians but are typically lower
than their upper 95th percentile values. 
Most of the wind particles launched using the new algorithm are now
able to reach $R_{25}$, even in the most massive systems. 

There are caveats when directly comparing the median velocities between
our simulations and FIRE. First, the nature of our winds differs from
theirs.  In \citetalias{muratov15}, the winds are explicitly accelerated by the
physical
processes that they adopt in their simulations \citep{hopkins12}. Their winds
are multi-phase by nature but they do not distinguish the cold and warm phases
when calculating the wind speed. In contrast, our winds are imposed
on the galaxies and represent only the cold, mass loaded outflow, which
cannot be self-consistently generated from the physics included in our
simulations. Before reaching $R_{25}$ where the wind speeds are measured
and compared, the wind particles in our simulations have been slowing down
owing to hydrodynamic interactions and gravity, while in their simulations the
wind particles keep being accelerated by radiation pressure and it is not clear
whether or not they have started to slow down at that radius. Therefore, the
kinematic structure and the evolution of winds in our simulation are likely
very different from theirs. Hence, we do not know if the comparison to FIRE
would be substantially different at a different radius, e.g. $R_{10}$ or $R_{40}$.

Second and more importantly, we measure the wind speeds in fundamentally
different ways. \citetalias{muratov15}
measure the wind speed using the flux-weighted average of all outflowing
particles over a given epoch. This measurement preferentially selects particles
where the outflowing flux is maximal, i.e, when the winds are strongest.
Furthermore, their definition of outflowing particles includes all gas 
particles in the halo at that radius as long as their radial velocities are 
above the halo velocity dispersion, while in our simulations we only include
the actual ejected wind particles in the measurement. Since in lower mass
haloes the typical wind speed is much larger than the random motions of the halo
component, their measurement likely underestimates the speed of the outflowing
material that is actually accelerated from the galaxies. 
In contrast, in larger mass haloes their method could measure winds even for
gas that is roughly in hydrostatic equilibrium within the halo and hence may
overestimate the wind velocities (and $\eta$). Our measurement reflects the
speed of the fastest outflowing particles within each halo and, therefore,
should be
more comparable to their 95th percentile values. In fact, if we try to measure
our winds in a way more similar to that in in \citetalias{muratov15}, it lowers
our median wind velocities to agree better with their median values.

Unlike \citetalias{muratov15}'s finding that the $v_{25} - \vcir$ relation is independent of
redshift, in our simulations the $v_{25}$ slightly decreases for a given
$\vcir$ at lower redshifts, especially in massive haloes, even though we launch
our winds with a redshift-independent initial velocity (\eqn{eqn:vwind}). The
winds lose more momentum as they move from the launching radius to $R_{25}$ at
low redshifts, likely owing to the combined effects of a deeper potential,
enhanced
hydrodynamic forces, and an underestimate of $\sigma$ for massive haloes using
the on-the-fly FoF group finder.

The significant differences between the wind behaviours in our simulations
are particularly interesting considering that the initial wind velocities are
not very different from each other. A relatively small difference in the
initial wind speed at launch has a significant impact on the wind
behaviour at larger radii. This indicates that this kinetic wind algorithm, 
which is adopted in many cosmological simulations
\citep[e.g.][]{springel03, oppenheimer06, agertz13, schaye15, mufasa, pillepich18a, fire2}
as a sub-grid model for stellar
feedback, is very sensitive to the details of its implementation and the
choice of wind parameters. We will discuss this sensitivity more in
\sect{sec:sensitivity}.

%
%

\section{Sensitivity to the wind model}
\label{sec:sensitivity}
In this section, we explore the scaling laws that determine the mass
loading factor and the wind launch speed in our new wind algorithm, and
study the sensitivity of our simulations to these parameters. All the test
simulations we use in this section 
are listed in \tab{tab:simulations} and all have the same numerical
resolution. We will explore the effects of numerical resolution in
\sect{sec:fiducial_simulation}.

We identify galaxies using \skid and \so as in \citet{huang19} and 
measure the stellar mass and halo mass for every galaxy that we identify. 
First, we will focus on comparing the galactic stellar mass functions at four
different redshifts and discuss the effects of changing the mass loading
factor and the wind speed separately. In addition, we will also examine the
growth of individual galaxies and how they differ among the simulations,
since their different star formation histories are an immediate consequence
of the wind algorithm, which controls their gas supply. To make direct
comparisons between individual galaxies, we cross-match galaxies from different
simulations to those in the Ref simulation based on their phase-space
information. We also require matched galaxies to have a stellar mass difference
smaller than 1 dex to avoid matching satellite galaxies to centrals. We define
the differences of stellar masses between matched galaxies as $\Delta\log(M_*)
\equiv \log(M_{*}) - \log(M_{*,f})$, where $M_{*}$ is the stellar mass of a
galaxy in a simulation and $M_{*,f}$ is the stellar mass of its matched galaxy
in the fiducial simulation.

Finally, we will look at how the cosmic mean stellar density evolves with time in
each test simulation. As an integrated quantity, the cosmic stellar density at
different redshifts indicates whether or not a simulation produces the right
amount of stars in total.

\subsection{The Galactic Stellar Mass Functions}
The GSMFs are one of the most robust measurements from observations and have
been used as an important constraint for calibrating sub-grid models in
cosmological simulations. To compare our simulated GSMFs with observations, we
use results from multiple large galactic surveys. All of these observations
assume a \citet{chabrier03} IMF for their stellar mass estimate as in our
simulation. Different measurements of the $z = 0$ GSMFs \citep[e.g.][]{li09,
baldry12, bernardi13, moustakas13} generally agree at the faint end up to
$\logMstar \sim 10.5$ and start to deviate slightly from each other at higher
masses. Both the choice of the aperture \citep{bernardi13} and the choice of
the stellar population synthesis template \citep[e.g.][]{mitchell13, tomczak14}
contribute to relatively large uncertainties in stellar masses at the massive end.
\citet{conroy09} estimated that the systematic error on stellar masses ranges
from 0.3 dex at $z \sim 0$ to 0.6 dex at $z \sim 2$. Our choices of the
observed GSMFs at $z = 0$ reflect these uncertainties. The \citet{baldry12}
result is based on single-Sersic fits to the light profiles of galaxies at $z <
0.06$ from the Galaxy And Mass Assembly (GAMA) survey, while \citet{bernardi13}
use a Sersic-bulge + exponential disc model that results in larger stellar
masses at the bright end. For $z = 1$ and $z = 2$ GSMFs, we use the data from
\citet{tomczak14}, who compiled GSMFs over a broad redshift range $0.2 < z <
3$ using deep observations from the FourStar Galaxy Evolution Survey (ZFOURGE)
and the Cosmic Assembly Near-infrared Deep Extragalactic Legacy Survey
(CANDELS), obtaining a completeness limit of $\logMstar \sim 9.5$. We use
the \citet{song16} results from the CANDELS survey for comparison at $z = 4$.

First, we will show how the GSMFs are sensitive to the mass loading factor. The
mass loading factor $\eta$ is controlled by three parameters (\eqn{eqn:eta_w}):
a normalisation factor $\alpha_\eta$, a power law index $\beta_\eta$ that
determines how it scales with $\sigfof$, which is the velocity dispersion of
the host halo identified from the on-the-fly group finder, and the characteristic
velocity $\sigma_{ezw}$ above which the scaling with $\sigfof$ changes from
$\eta \propto \sigma^{-(1+\beta_\eta)}$ to $\eta \propto \sigma^{-1}$.

All the simulations that we use for this comparison are listed in
\tab{tab:simulations} and use the same
parameters for the wind speed as the fiducial simulation, but have different
parameters for $\eta$. Here we describe the features of each simulation using
the low resolution fiducial simulation Ref as a reference. StrongFB increases the
overall mass loading by a factor of 2 and WeakFB reduces it by a factor
of 2. Sigma75 uses a smaller $\sigma_{ezw}$ of 75 $\kms$ than the
fiducial 106 $\kms$, but also renormalises $\alpha_\eta$
so that it has the same scaling
with $\sigfof$ for small haloes below $\sigma_{ezw}$. Ref$\sigma$4 and
Ref$\sigma$3 use shallower scalings with $\sigfof$ for haloes smaller than the
characteristic $\sigma_{\ezw}$, with a power index parameter $\beta_\eta$
equal to 3 and 2, respectively, instead of the fiducial value of 4. 
ezwFast has the same mass loading as ezw but has the wind launch speed scalings
of Ref, which produce faster winds. 
The bottom panel of \fig{fig:massloadings} shows
the relation between $\eta$ and $\sigfof$ for these simulations.

\subsubsection{Dependence on $\beta_\eta$ and $\sigma_{ezw}$}
\label{sec:beta_dependence}
\begin{figure*} 
\centering
\includegraphics[width=1.90\columnwidth]{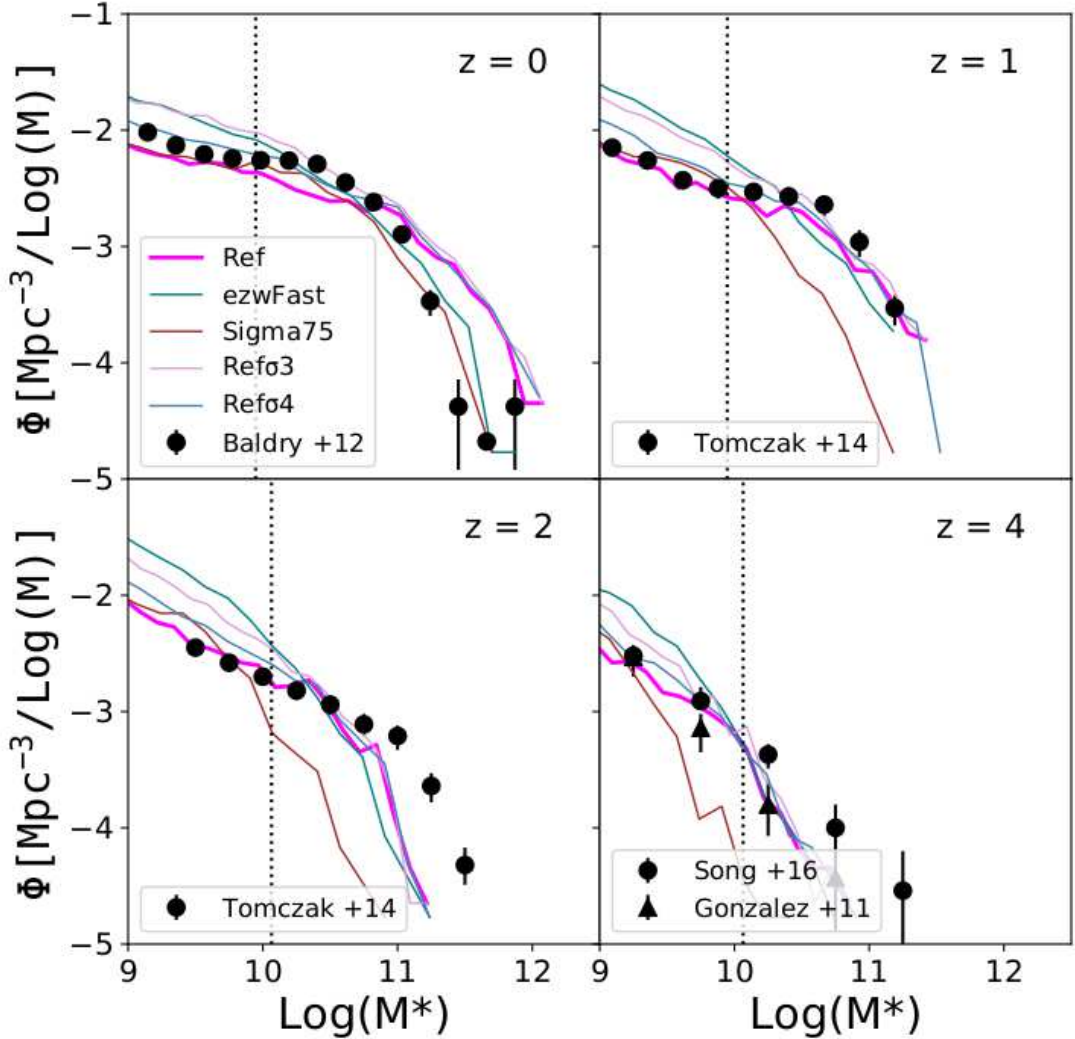}
\caption{The galactic stellar mass function at different redshifts. At each
redshift, we compare the GSMFs from a set of test simulations, which are
shown in different colours according to the colour scheme defined in
\tab{tab:simulations}. The dotted vertical line in each panel indicates the
resolution limit for galaxies corresponding to a total mass of 128 SPH
particles in these low-resolution simulations. The observational data for these
redshifts are described in the text.}
\label{fig:gsmfs_calib_etas} 
\end{figure*}

\fig{fig:gsmfs_calib_etas} shows how changing the power index $\beta_\eta$ and
the characteristic velocity $\sigma_{ezw}$ affects GSMFs at different times. Not surprisingly, the
faint end of the GSMFs is particularly sensitive to $\beta_\eta$. Since all
these simulations except for ezwFast have the same $\eta_w - \sigfof$
relation above $\sigma_{ezw}$, a higher $\beta_\eta$ means more mass in outflows from
smaller haloes and less star formation. Ref and
Sigma75 are the only simulations that successfully reproduce the
observed faint end at all redshifts, and both have a strong scaling with $\eta
\propto \sigfof^{-5}$ ($\beta_\eta = 4$). Simulations with a shallower
dependence on $\sigfof$ tend to produce more low mass galaxies at $z > 1$,
creating a stronger tension with the observational data. For example, the
ezwFast simulation shows that the shallow scaling $\eta_w \propto
\sigfof^{-2}$ predicted from the analytic momentum/energy-driven model results
in too many faint galaxies formed at high redshifts.

The above result shows that a steep scaling between the mass loading factor and
the circular velocity, or equivalently the halo mass, is essential to explain
the flat faint end of GSMFs at $z = 1$ and $z = 2$ when one uses a kinetic
feedback model such as ours. This requirement for a strong dependence 
between $\eta$ and $\sigma$ has also been recently found in other work 
\footnote{In other work, the mass loading factor is often correlated with 
either the halo mass
$\mh$ or a characteristic velocity that scales with $\mh^{1/3}$, though the
specific definitions for the mass and the velocity are slightly different. For
consistency, we will use $\mh$ for halo mass and $\sigma$ for the
characteristic velocity.}. In the IllustrisTNG simulations, \citet{pillepich18b}
reported a scaling with $\eta \propto \mh^{-1} \propto \sigma^{-3}$ (see their
Figure 7) as their fiducial choice to fit a wide range of observables. The FIRE
simulations generate galactic outflows self-consistently instead of using a
simple scaling law and found $\eta \propto \sigma^{-3.3}$ \citepalias{muratov15}.
Semi-analytic studies (SAM) also require steep scalings to fit the GSMFs at
different redshifts. \citet{lu14} reported a rather steep scaling with
$\eta \propto \sigma^{-6}$. \citet{somerville12} parametrise their mass loading
as $\eta \propto \sigma^{-\beta_{LD}}[1+\sigma^{\beta_{EJ}}]^{-1}$ with
fiducial parameters $\beta_{LD} = 2.25$ and $\beta_{EJ} = 6$. This scaling,
also shown in \fig{fig:massloadings}, is similar in form to our scalings, though the
normalisation is lower. \citet{peeples11} also find with their chemical evolution model
that very steep mass loading scalings ($\eta \propto \vcir^{-3}$ or steeper)
are required to explain the steep slope of the observed MZR at $z=0$. Even
though the specific wind models used in these other works have important 
differences, such as whether or not they add additional heating, it is clear
that the efficient
suppression of star formation in less massive galaxies requires stronger winds
than those predicted from the energy-driven model ($\eta \propto \sigma^{-2}$)
or the momentum-driven model ($\eta \propto \sigma^{-1}$), which were
previously assumed in many cosmological simulations. Note that the GSMF at $z =
0$ alone is insufficient to differentiate between these different scalings. Accurate
measurement of the stellar content at higher redshifts could, therefore, place
important constraints on the wind models.

For more massive galaxies, $\beta_\eta$ has a less significant effect than the
parameter $\sigma_{ezw}$, which determines the mass scale where the steep
scaling $\eta \propto \sigma^{-(1+\beta_\eta)}$ changes to the momentum-driven
wind scaling $\eta \propto \sigma^{-1}$. The Ref, Ref$\sigma$4, and
Ref$\sigma$3 simulations have indistinguishable GSMFs above $\logMstar = 10.5$
in spite of their different $\beta_\eta$ values. In contrast, the GSMFs from
the ezwFast and the Sigma75 simulations start showing clear
differences at the massive end from the other three simulations as early as $z
= 2$, indicating that the growth of massive galaxies is very sensitive to the
choice of $\sigma_{ezw}$. For example, \fig{fig:massloadings} shows that the
mass loading factor in the Sigma75 simulation is the same as that in the
fiducial simulation at $\sigfof < 75$ $\kms$, but it becomes larger by a steadily
increasing factor for $\sigma_{ezw} > 75\kms$ and is a factor of $\sim 4$ higher
in all haloes with $\sigma_{ezw} \ge 106\kms$, our fiducial value of $\sigma_{ezw}$.
As a result, the growth of massive galaxies in the Sigma75 simulation
is significantly suppressed at all redshifts.

Interestingly, the ezwFast and Sigma75 simulations have the best
overall agreement with the observed $z = 0$ GSMF among all these
test simulations. However, they significantly
under-produce the number of massive galaxies at higher redshifts. Some of the
other simulations, including the fiducial simulation, agree with observations
better at higher redshifts at the cost of a slight excess of the most massive
galaxies at $z = 0$. We find that matching the massive end of the GSMFs at both
$z = 0$ and $z = 2$ simultaneously to be very challenging within our current
framework for feedback. A successful feedback model must allow a rapid build-up
of massive galaxies at $z = 2$ but also must account for the slow evolution of
the massive end from $z = 0$ to $z = 2$ as suggested by observations. The
success of our Ref model at $z \ge 1$ but failure at $z = 0$ suggests that an
additional mechanism such as AGN feedback suppresses the growth of massive
galaxies at low redshift, or else that the galaxy scalings of stellar feedback change
sharply at $z < 1$.

\subsubsection{Dependence on the normalisation $\alpha_\eta$}
\begin{figure*} 
\centering
\includegraphics[width=1.90\columnwidth]{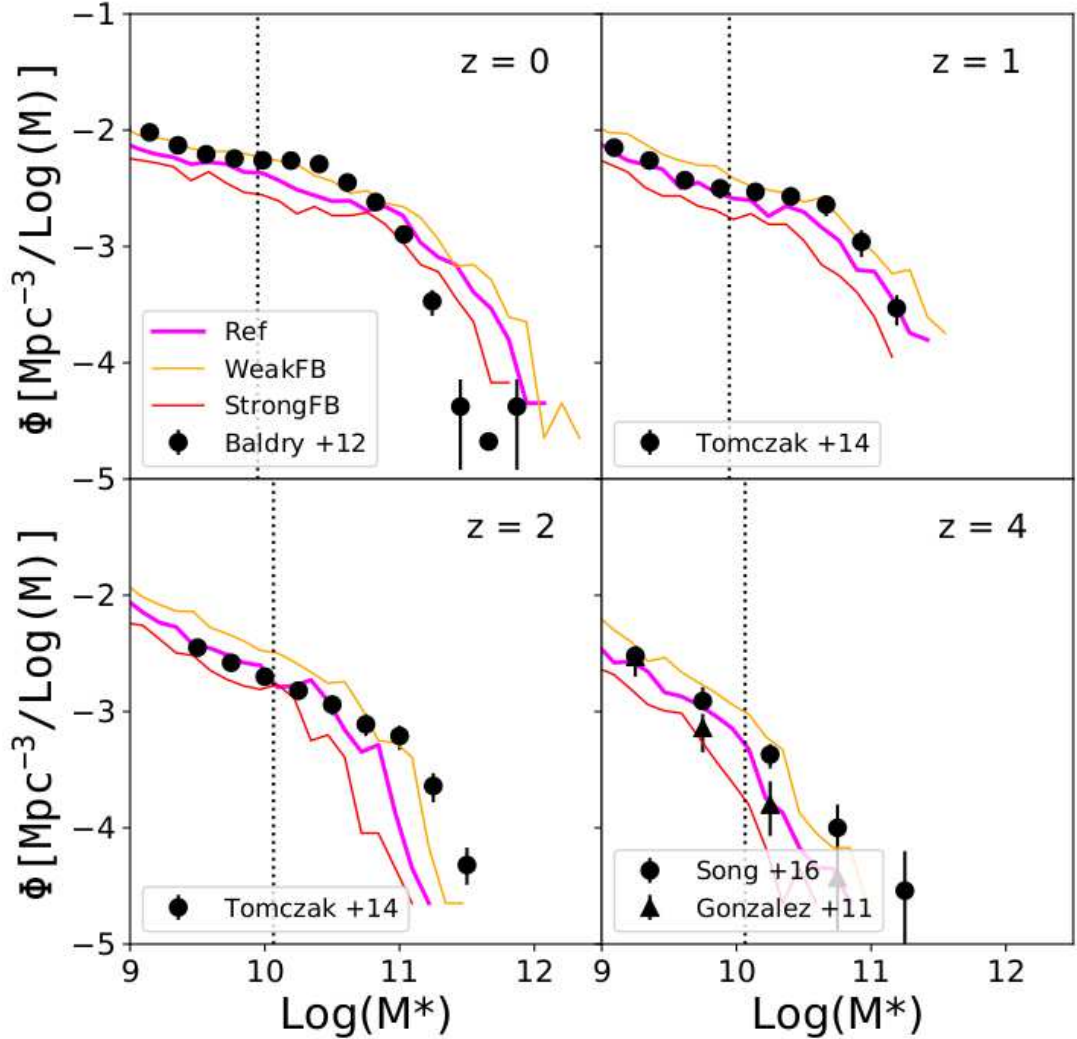}
\caption{Same as \fig{fig:gsmfs_calib_etas}, except that here we focus on the
effect of
the linear factor $\alpha_\eta$ of the mass loading factor. The $\eta$ in these
simulations are different by a factor of 2 so that for the same galaxy,
the StrongFB model (red) produces $2\times$ more massive winds than the fiducial
Ref model (magenta), while the StrongFB model (orange) produces
$2\times$ less massive winds.}
\label{fig:gsmfs_calib_linfac} 
\end{figure*}

\fig{fig:gsmfs_calib_linfac} shows the effects of changing the linear
normalisation factor $\alpha_\eta$ by comparing three simulations with
$\alpha_\eta$ incrementally varying by a factor of 2. In general, a higher mass
loading normalisation results in less stars being formed because more cold gas is ejected in
galactic winds. To a rough approximation, the GSMFs of the WeakFB and
StrongFB simulations are offset horizontally by a factor of $2 - 4$ at all redshifts,
with the Ref simulation mid-way between them.

\begin{figure*} 
\centering
\includegraphics[width=1.90\columnwidth]{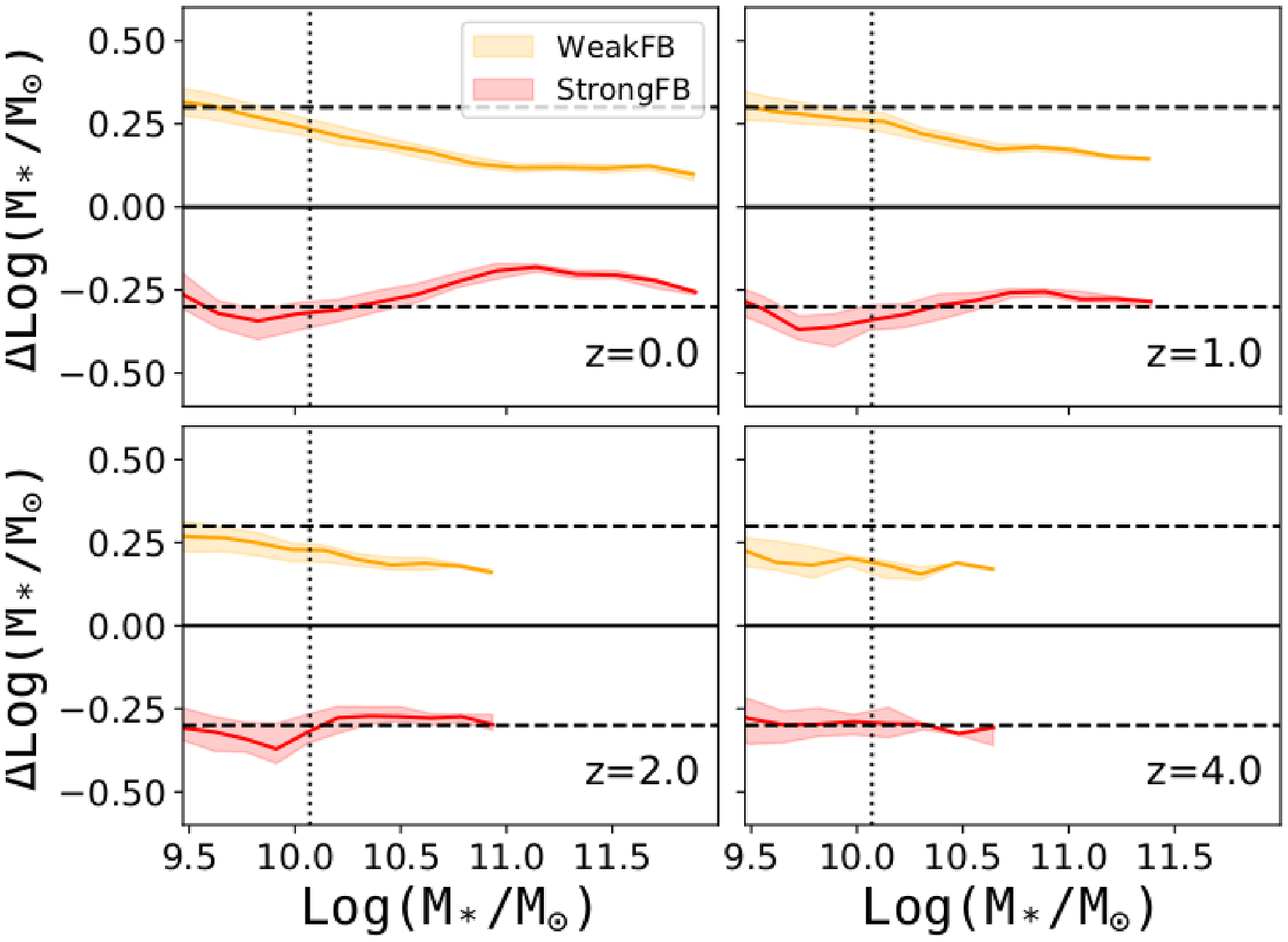}
\caption{The stellar mass differences between galaxies that are cross matched
between the different simulations and the Ref simulation at the given redshifts.
The shaded region indicates the $1\sigma$ scatter in each $M_*$ bin.
In each panel, the
dotted vertical lines indicate the resolution limit of galaxies corresponding to
a total mass of 128 SPH particles. The dashed horizontal lines indicate the
offset in stellar masses
($\pm 0.3$ dex) predicted by \eqn{eqn:simple_model_prediction}.}
\label{fig:pairs_mstar_linfac}
\end{figure*}

\fig{fig:pairs_mstar_linfac} compares the stellar mass differences between
individual, matched galaxies from these simulations. We use the fiducial
simulation as the reference so that all galaxies from that simulation lie on
the black horizontal line. For each galaxy in the fiducial simulation, we find
its matched galaxy in the other two simulations and calculate the stellar mass
differences. The medians are shown as coloured lines.

Using a simple analytic model we could predict the stellar mass of an isolated
galaxy whose growth is governed by gas outflow, star formation, and the 
inflow of both pristine gas and recycled winds. The equilibrium condition is
\begin{equation}
  \label{eqn:equilibrium_model}
\dot{M}_{in} + \dot{M}_{rec} = \dot{M}_{out} + \dot{M}_{*}
\end{equation}

This equilibrium equation is typically a good approximation in hydrodynamic simulations
\citep{finlator08}. Assuming that a fraction $\frec$ of the outflow recycles, so $\dot{M}_{rec} \approx \frec\dot{M}_{out}$, the final stellar mass of the galaxy would be:
\begin{equation}
M_* = M_{in}\left( \frac{1}{1+\eta} + \frec\frac{\eta}{(1+\eta)^2}\right),
\end{equation}
where $M_{in}$ is the time-integrated mass of unrecycled gas that accretes onto
the galaxy. This derivation assumes a constant $\eta$ and $\frec$ for the 
galaxy.
However, since both of these are functions of halo mass, this assumption breaks
down for galaxies that undergo a strong evolution in their halo mass such as
in a major merger.

Since two matched galaxies in different simulations have similar assembly histories,
gravitational potential, and outflow speeds, their $M_{in}$ and $\frec$ should remain
approximately the same, provided that the outflows do not themselves
disrupt accretion
or recycling. The ratio of final stellar masses between two simulations is thus
determined by the different $\eta$:
\begin{equation}
  \label{eqn:simple_model_prediction_full}
\frac{M_{*,1}}{M_{*,2}} = \left[ \frac{1+(1+\frec)\eta_2}{1+(1+\frec)\eta_1}
\right].
\end{equation}

For low mass galaxies, we can assume $\eta >> 1$ and $\frec \sim 0$ because
winds can easily escape from their shallow gravitational potential. Hence,
the ratio above will asymptotically approach:
\begin{equation}
  \label{eqn:simple_model_prediction}
\left( \frac{M_{*,1}}{M_{*,2}}\right)_{\eta \gg 1} = \frac{\eta_2}{\eta_1}.
\end{equation}
Therefore, this simple model predicts that the stellar mass of small galaxies
in the WeakFB and StrongFB simulations should differ from their
corresponding galaxies in the fiducial simulation also by a factor of 2 (0.3
dex). \fig{fig:pairs_mstar_linfac} shows that this prediction agrees with our
simulations reasonably well. 
It works almost perfectly when we increase $\eta$
by a factor of two. When we decrease $\eta$ by the same factor, it does not
work as well and the error increases with galaxy mass. This is because the approximation
that $\eta \gg 1$ is less robust with decreasing $\eta$.
For example, \fig{fig:massloadings} shows that $\eta \sim 1$ at
$\logMstar \sim 10$ in the WeakFB simulation.
In fact, in the opposite limit, i.e., $\eta \ll 1$, \eqn{eqn:simple_model_prediction_full}
predicts that $M_{*,1} = M_{*,2}$, consistent qualitatively with the convergence of
curves at high mass in \fig{fig:pairs_mstar_linfac}.

The degree to which this toy model works is perhaps surprising, since it makes many
over-simplified assumptions.
First, the equilibrium equation (\eqn{eqn:equilibrium_model}) assumes that the total
amount of cold gas in galaxies does not change with time while in the
simulations this is not guaranteed. Second, it treats galaxies in isolation,
neglecting any interactions with other galaxies, but
in reality a significant fraction of gas accretion may have come from previous
outflows from other galaxies \citep[e.g.][]{ford14, angles-alcazar17}. Third,
it assumes that any outflow will have no subsequent effect on the galaxy
except through a nearly constant fraction of immediate wind recycling, but the
outflowing gas may interact with the surrounding gas and thus affect further
gas accretion. The success of the toy model
suggests that these concerns are likely not dominant in our cosmological
simulations.

\subsubsection{Wind Speed}
\begin{figure*} 
\centering
\includegraphics[width=1.90\columnwidth]{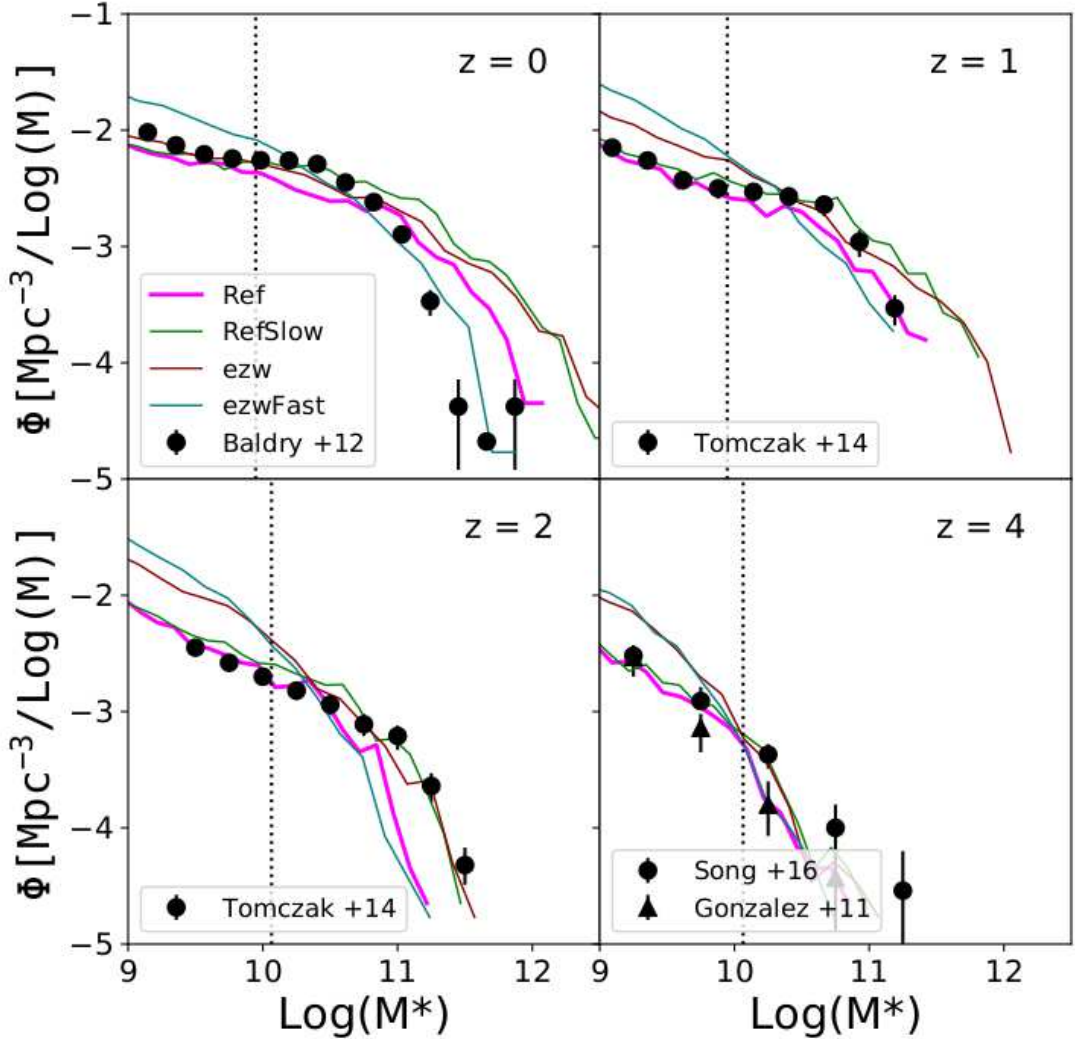}
\caption{Same as \fig{fig:gsmfs_calib_etas}, except that here we focus on the 
effect of different initial wind speeds on the GSMFs.
The fiducial Ref (magenta)
and the RefSlow (green) simulations use the new wind launch algorithm and
are only different in $v_w$. The other two simulations use the original \ezw
wind model. However, the ezwFast (teal) simulation has fast winds
as in the fiducial simulation while the ezw (dark red) simulation uses
the \ezw wind speed.}
\label{fig:gsmfs_calib_vwind}
\end{figure*}

In this section we will show how the GSMFs are sensitive to the initial wind speed.
The initial wind speed at launch depends on two parameters (\eqn{eqn:vwind}): a
normalisation factor $\alpha_v$ and a power law index $\beta_v$ that determines
how wind speed scales with $\sigfof$. The fiducial simulation adopts $\alpha_v
= 3.5$ and $\beta_v = 0.6$. These values are tuned to match the $v_{25}$ -
$v_c$ scaling from the FIRE simulation \citepalias{muratov15} at $z = 2$. The ezw simulation
uses the original \ezw formula for the wind speed (\eqn{eqn:vwind_ezw}). The
ezwFast simulation uses the fiducial wind speed scaling but the mass loading 
of the ezw simulation. The RefSlow simulation uses a
slightly shallower slope $\beta_v = 0.5$ than the fiducial Ref
simulation. Therefore, we will demonstrate the effects of wind speed by
comparing the ezw and the ezwFast simulations under the original
$\eta$ formula, and comparing the Ref and the RefSlow simulations
under the fiducial $\eta$ formula.

\fig{fig:gsmfs_calib_vwind} shows that the massive end of the GSMFs is very
sensitive to the initial wind speed. First compare the two \ezw simulations. At
$z = 4$, the GSMFs are still very similar, because even the slower winds are
above the escape velocities of haloes at this redshift. But the massive ends of
the GSMFs start to show clear differences after $z = 2$. As \fig{fig:v25vc} has
shown, with the original \ezw wind speed a significant fraction of wind
particles fall back towards the galaxy before reaching $R_{25}$ and
become star forming again very
soon after being launched. The new wind speed in the ezwFast allows wind
particles to travel much further, and they return much later,
if at all. This reduces the amount of stars formed in intermediate mass haloes
and at least delays stellar growth in massive haloes. As a result, galaxies in
the ezwFast simulation are less massive,
with the mass difference increasing towards more massive systems. 

Now compare
the fiducial simulation to the RefSlow simulation. The only difference
between them is that the wind speed in the fiducial simulation increases
slightly faster with halo mass. Even in the most massive haloes, the
difference in wind speed is only a factor of 2. However, the massive galaxies in the RefSlow
simulations are much more massive than those in the Ref simulation. Even more
strikingly, the massive ends of the Ref and ezwFast simulations are quite close,
even though their mass loading factors at $\sigma > 106\kms$ differ by a factor of
$\sim 10$. This similarity shows that the wind speed (matched between these simulations) is a crucial governor of high mass galaxy growth, probably
because of its impact on recycling rates.

\subsection{Stellar Density Evolution}
\begin{figure} 
\centering
\includegraphics[width=0.96\columnwidth]{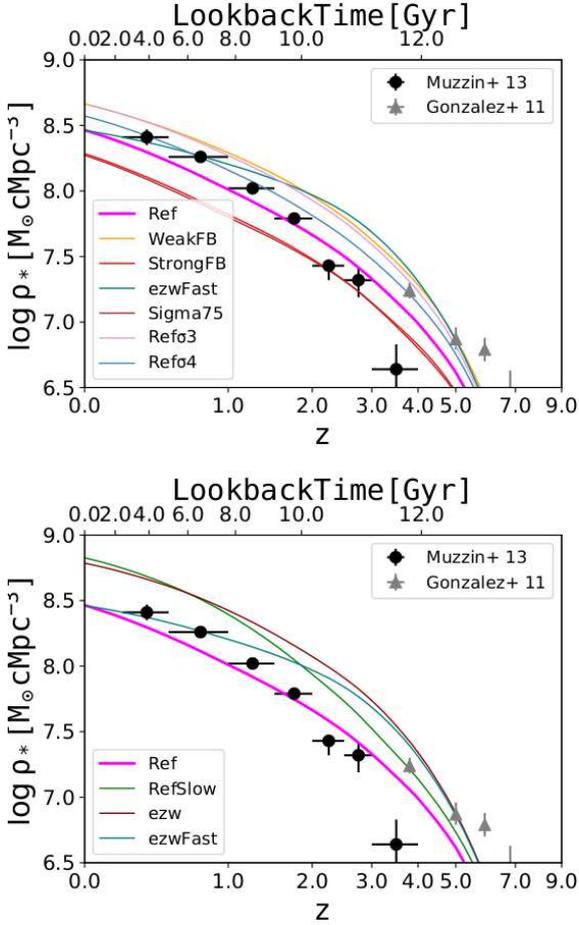}
\caption{The evolution of the comoving stellar mass density with redshift from
the test simulations. We use results from \citet{muzzin13} and
\citet{gonzalez11} as observational constraints, though other measurements in
the literature agree with each other in general. The colour scheme for the
different lines is defined in \tab{tab:simulations}. Only the fiducial
Ref and the Ref$\sigma$4 simulations agree with the observations.
The other simulations either over-produce stars at high redshifts or fail to
match the evolution at low redshifts. The \textit{upper} panel compares 
simulations with different mass loading factors. The \textit{lower} panel 
compares simulations with different wind speeds.}
\label{fig:sde_calib}
\end{figure}

In \fig{fig:sde_calib} we show the stellar density evolution (SDE) as a
function of redshift. This has been measured observationally in many studies
\citep{li09, gonzalez11, baldry12, ilbert13, muzzin13, moustakas13, tomczak14}.
Most of these measurements agree well within 0.1 dex at redshifts below $z =
2$, but the differences become larger at higher redshifts. Here we use the data
from \citet{muzzin13} for redshifts $0$ to $3$ and
\citet{gonzalez11} for higher redshifts. The \citet{muzzin13} sample has a mass
completeness limit of $\log(M_*) = 10.76$ at $z = 2.5 \sim 3.0$ and $\log(M_*)
= 10.94$ at $z = 3.0$ to $4.0$ and hence the data have to be extrapolated to
estimate the stellar mass densities at these redshifts. The \citet{gonzalez11}
data can be interpreted as upper limits since they did not correct for nebular
emission when using the UV data to derive the stellar densities
\citep{smit14}.

Only the Ref$\sigma$4 simulation and the fiducial simulation are consistent with the observations at all redshifts. These two simulations differ only
in $\beta_\eta$, with the Ref$\sigma$4 simulation
having a shallower $\eta - \sigma$
slope $\beta_\eta = 3$ that launches less winds in low mass
galaxies. The slight excess of low mass galaxies in the Ref$\sigma$4
simulation (\fig{fig:gsmfs_calib_etas}) explains its overall higher stellar densities in
\fig{fig:sde_calib}.

The simulations with a lower $\beta_\eta$, i.e., the ezw, ezwFast
and Ref$\sigma$3 simulations, over-produce low mass galaxies at high
redshift, leading to much higher stellar densities at $z > 2$. This discrepancy supports
the claim in \sect{sec:beta_dependence} that one requires a strong dependence
of the mass loading factor on the halo velocity dispersion.

The lower panel of \fig{fig:sde_calib} shows that the wind speed has a strong
effect on the evolution at lower redshifts. The wind dynamics is more sensitive
to the initial wind speed in massive haloes as shown in \fig{fig:v25vc}. The
fast wind in the Ref and the ezwFast simulations significantly limits the
growth of massive galaxies compared to the RefSlow and the ezw
simulations, resulting in a slower growth of $\rho_*$ at $z < 2$. However, 
ezwFast still over predicts the total stellar mass at $z=0$ because of
too much star formation at earlier redshifts owing to a relatively smaller mass
loading factor in low mass galaxies.

Here we have shown how the stellar content of galaxies in our simulations are
sensitive to wind parameters, namely $\eta$ and $v_w$. To make more robust
comparisons to observations, we will need to further transform our simulation
data into mock observations and take into account various observational effects,
such as corrections for aperture \citep{schaye15, pillepich18b} and
completeness \citep{furlong15}. We did not make these corrections in this work
as these comparisons are not meant to be interpreted as rigorous tests of
our galaxy formation model, but rather to demonstrate the sensitivity of the
numerical results to the wind implementations and their associated
parameters.

\section{DISCUSSION}
\label{sec:discussion}
In the previous section we demonstrated that the properties of galaxies in our
simulations
are sensitive to the sub-grid model for galactic winds. To summarise, first,
the low-mass end slopes of the GSMFs are sensitive to the mass
loading factor $\eta$, especially to the power-law index $\beta_\eta$ that
determines how strongly $\eta$ scales with $\sigma$ in low mass galaxies; second, the stellar masses
of massive galaxies are sensitive to the initial wind launch speed $v_w$. In
this section we study in detail how galaxies build up their stellar masses in
our simulations and how they are affected by the wind algorithm. 

The stellar content in a given halo at any redshift
is closely related to the baryon cycles (inflow/outflow) that it has experienced
over cosmic time. The sub-grid wind algorithm controls outflow in our
simulations while the amount of cosmic accretion through filaments (cold accretion) or cooling
flows from the shocked halo gas (hot accretion) governs the inflow. In addition, galactic winds that were
launched in the past can also fall back onto the galaxies after they lost
their initial momentum. This wind recycling often dominates at low
redshifts \citep{oppenheimer10, angles-alcazar17}. In this section, we will study the accretion history of the gas that ultimately forms
stars. The particle nature of our SPH simulations makes it convenient to track
the evolution of individual gas particles. In \sect{sec:accretion_tracking},
we will describe how we differentiate between
cold and hot primordial accretion and wind recycling through cosmic time.

We will focus on analysing and comparing three
simulations: the fiducial Ref, the RefSlow and the Ref$\sigma$3
simulations.  These simulations show clear differences in the resultant GSMFs,
SDEs and Stellar Mass-Halo Mass functions (SMHMs) at different redshifts,
even though they use the same wind algorithm albeit with different wind
parameters. Using the fiducial simulation as a baseline for
comparison, the RefSlow simulation represents test simulations
that explore the effects of the wind speed, which we will show affects not only
wind recycling but also pristine gas accretion. On the other
hand, the Ref$\sigma$3 simulation represents a test simulation with varying
parameters for the mass loading factor, which directly controls the amount of
outflows from haloes of different masses.

In \sect{sec:discussion_smhm} we will first show how the SMHMs evolve from
$z=4$ to $z=0$ in these simulations. In \sect{sec:discussion_z2}, we will
analyse how the wind algorithms shape the SMHMs at $z=2$ and address the
differences between the simulations. Galaxy evolution at higher redshifts ($z>2$) is much less
complicated than later evolution for lower mass galaxies, since it
involves various important additional processes, such as mergers, 
cold halo - hot halo dichotomy, and group and cluster formation.
They play less significant roles at higher redshifts. However, galaxies at $z=2$
are the building blocks for those at lower redshifts and must also agree with
the observational constraints. In \sect{sec:discussion_z0}, we will focus on the
late evolution after $z=2$ and the formation histories of galaxies at $z=0$. In
\sect{sec:discussion_mergers} we will discuss the importance of mergers to the
assembly of stellar mass, and in \sect{sec:discussion_recycling} we will study
in detail the properties and the effects of wind recycling. Finally in
\sect{sec:discussion_agn} we will discuss what might be missing from our
feedback prescriptions and what might be needed to remove the remaining
discrepancies between the simulations and the observations.

\subsection{Tracking the Accretion History}
\label{sec:accretion_tracking}
To understand how galaxies acquire the gas that ultimately forms their stars, we
track the evolution of individual SPH particles that at some point become star
forming\footnote{See \sect{sec:simulations} for the definition of star-forming
particles in our simulations.}. At each time-step, we track all the
accretion events, i.e., whenever a gas particle changes from non-star-forming to
star-forming at that time-step, and output the properties of the accreted
particle and the galaxy onto which it accretes. To distinguish these accretion
events, we introduce a parameter $T_{max}$ to characterise the thermodynamic
history of the accreted particle as in \citet{keres05} and 
\citet{oppenheimer10}. We
define $T_{max}$ as the maximum temperature the particle ever reaches before
becoming star-forming. We define an accretion event as \textit{hot mode
accretion} if $\log(T_{max}) > 5.5$ or \textit{cold mode accretion} otherwise.
Both of these accretion modes are also referred to as \textit{pristine gas
accretion}. On the other hand, if a particle is launched as a wind and
subsequently re-accretes unto a galaxy, we define this 
accretion as \textit{wind recycling}. Unlike in our previous work, 
we reset $T_{max}$ to 0 once a particle is
launched as a wind so that at the time it recycles, it will have a different
$T_{max}$. In this way we can further divide a wind recycling events into
\textit{hot wind accretion} and \textit{cold wind accretion} based on the same
temperature criteria. In addition, once a gas particle spawns or turns into a
star particle, we associate this star-forming event with
the last accretion event
of that SPH particle. Therefore, for each star particle in the simulation we
can tell when, where, and in which mode its progenitor gas particle accretes.
In the following sections we will study the formation history of galaxies by
looking at their star particles.

\subsection{The Stellar Mass - Halo Mass Functions}
\label{sec:discussion_smhm}
\begin{figure*} 
\centering
\includegraphics[width=1.90\columnwidth]{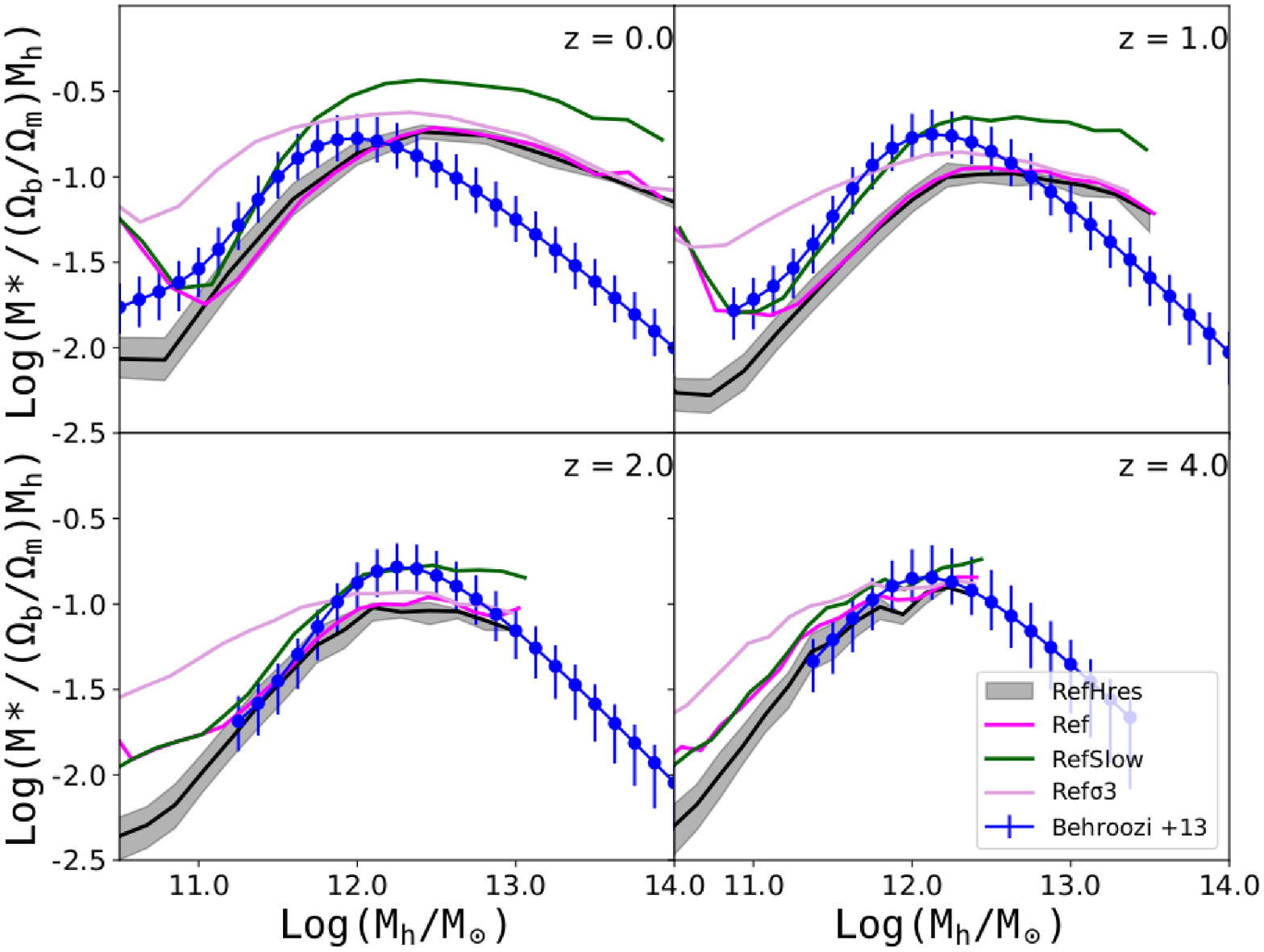}
\caption{The stellar mass - halo mass functions (SMHMs) 
at $z=0$,1,2, and 4. We compare the
SMHMs from the same set of simulations as in \fig{fig:gsmfs_calib_etas}. The solid lines
are the running medians of the relation. We also show the scatter of the
relation for the RefHres simulation as shaded regions that enclose 68\% of
all galaxies within each $M_h$ bin. The upturn in the SMHMs below
$M_h < 10^{11}M_\odot$ at $z=0$ and $z=1$ is a selection effect owing to an artificial
stellar mass cut for under-resolved galaxies. The blue lines show the empirical best-fit
models from \citet{behroozi13} as observational constraints.}
\label{fig:smhms_calib}
\end{figure*}

The stellar mass - halo mass function (SMHM) complements the GSMFs by
showing how efficiently baryons turn into stars in haloes of different masses. Instead of
directly plotting the ratio of stellar mass to halo mass, in \fig{fig:smhms_calib} we instead plot the baryon conversion efficiency, i.e., $\epsilon_b \equiv M_* / \mh
(\Omega_b/\Omega_\mathrm{m})^{-1}$ to visually
capture the small differences between the models more easily. Observationally one determines this
relation using empirical models that connect observed galaxies to
dark matter haloes from N-body simulations. The empirically constrained SMHMs
depend on the method used, but overall they agree with each other fairly well (see
\citet{moster18} for a recent compilation). In \fig{fig:smhms}, we compare the
$z = 0$ SMHMs for the central galaxies from our simulations to the SMHM that is
obtained in \citet{behroozi13} using subhalo abundance matching.

The Ref and RefHres simulations agree reasonably well with the observationally
inferred \citep{behroozi13} SMHM up to the peak at $\logmvir \sim 12$, with
the largest difference at $z=1$. The $288^3$ and $576^3$ simulations of this model
predict similar results for $\logmvir > 11.2$, but the lower resolution simulation
artificially boosts $M_*/\mh$ at lower masses. The turnover of the SMHM is much
sharper in the observations than in any of the simulations, and all models
drastically overpredict $M_*/\mh$ for $\logmvir > 13$ at $z=0$. The Ref$\sigma$3
simulation, with weaker outflows in low mass haloes, overpredicts the observed
$M_*/\mh$ in low mass haloes with $\logmvir < 11.5$ at all redshifts and converges to
the Ref model at high masses. The RefSlow simulation, with lower wind
velocities, makes similar predictions to the Ref model at $z=4$, but at $z=2$,
and increasingly at lower redshifts,
it predicts higher $M_*/\mh$ in haloes near or above the SMHM
turnover. The agreements and disagreements in \fig{fig:smhms_calib}
closely track those seen previously in the GSMF (\fig{fig:gsmfs_calib_etas}).

We now examine the contributions to galaxy stellar masses in more detail,
focusing first on $z=2$ and then on $z=0$.

\subsection{Galaxies at Redshift $z=2$}
\label{sec:discussion_z2}

\begin{figure} 
\centering
\includegraphics[width=0.95\columnwidth]{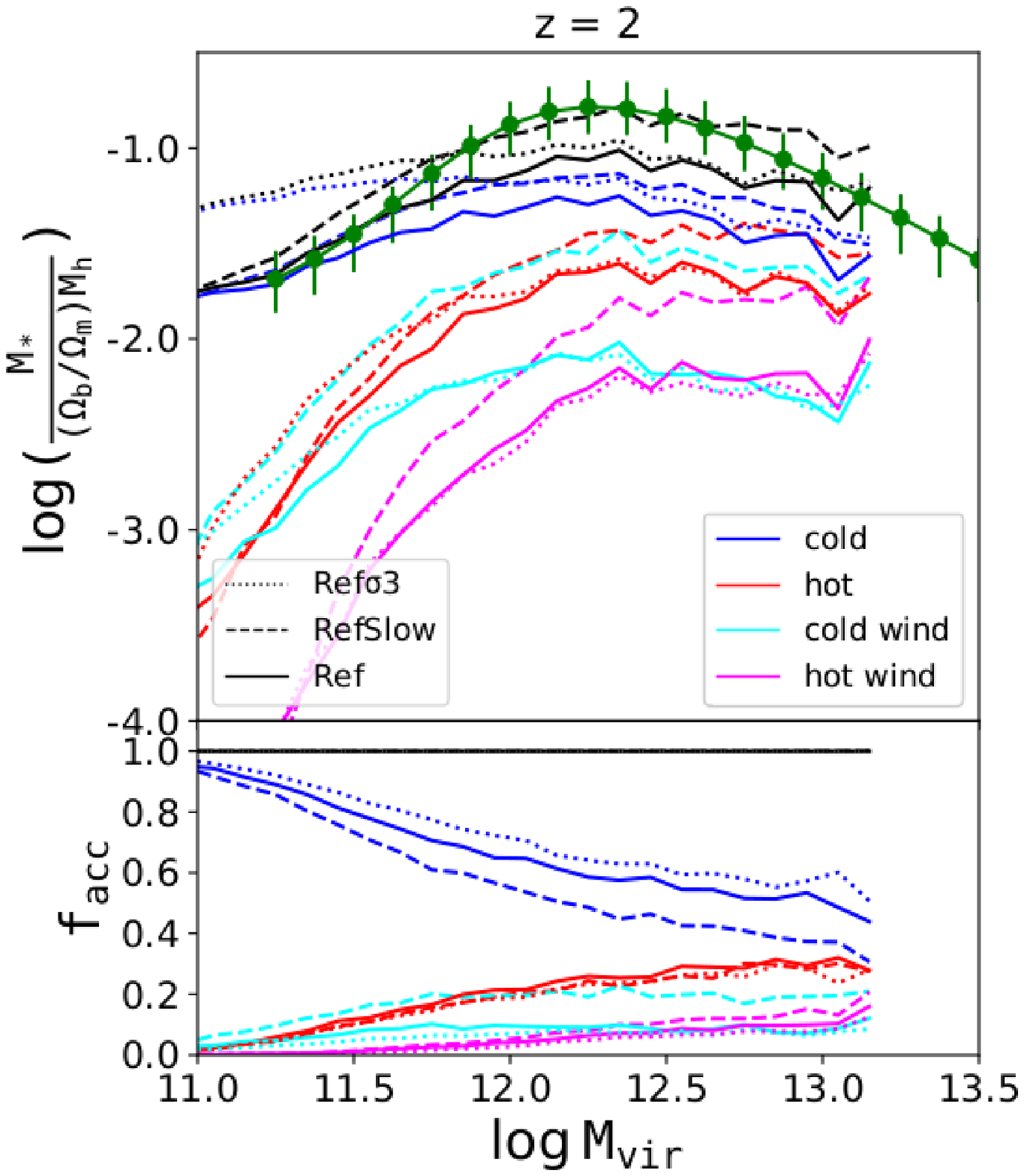}
\caption{A closer look into the SMHM at $z=2. $\textit{Upper panel: }The black lines show the total mass of stars,
averaged over all central galaxies from each halo mass bin, as a function of
the halo mass, i.e., SMHM. The stellar masses are further divided into four categories,
based on the accretion histories of their progenitor gas particle. Blue, red,
cyan, and magenta lines show stellar mass formed from cold, hot,
cold wind and hot wind accretion, respectively. We also plot the empirically
determined relation between stellar mass and halo masses from
\citet{behroozi13} as the green line.\textit{Lower panel}: The fraction of
stellar mass that falls into each category. In each panel, the line styles
indicate the three simulations used in this comparison.}
\label{fig:sfhistoryz2}
\end{figure}



\fig{fig:sfhistoryz2} shows the contribution of cold, hot, cold wind,
and hot wind accretion to the stellar mass content of galaxies at $z = 2$ in the
Ref, RefSlow, and Ref$\sigma$3 simulations.
The qualitative trends are similar between the simulations.
Cold mode accretion dominates,
contributing to nearly 100\% of all star formation in small haloes with
$\logmvir < 11.0$ and over half of all stars in the most massive haloes. The
hot mode fraction grows with halo mass and becomes comparable to the cold mode
in the most massive haloes. Wind recycling is not yet important at $z = 2$,
especially in less massive haloes where most winds are able to escape the halo
potential and not return.

Comparing the dotted lines and the solid lines shows the effect of changing the
mass loading factor. The Ref$\sigma$3 simulation has a smaller $\eta$ in
low-mass galaxies compared to the fiducial simulation and, therefore, allows more
gas to turn into stars. As a result, there is much more stellar mass formed from cold
accretion in these galaxies, while in the other two simulations this gas is
more likely to be launched as a wind before forming stars.
These simulations have the same $\eta$ values in massive haloes, and the stellar mass
production converges at $\logmvir > 12$. This convergence implies that the winds from
low mass galaxies are not affecting the pristine gas accretion unto high mass galaxies.
In principle, one expects the Ref simulation to have more wind recycling than the Ref$\sigma$3 simulation owing to the larger
amount of wind ejection, but this is not seen because the ejected particles
have not yet had enough time to recycle. Hence, wind recycling remains a
small fraction up to $z=2$ in both models.

Comparing the dashed lines and the solid lines shows the effect of changing the
wind speed. The most significant effect is that the slower winds in the
RefSlow simulation result in much more wind recycling, especially the cold
wind accretion in haloes of all sizes at $z=2$. This is a direct consequence of
the shorter recycling time. Another clear effect is that the fiducial
simulation has less cold and hot accretion than the RefSlow simulation,
indicating that
the fast wind speed not only suppresses wind recycling but also plays a role in
preventing pristine accretion through hydrodynamic interactions with the fresh,
in-falling gas. Since these two simulations have the same outflow rate for a
given halo mass, the higher stellar mass in the RefSlow galaxies can be
explained by the enhanced accretion rate owing to the slow wind speed.

In summary, two major factors contribute to the different $z=2$ SMHMs from our
test simulations. The mass loading factor controls the amount of outflow for a
given halo but has little effect on the total amount of cold or hot
accretion, which dominates at that redshift and above. The wind speed affects
the amount of inflow. Faster winds reduce cold and hot accretion and also
reduces the wind recycling by a similar amount. 

\subsection{Galaxies at Redshift $z=0$}
\label{sec:discussion_z0}
\begin{figure*} 
\centering
\includegraphics[width=1.95\columnwidth]{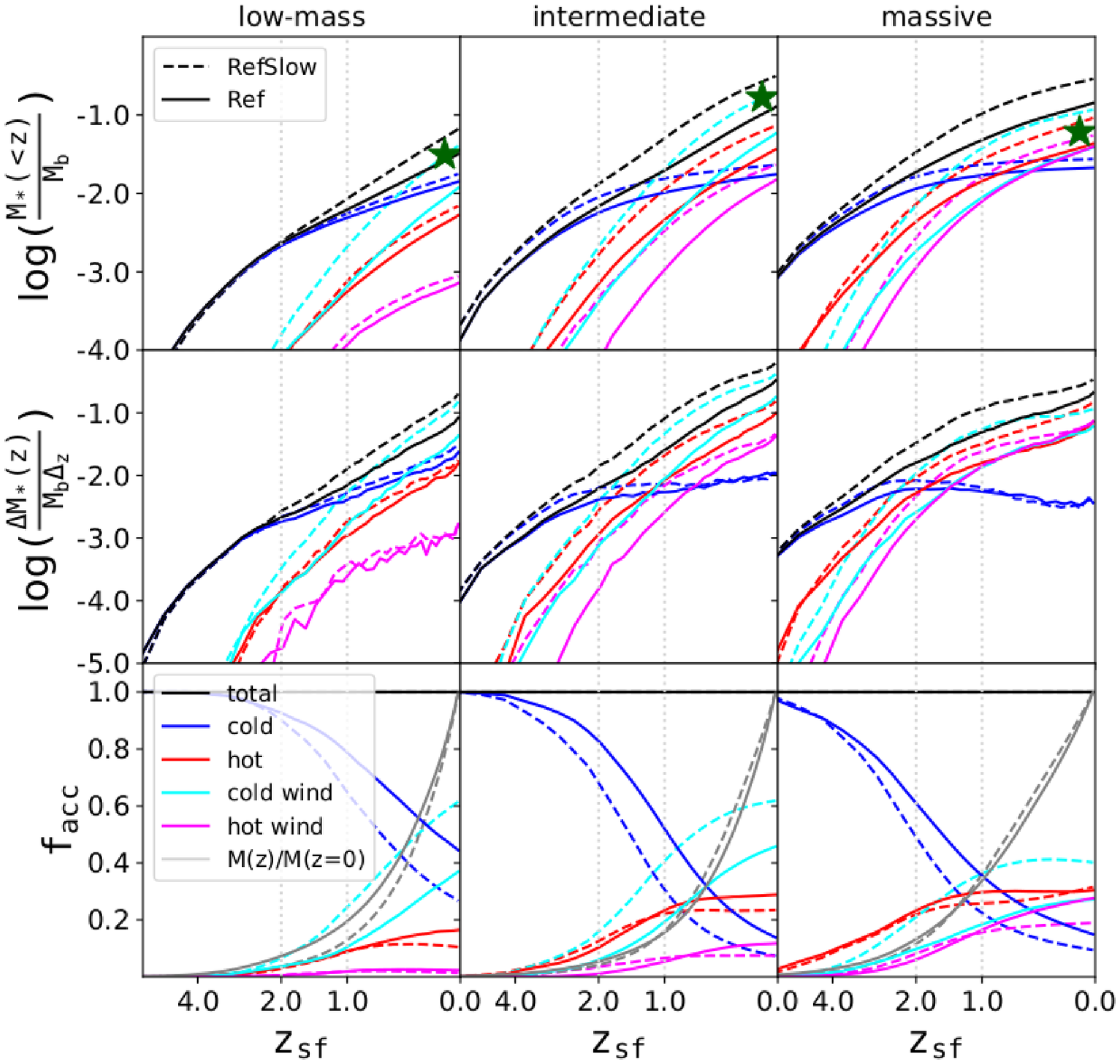}
\caption{We select and divide $z=0$ central galaxies into three groups based on
their halo virial mass. Columns from \textit{left} to \textit{right} show how
stellar mass on average grows with time in \textit{low-mass} ($11.0 < \logmvir < 11.5$),
\textit{intermediate-mass} ($11.85 < \logmvir < 12.15$)
and \textit{massive} ($12.85 < \logmvir < 13.15$) haloes. Similar to
\fig{fig:sfhistoryz2}, at each redshift, we divide star particles in these
galaxies into four channels based on their accretion history: blue, red, cyan,
and magenta lines indicate cold, hot, cold wind, hot wind accretion,
respectively. \textit{Top row} shows the cumulative mass growth history.
\textit{Middle row} shows differential stellar mass growth within a constant
redshift interval $\Delta_z$. In the upper and middle rows, we have normalised
the stellar mass by the halo mass to indicate the baryon conversion factor. The
green stars are the empirical results from \citet{moster18} for comparison.
\textit{Bottom row} shows the fraction of stars formed
within each sub-category, with
the grey line showing the fraction of total stellar mass at $z=0$ that has
already formed at a certain redshift. In each panel, we compare two
simulations: the Ref and the RefSlow simulations,
indicated by solid and dashed lines, respectively.}
\label{fig:sfhistory1}
\end{figure*}

\begin{figure*} 
\centering
\includegraphics[width=1.95\columnwidth]{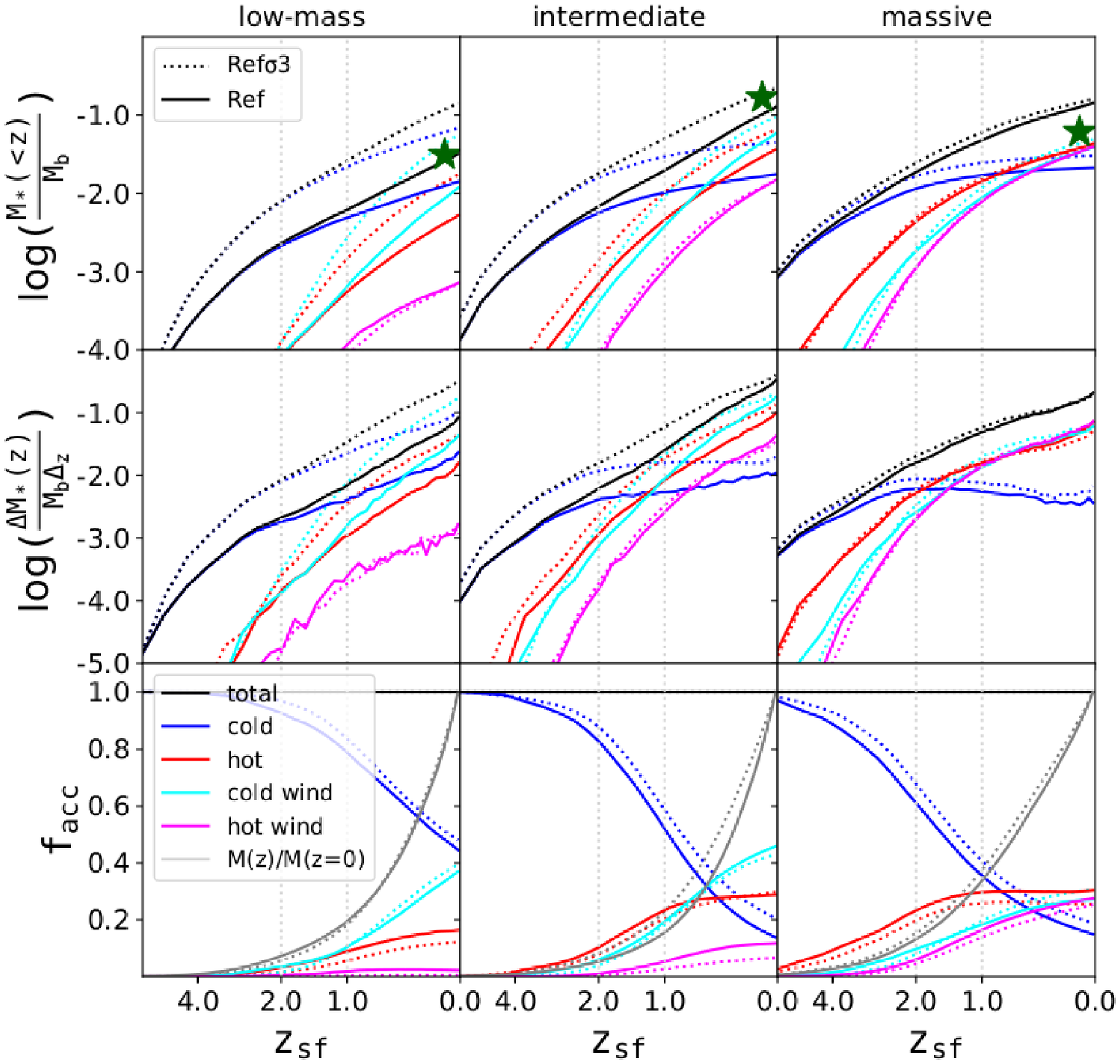}
\caption{Same as \fig{fig:sfhistory1}, except that here we compare the Ref and the Ref$\sigma$3 simulations, indicated by solid and dotted lines, respectively.}
\label{fig:sfhistory2}
\end{figure*}

The observed SMHMs at $z=0$ show a characteristic $\Lambda$-shape with the
intermediate-mass haloes ($\logmvir \sim 12$) having the peak baryonic
conversion efficiency. The efficiency declines in more massive haloes as well
as in less massive ones, although the reasons are likely very different:
theoretical models of galaxy formation suggest that the formation of massive
galaxies is characterised by late assembly of smaller systems that formed at early
times and by having little \textit{in situ} star formation at low redshifts.
On the other hand, the low mass haloes in the local Universe followed more
linear growth histories and formed many of their stars more recently. As shown
in \fig{fig:smhms_calib}, our simulations in general fail to match the observed
SMHMs at $z=0$ over the entire mass range. The discrepancies are most
prominent in the most massive haloes, motivating us to perform separate
analyses on galaxies that form in haloes of
different masses. Here we select haloes from three different mass bins and
study the formation histories of their central galaxies. The
low-mass bin consists of $\sim 2000$ haloes with $11.0 < \logmvir < 11.5$. The
intermediate-mass bin consists of $\sim 400$ haloes with $11.85 < \logmvir <
12.15$ and the massive bin consists of $\sim 60$ haloes with
$12.85 < \logmvir < 13.15$. The exact number of haloes within
each mass bin varies slightly among
the simulations, but we focus on comparing the average baryonic conversion
efficiency, which is normalised by the total virial mass of all haloes within
each mass bin.

In \fig{fig:sfhistory1} and \fig{fig:sfhistory2},
we show how stellar mass grows with time in haloes
selected from the three mass regimes and divide the stellar mass at any time
into categories based on their accretion histories as in the previous section. 

\textit{Low-mass haloes.} The stellar mass from the Ref and the
RefSlow simulations matches the observed value at $z=0$, but the
Ref$\sigma$3 simulation
over-produces stellar mass by 0.6 dex. At $z=0$, cold accretion
and cold wind recycling each contribute roughly half of the total stars
formed, while hot mode accretion contributes $\sim 10 - 15\%$ of star formation.
Cold accretion dominates the supply of star-forming gas in all three
simulations until $z=2.0$, after which cold wind recycling and hot mode
accretion start to be important. Compared to the other simulations,
the haloes in Ref$\sigma$3 form many more stars from both cold accretion and wind
recycling, not because of more inflow but because they have less outflow
as a result of the smaller mass loading factors. The slower wind speed in the
RefSlow model increases cold wind recycling by a large amount compared to the
fiducial simulation. As a result, the galaxies at $z=0$ are slightly more
massive as wind recycling gains importance after $z=2$, but their stellar
masses are still consistent with the observations.

\textit{Intermediate-mass haloes.} The $z=0$ stellar mass from all three
simulations are consistent with the observations within a small factor. The evolution of
stellar content in these haloes is qualitatively similar to the small-mass
haloes but with several major differences. First, stars from all accretion
channels formed earlier in these more massive haloes, as is expected from the
hierarchical assembly of galaxies. Second, both hot accretion and hot
wind recycling, though still sub-dominant over most of the time, become more
important at low redshifts, and together contribute $\sim 30\%$ of the
total stars
formed at $z=0$. Third, cold accretion still dominates star formation at $z > 2$
but nearly stops after $z=2$ when the shock heated gas starts to develop
a hot corona in these haloes. In the end, cold accretion only accounts for
$\sim 20\%$ of the total stars formed. Cold wind recycling still plays a
critical role in determining the final stellar mass of the galaxies, and
largely accounts for the differences among the three simulations. Its
contribution is more prominent in the RefSlow simulation.

\textit{Massive haloes.} At the massive end, galaxies from the Ref and
the Ref$\sigma$3 simulations evolve very similarly and over-produce stars by a
factor of 3 at $z=0$ (\fig{fig:sfhistory2}). The Ref simulation
has larger $\eta$ in small haloes, but the
differences in $\eta$ decrease with $\sigma$ and become the same when
$\sigma > 106\kms$. Therefore, the larger mass loading factor in the Ref simulation
only affects the progenitor galaxies during the earliest stages of their
assembly when they were still small. Since these haloes assembled fast at high
redshift, the different scalings of $\eta$ and $\sigma$ in the low-mass haloes
have little effect on the massive galaxies in our simulations. Compared
to the intermediate-mass haloes, they have even earlier star formation and
a higher fraction of hot accretion and hot wind recycling. Except for the
RefSlow simulation, where cold wind recycling is clearly more important for
stellar growth than the other channels, all four accretion channels contribute
comparable amounts
in the other simulations, with cold accretion + cold wind recycling
and hot accretion + hot wind recycling each responsible for half of the stars
formed. The RefSlow simulation over-produces stellar mass by a factor of 5,
more than the other simulations. \fig{fig:sfhistory1} shows that this owes not
only to more cold wind recycling because of the slower wind speed, but
also because of a significantly higher amount of hot accretion and hot wind
recycling than the other simulations. Furthermore, hot accretion and hot wind
recycling are also higher in the low-mass and intermediate-mass regimes, but
unlike in the massive haloes, they are always sub-dominant to the total mass
budget in less massive haloes. Naturally, any feedback mechanism designed
to suppress star formation at these masses would strongly impact these trends.

\subsection{The Importance of Mergers}
\label{sec:discussion_mergers}
\begin{figure} 
\centering
\includegraphics[width=0.95\columnwidth]{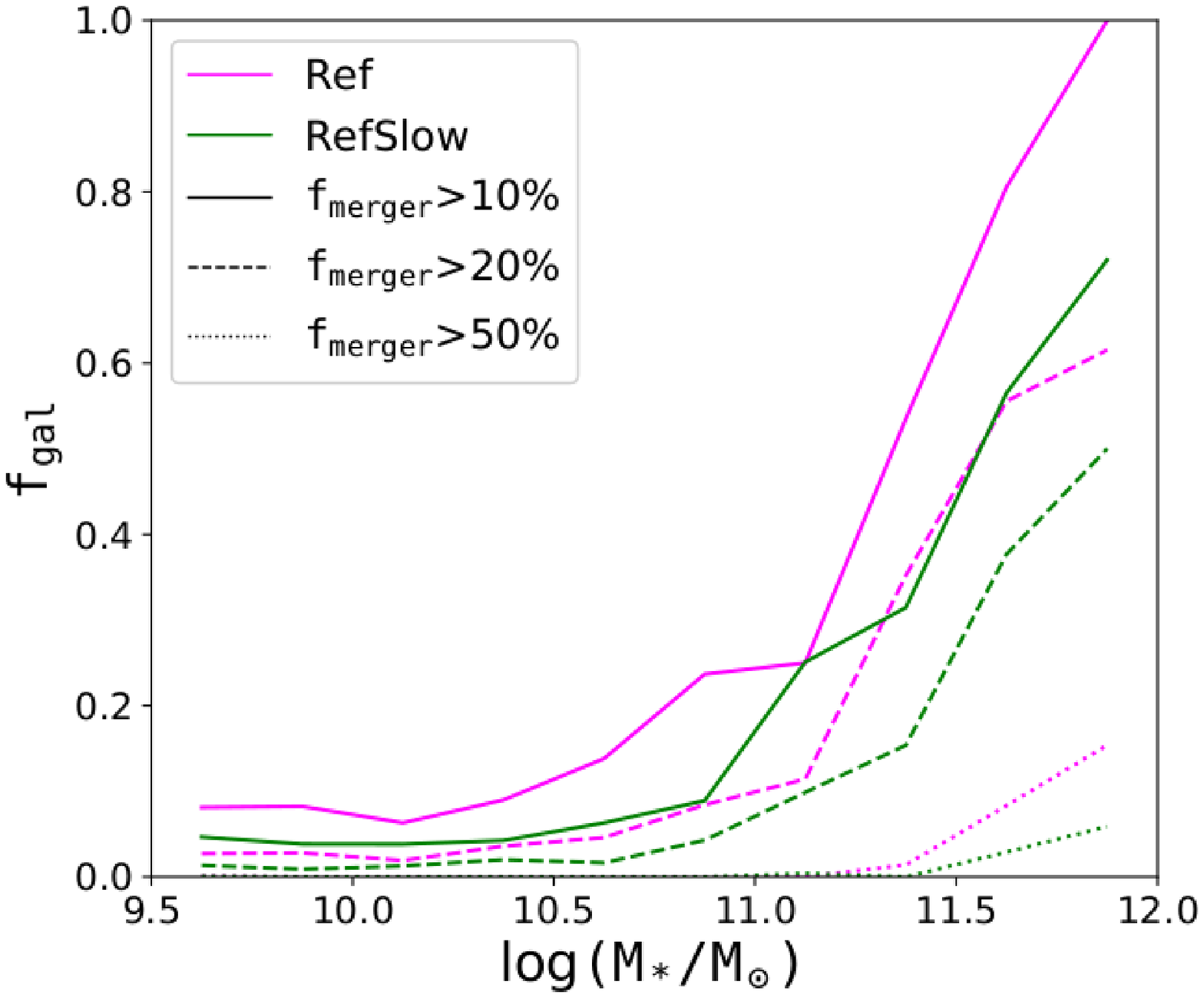}
\caption{Dotted, dashed and solid lines indicate the fraction of galaxies that
have more than 10\%, 20\% and 50\% of their $z=0$ stellar mass gained by major
mergers. The Ref and RefSlow simulations are shown in magenta and
green, respectively. In general, the importance of mergers increases with
$M_*$, but even in the most massive galaxies, the fraction of stars from
mergers are less than those formed \textit{in situ}.}
\label{fig:fmergers}
\end{figure}

In the above discussions we focused on studying the histories of star particles
that end up in certain galaxies at a certain redshift. However, galaxies in a
hierarchical Universe are often the result of many merging events. In particular,
massive galaxies are often assembled from many smaller galaxies that formed in
a wide range of haloes and environments. Since the \textit{in situ} star
formation efficiency, which is regulated by feedback, strongly depends on the
halo mass, the final mass of a galaxy could be sensitive to feedback in those
haloes where star formation was most efficient.  Therefore, to
understand how the feedback algorithms affect the final stellar mass of massive
galaxies at $z=0$, it is necessary to study when and where their progenitors
formed.  

To evaluate the importance of mergers, we need to trace the evolution
of galaxies in our simulations over time. At each output, if most stars within
a galaxy are found in some galaxy at the next output, we define the first
galaxy as a progenitor of the second galaxy. A galaxy could have more than one
progenitor at any time, and we define its main progenitor as the most massive
progenitor. We consider any other progenitor of this galaxy as a merger into
this galaxy between the two outputs. Therefore, we can define the main
evolutionary path of a galaxy at $z=0$ by sequentially tracking its main
progenitors over time. At any time when a merger occurs, we calculate the mass
ratio between the two galaxies. In this work, we define major mergers as those
that involve two galaxies with a mass ratio over $1/5$. One caveat is that some
galaxies take longer than a few outputs to completely merge with their host
galaxies. In some situations they were first grouped with the host galaxy
during the first pass-by but left and became a separate galaxy later on, before
they finally merged again. To avoid counting these galaxies as individual
mergers multiple times, we consider only the first merging event by requiring
that the mass of the host galaxy be at its maximum up to the merging event.
Therefore, if the merging galaxy later left, the mass of the host galaxy would
decrease and any subsequent pass-by will not be counted until the merger is
complete. This criterion effectively removes most of the spurious mergers
without missing any real mergers.

To evaluate the importance of mergers, we look at galaxies at $z=0$ and
determine what fraction of stars each galaxy accreted through major mergers and
where and when the stars present at $z=0$ form. \fig{fig:fmergers} shows that
for most galaxies, the fraction of stars acquired through major mergers
is less than 10\%. In general, more
massive galaxies have a higher fraction of their stars formed in other galaxies
and merged with it at later times,
but even in the most massive bins, only 30\% of galaxies
have more than half of their stars added through major mergers. The major merger
fractions are also similar between the Ref and the RefSlow
simulations. Galaxies in the RefSlow simulation in general have a higher
fraction of stars formed \textit{in situ} because of more wind recycling onto
the main progenitors.
These results do not change very much if we include mergers with a
mass ratio less than $1/5$.


In summary, the stellar growth of galaxies in our simulations is dominated by
\textit{in situ} star formation, with major mergers contributing a small
fraction, except in the most massive galaxies.
The final stellar mass of a galaxy is in most cases determined by the
growth of its most massive progenitor, which is in turn regulated by how
efficiently feedback suppresses star formation during the entire time of the
evolution of the progenitor and its host halo.
However, the relative importance of mergers in galaxy growth could increase
if one added additional feedback to remove
all the late time star formation in massive galaxies, as required to match 
observations.

\subsection{Wind Recycling}
\label{sec:discussion_recycling}
Wind recycling dominates the supply of star-forming gas at lower redshifts and
is responsible for a considerable fraction of the total stellar mass in most
haloes. In this section we will show that wind speed strongly affects the
recycling timescale $\trec$ of the launched winds. 

\begin{figure*} 
\centering
\includegraphics[width=1.90\columnwidth]{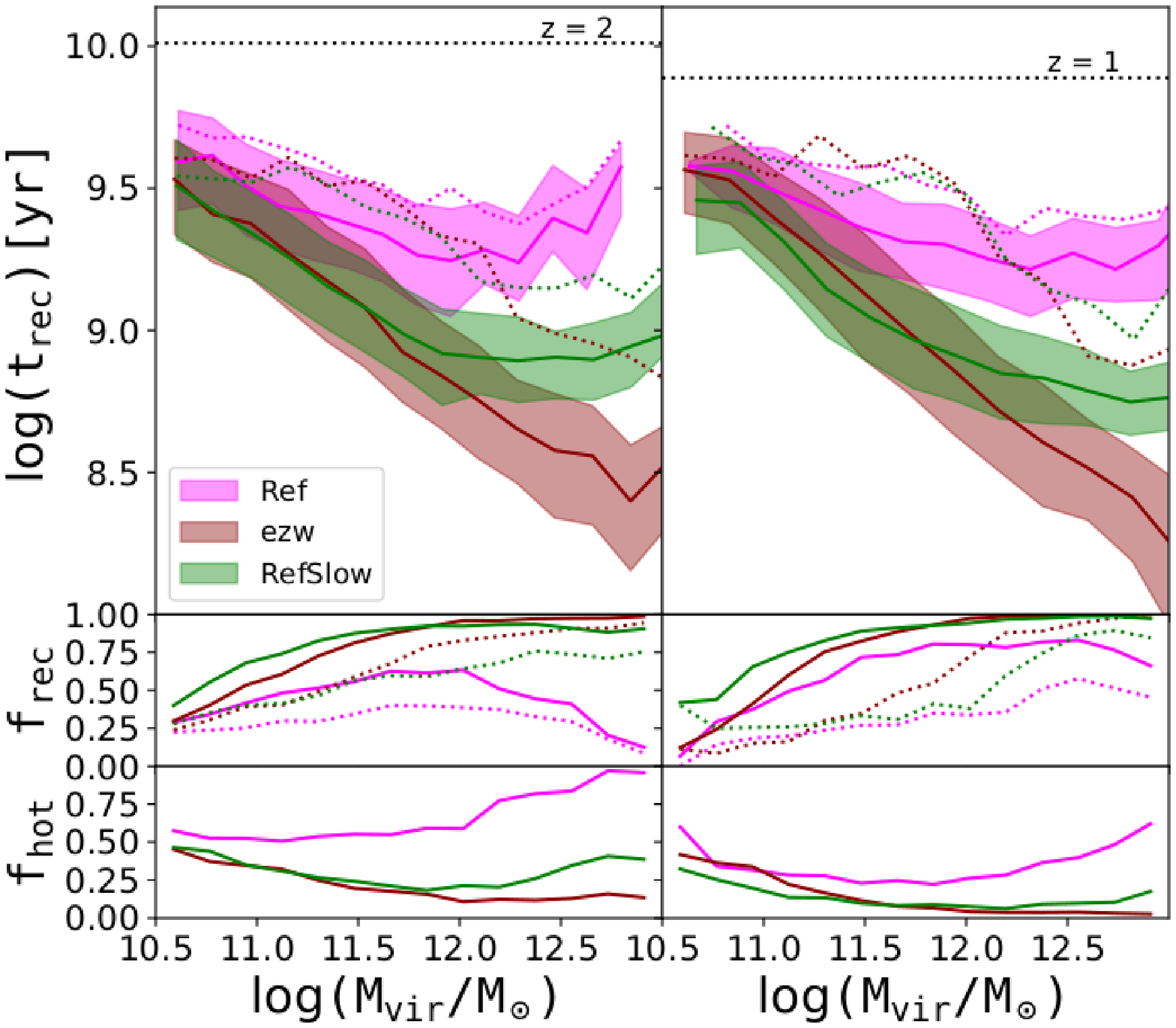}
\caption{\textit{Upper panels}: The median recycling time $\trec$ of winds that
have recycled by $z = 0$ as a function of the virial mass $\mvir$ of the halo
from which the winds were launched. We only include winds launched
from central galaxies. The dotted lines show $\trec$ only for winds that become hot.
The \textit{left} and \textit{right}
panels show results for winds launched at $z=2$ and $z=1$, respectively. We
include all wind particles that are launched during a small redshift window
with $\Delta_z = 0.002$ at these redshifts. The dotted horizontal line in each
panel indicates the lookback time at that redshift and is the upper limit of
$\trec$ for those winds. The shaded area shows the $1\sigma$ scatter in each $\mh$
bin. \textit{Middle panels}: Solid lines indicate the fraction of all winds that have ever re-accreted onto any galaxy at least once by $z = 0$. Dotted lines include only those
winds that become hot. In general, the fraction of hot winds that recycle is lower.
\textit{Bottom panels}: The fraction of all winds that become hot, regardless of
whether or not they have recycled by $z=0$.}
\label{fig:trec}
\end{figure*}

The amount of winds that re-accrete after being launched is closely
related to the recycling timescale $\trec$ \citep{oppenheimer10}, defined as
the time between a particle being launched as wind and it becoming star forming
again. \fig{fig:trec} compares the $\trec$ of wind particles from the three
simulations with different wind speeds (see \fig{fig:v25vc}).
The wind particles from the ezw
simulation have $\trec$ that strongly depend on the halo mass. The deep
gravitational potential of massive haloes causes wind particles to fall back
shortly after being launched, creating a galactic fountain that is
categorically different from the galactic scale winds in smaller galaxies.
\citet{oppenheimer10} use the same wind algorithm in their simulations and find
a similar trend. They refer to it as ``differential recycling'', which is key to
regulating star formation as a function of halo mass and thereby shaping the
galactic stellar mass functions. However, the recycling times from
\citet{oppenheimer10} are greatly affected by the fact that their wind speeds do
not scale as $v_w \propto \sigma$ after they leave the galaxy (\fig{fig:v25vc})
as they were originally intended, which was the motivation for our new wind model. 

The other two simulations have a mass dependent enhancement to the wind speed.
Increasing wind speed with mass has a direct effect on $\trec$, with a
stronger enhancement (Ref) leading to a longer $\trec$ in massive haloes.
Similar to what \fig{fig:v25vc} indicates, the wind dynamics inside these
haloes are sensitive to their initial speed. For example, in the most massive
galaxies at $z = 2$, the average wind speed in the Ref simulation is
$\sim 2$ times faster than in the RefSlow simulation while the recycling
time is $\sim 3$ times longer.

We have further divided wind recycling into cold wind and hot
wind recycling based on whether or not the wind particle heats up to
$10^{5.5}\ K$. In the middle row of \fig{fig:trec}, the dotted lines
show that the hot winds are less likely to re-accrete unto galaxies by $z=0$
than cold winds. For the winds that did recycle,
the upper rows of \fig{fig:trec} show that the recycling
timescales are significantly longer for the hot winds (dotted) on average.
Most cold winds
that formed stars at $z\sim0$ were launched well below $z=1$, while most hot
winds were launched around $z=2$. Moreover, \fig{fig:sfhistory1} and
\fig{fig:sfhistory2} show that even though hot wind recycling is nearly
negligible in low-mass and intermediate-mass haloes, it is important to the
late star formation in massive galaxies. Because of the long $\trec$ of hot
winds, it also indicates that a considerable fraction of stars in massive
galaxies formed from outflow material launched long ago, at least in our
simulations.

The wind particles in our simulations are shock heated immediately after they
hydrodynamically recouple to the ambient SPH particles. The initial wind speed
plays a critical role in determining the post-shock temperature of the wind
particles. Once they heat up to a temperature where cooling becomes
inefficient, they will likely stay hot and become indistinguishable from a
normal gas particle of the hot corona gas. The evolution of the hot wind
particles thus depends more on the cooling physics than the dynamics that
governs the recycling of cold wind particles. The bottom row of
\fig{fig:trec} shows that the fraction of winds that became hot
is very sensitive to wind speed. The hot wind fraction is significantly higher in the
Ref simulation, where wind heating is
more efficient owing to the faster wind speeds. Note that with a sufficiently
fast wind speed, our wind algorithm naturally results in a multi-phase outflow,
without the need to artificially add a hot component to the winds
at launch as in MUFASA \citep{mufasa}. We also find that the hot wind
fraction is negligible in even the most massive galaxies in the
ezw simulation, where the wind speeds are even lower.
This was the wind model used in \citet{dave13}. 

However, we must caution that the interactions between the winds and the halo
gas likely involves processes such as hydrodynamic instabilities and thermal
conduction that are unresolved in our our simulations and likely
even in galaxy zoom-in simulations with the highest
resolution today \citep[e.g.][]{schneider17}. The evolution of winds inside and outside galactic haloes
in galaxy simulations, therefore, likely depends as much on numerics as on the
true underlying physics. Hence, the behaviour and the effects of wind recycling
must be re-examined with future simulations that have higher resolution or
accurate and numerically robust sub-grid models that incorporate
necessary physics that has been neglected or incorrectly modelled in simulations
up until now. Simulations that concentrate resolution in gaseous haloes \citep{peeples19, vandevoort19, hummels19}
can improve modelling of physics in the circumgalactic medium, though even with
this approach the resolution may not be sufficient to accurately model interactions within the multiphase CGM 
\citep{scannapieco15, schneider17} and in addition
it may be difficult to quantify recycling for ensembles
of galaxies with a range of properties.

\subsection{Implications for Additional Feedback}
\label{sec:discussion_agn}
Feedback processes are essential in cosmological simulations to successfully
reproduce the observed stellar content of the Universe. Stellar feedback such
as galactic winds generated from the brightest stars and supernovae have been
widely applied to explain the growth of small galaxies, but these processes
alone are usually insufficient to suppress the growth of massive galaxies. The
stellar feedback models in simulations are usually tuned to match observational
constraints at the low mass end, while one often invokes additional feedback such as AGN feedback to produce more realistic massive galaxies. It is also unclear what exact role AGN feedback plays in suppressing
star formation. It could work as preventative feedback that limits the amount
of inflow, or as kinetic feedback that drives additional outflows from
galaxies. In the previous sections we have shown that changing the parameters
of our particular stellar feedback model within our explored range could
significantly affect galaxy growth, even in the most massive haloes. This hints
at the possibility that a combination of carefully tuned galactic 
wind parameters might be
able to simultaneously reproduce the stellar content on all mass scales at any
redshift. Even if the wind model is unable to meet all the constraints, it is
important to understand the shortcomings of the current model that have to be
solved with additional feedback processes.

A successful galaxy formation model that reproduces $z=0$ results must also be
able to match observations from higher redshifts. At $z=2$, we have shown that
a strong halo mass dependence of the mass loading factor is key to matching the
faint end slope of the SMHM, while different wind speeds are responsible for
variations of the stellar mass in massive haloes. The Ref and RefSlow
simulations both reasonably match the observed relation at $z=2$. The
RefSlow simulation produces more stars in intermediate to massive haloes owing to
slower winds and a short recycling time, and agrees better with observations at
the ``knee'' but worse at the massive end.

At $z=0$, the discrepancy at the massive end grows, resulting in a factor
of 3 times more stars in the massive bin and even larger discrepancies in more
massive haloes. The grey lines in the bottom panels of \fig{fig:sfhistory1}
and \fig{fig:sfhistory2} show
the build up of stars that end up in the massive galaxies. They are
consistent among the simulations, with more than $80\%$ of stars formed after
$z=2$ and $60\%$ of stars formed after $z=1$. Therefore, in our simulations it is
the late star formation in massive galaxies that must be greatly reduced to
match the $z=0$ observations.
\sect{sec:discussion_mergers} confirms that most of these stars formed
\textit{in situ} instead of through merging. A successful model must maintain the
level of star formation up to $z=2$ as in the RefSlow simulation but
significantly reduce the amount of stars formed afterwards. In fact, the top
right panels of \fig{fig:sfhistory1} and \fig{fig:sfhistory2}
shows that galaxies in the massive bin have
already formed by $z=2$ as many stars as required to match the $z=0$ observations.
Hence, to make these galaxies agree with the $z=0$ constraint,
nearly all \textit{in situ} and \textit{ex situ} star formation after $z=2$
needs to be suppressed. Observationally, this phenomenon is known as downsizing,
that massive galaxies at $z=0$ formed earlier but have little late time star
formation. However, it is challenging to reproduce this effect in our
simulations without additional feedback. 

Stars from cold accretion mostly formed at high redshifts in small haloes that
later assembled into the massive galaxies. The most efficient way to remove
them from our simulations is to have stronger winds, i.e., stronger mass
loading factors in those haloes. However, having too much winds early will
unavoidably fail to match observations at higher redshifts. The right panels of
\fig{fig:sfhistory1} and \fig{fig:sfhistory2} show that cold accretion has
nearly stopped after $z=2$
for these galaxies but hot accretion and wind recycling continues growing
rapidly at low redshifts, and is responsible for most of the excess stars
formed.
After $z=2$, the stars in massive galaxies that must be prevented from forming
come from almost equal parts: hot accretion, hot wind re-accretion, and cold
wind re-accretion.  Hence, preventing hot gas
from cooling and forming stars at these times will eliminate both the
hot mode accretion and hot wind reaccretion and will lessen the tension in
massive haloes. There are several potential mechanisms,
such as AGN and cosmic ray heating,
that could reduce the amount of this cooling gas but are not yet included in our
simulations. The hot wind recycling would be harder to affect
by extra heating, because the higher metallicity of this gas makes it cool
faster. In hydrodynamic simulations, mixing between ejected wind elements and
the surrounding gas may have a large impact on the amounts of hot and hot wind
accretion. It is, however, unclear whether
this type of feedback could prevent the $\sim 1/3$ of stars formed through cold
wind re-accretion after $z=2$ from forming, which is also necessary to match
the observations. It is possible that a more accurate treatment of the cloud-CGM interaction would allow a
larger fraction of these winds to become hot, alleviating this problem.

%
%

\section{The High Resolution Simulation of the Reference Model}
\label{sec:fiducial_simulation}
In this section we present key results from the high-resolution RefHres
simulation. It adopts the new wind launch algorithm as described in the
previous sections using our fiducial set of wind parameters listed in
\tab{tab:simulations}. It is also implemented with the numerical improvements
to the SPH hydrodynamics introduced by \citet{huang19}. We will focus
on those predictions that have changed significantly from \citet{huang19} 
and from our previously published work with our new wind algorithm.

\subsection{The Stellar Content}
\begin{figure*} 
\centering
\includegraphics[width=1.90\columnwidth]{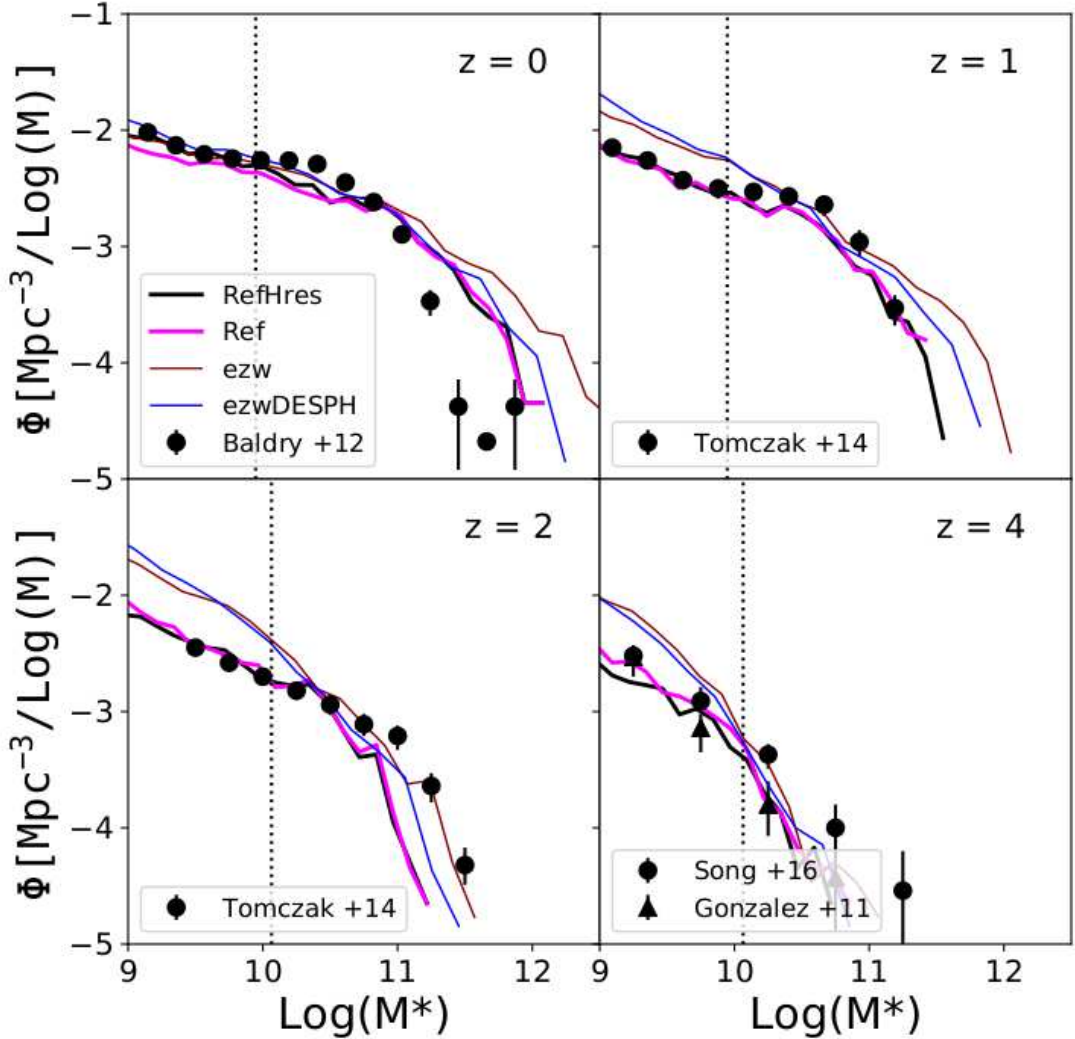}
\caption{Same as \fig{fig:gsmfs_calib_etas}, except that here we show a
different set of simulations, including the fiducial high resolution simulation
RefHres. See text and \tab{tab:simulations} for descriptions of these
models. The vertical dotted lines correspond to the mass of 1024 SPH
particles in the RefHres simulation and 128 SPH particles in the other simulations.}
\label{fig:gsmfs}
\end{figure*}

\fig{fig:gsmfs} shows that the GSMFs from our fiducial simulation, shown as
black lines in each panel, are mostly consistent with observations at all redshifts
from $z = 0$ to $z = 4$. The agreement is particularly good at the faint end except
for $z = 0$, where our simulation slightly underproduces the number of these low
mass galaxies. At the massive end, our GSMFs agree with observations at $z = 1$
and $z = 4$. However, our fiducial simulation produces too many massive galaxies
at $z = 0$ and too few massive galaxies at $z = 2$,
even after taking account for systematic uncertainties in the stellar mass
measurements at these redshifts.

The level of our agreement is comparable to other cosmological hydrodynamic
simulations such as EAGLE \citep{furlong15}, MUFASA \citep{mufasa} and
illustrisTNG \citep{pillepich18b}, except for massive galaxies at $z = 0$ where
AGN feedback incorporated in these other simulations more strongly suppresses
stellar mass growth. The \ezw model as implemented by \citet{dave13} includes
an ad hoc quenching scheme in massive galaxies to reproduce the $z = 0$
GSMF. Absent this mechanism, however, the \ezw wind formulation produces worse
agreement with observations: too many small galaxies at higher redshifts
($z = 2$ and $z = 4$) and too many massive galaxies at lower redshifts ($z = 1$ and
$z = 0$). We have shown in \sect{sec:sensitivity} that the success of reproducing
the faint end of GSMFs at $z > 1$ relies on a steeper scaling between the mass
loading factor $\eta$ and the halo mass. On the other hand, suppressing the
growth of massive galaxies relies on a higher wind velocity to effectively remove
cold gas from the galaxies.

\fig{fig:gsmfs} also shows that different hydrodynamic algorithms (comparing
ezwDESPH and ezw) have noticeable effects on the GSMFs principally at the
massive end, although these are much less significant than the changes
driven by the wind algorithm. Comparing the Ref simulation and the RefHres simulation
shows that the results are also robust to numerical resolution, but note that
in the higher resolution simulation we have increased the overall
wind speed by a small factor to obtain a similar $v_{25} - \vcir$ relation.

\begin{figure} 
\centering
\includegraphics[width=0.96\columnwidth]{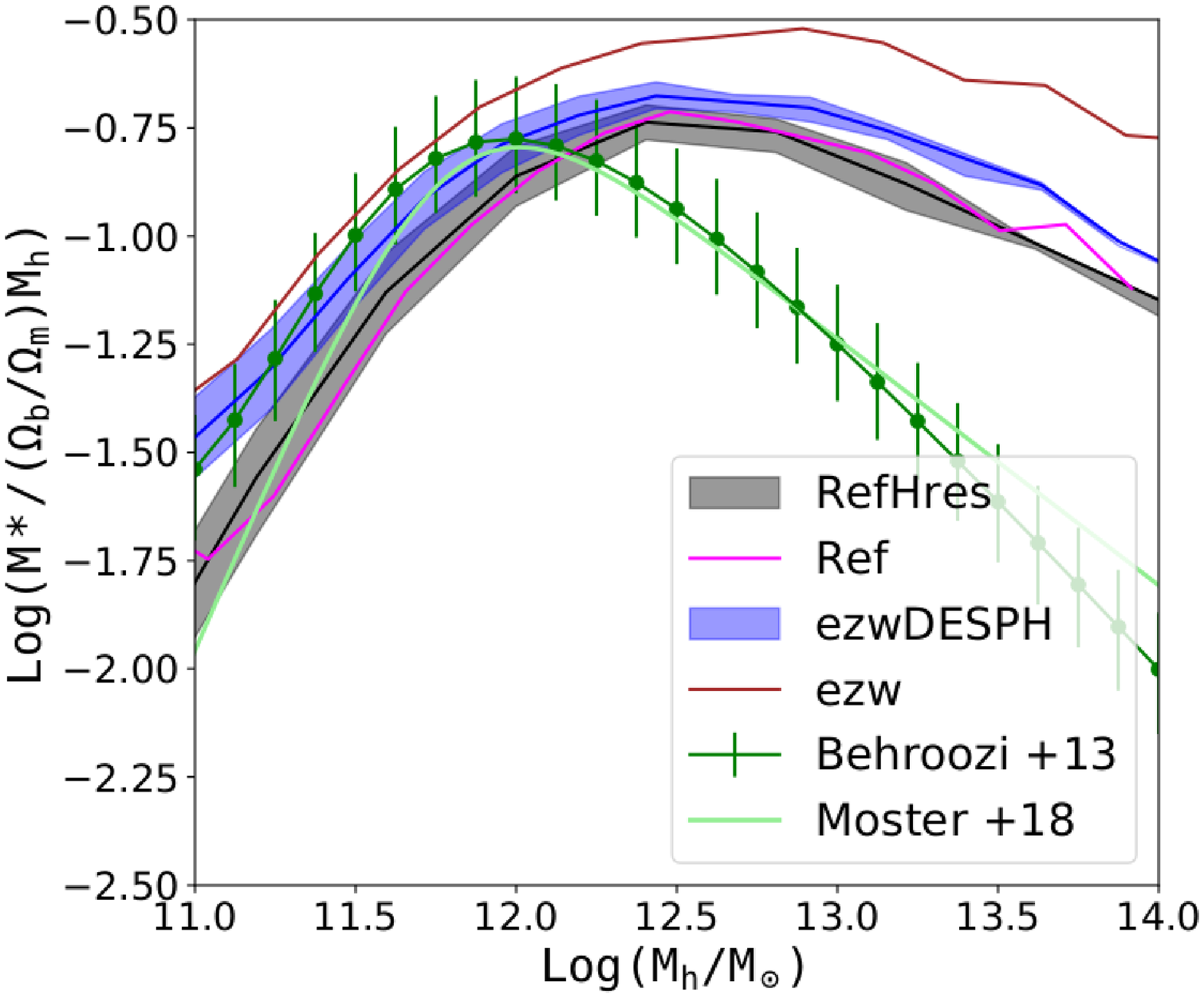}
\caption{The stellar mass - halo mass functions at $z = 0$. We compare the
SMHMs from the same set of simulations as in \fig{fig:gsmfs}. The solid lines
are the running medians of the relation. We show the scatter of the relation
for the RefHres and the ezwDESPH simulations as shaded regions that
enclose 68\% of all galaxies within each $M_h$ bin. The green lines show the
empirical best-fit model from \citet{behroozi13} and \citet{moster18} as
observational constraints.}
\label{fig:smhms}
\end{figure}

\fig{fig:smhms} shows that the baryon conversion efficiency from our fiducial
simulation agrees well with observations \citep{behroozi13, moster18} in small
haloes and
reaches a similar peak value, but becomes too high in more massive haloes.
This is a more clear illustration of the excess of stars in massive haloes than
that seen in the $z = 0$ GSMF. Comparing to the ezw simulation, which
uses the same SPH method but the \ezw wind model, the new wind algorithm
significantly reduces the stellar content in these massive galaxies but it is
still not enough to match the observations. Also, increasing
the numerical resolution has little effect on the SMHM.

\subsection{Stellar Density Evolution}
\begin{figure} 
\centering
\includegraphics[width=0.96\columnwidth]{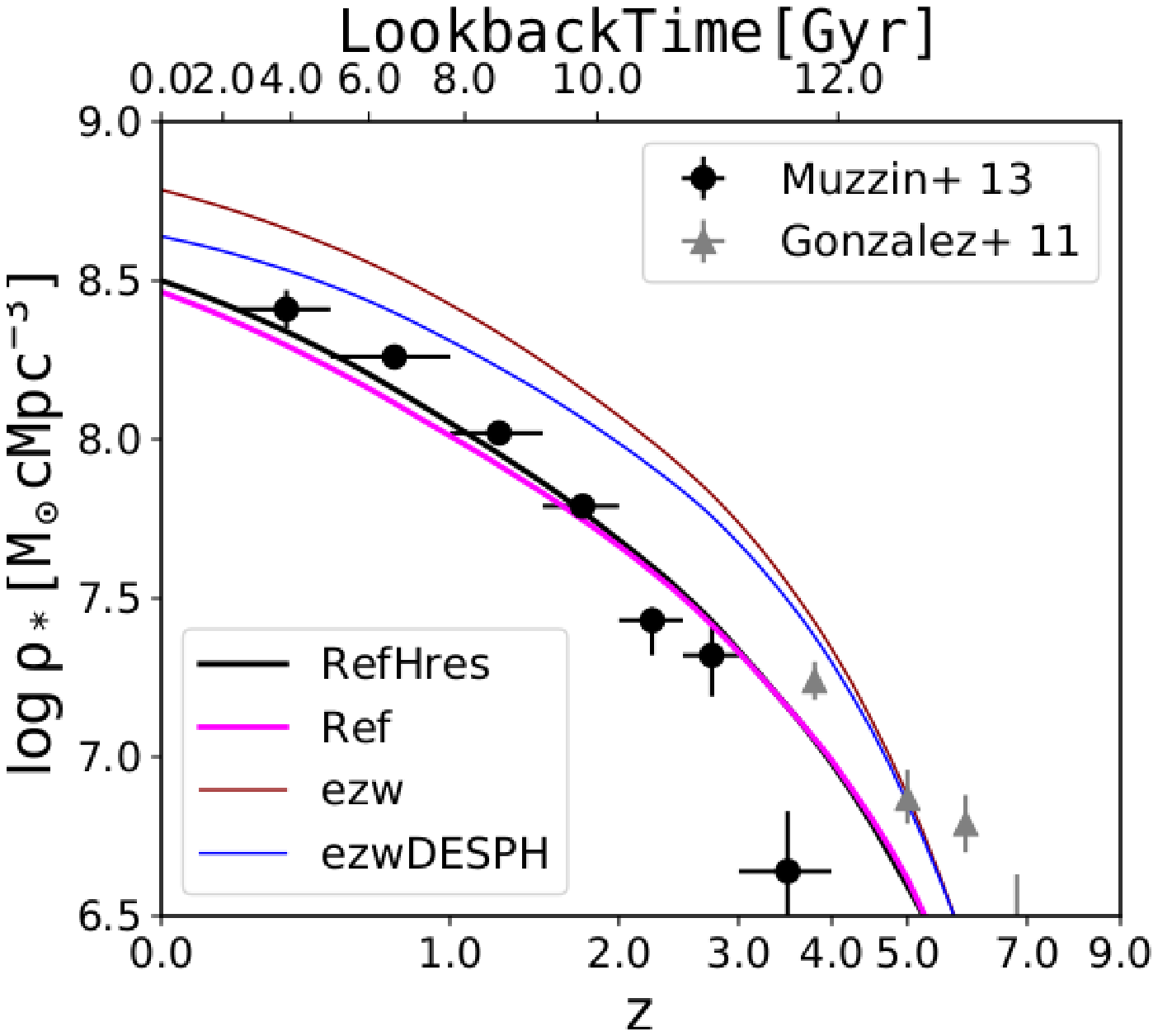}
\caption{Same as \fig{fig:sde_calib}, except that here we show results from the
fiducial high-resolution simulation compared with a few test simulations. Our
fiducial wind model (in the RefHres and Ref simulations) reproduces
the observations well, but the original \ezw wind model starts over-producing
stars at very early times.}
\label{fig:sde}
\end{figure}

\fig{fig:sde} shows the stellar density evolution of a few simulations. Our
fiducial simulation, shown as the thick, black line in \fig{fig:sde}, agrees
with the observational data to within 0.1 dex below redshift $z = 3$. At higher
redshifts, it falls in between the \citet{muzzin13} data and the upper limits
from \citet{gonzalez11}. Our simulation is capable of capturing the general
trend of the cosmic stellar density evolution. Since the stellar density at any
epoch is equivalent to the integration of the GSMFs at that redshift, the
success of matching the GSMFs at various redshifts is key to matching the
observed stellar density evolution.

The original \ezw model not only produces too many stars at $z \sim 0$, owing
mostly to the excess of stellar mass in massive galaxies, but also has
started over-producing stars since $z=5$ owing to an 
insufficiently large $\eta$. Changing the
numerics from the ezwDESPH simulation to the ezw simulation allows
more star formation at lower redshifts, but the effects are less significant
than the effects of changing the wind algorithm.

\subsection{Gas Fractions and Metallicity}
\begin{figure*} 
\centering
\includegraphics[width=1.90\columnwidth]{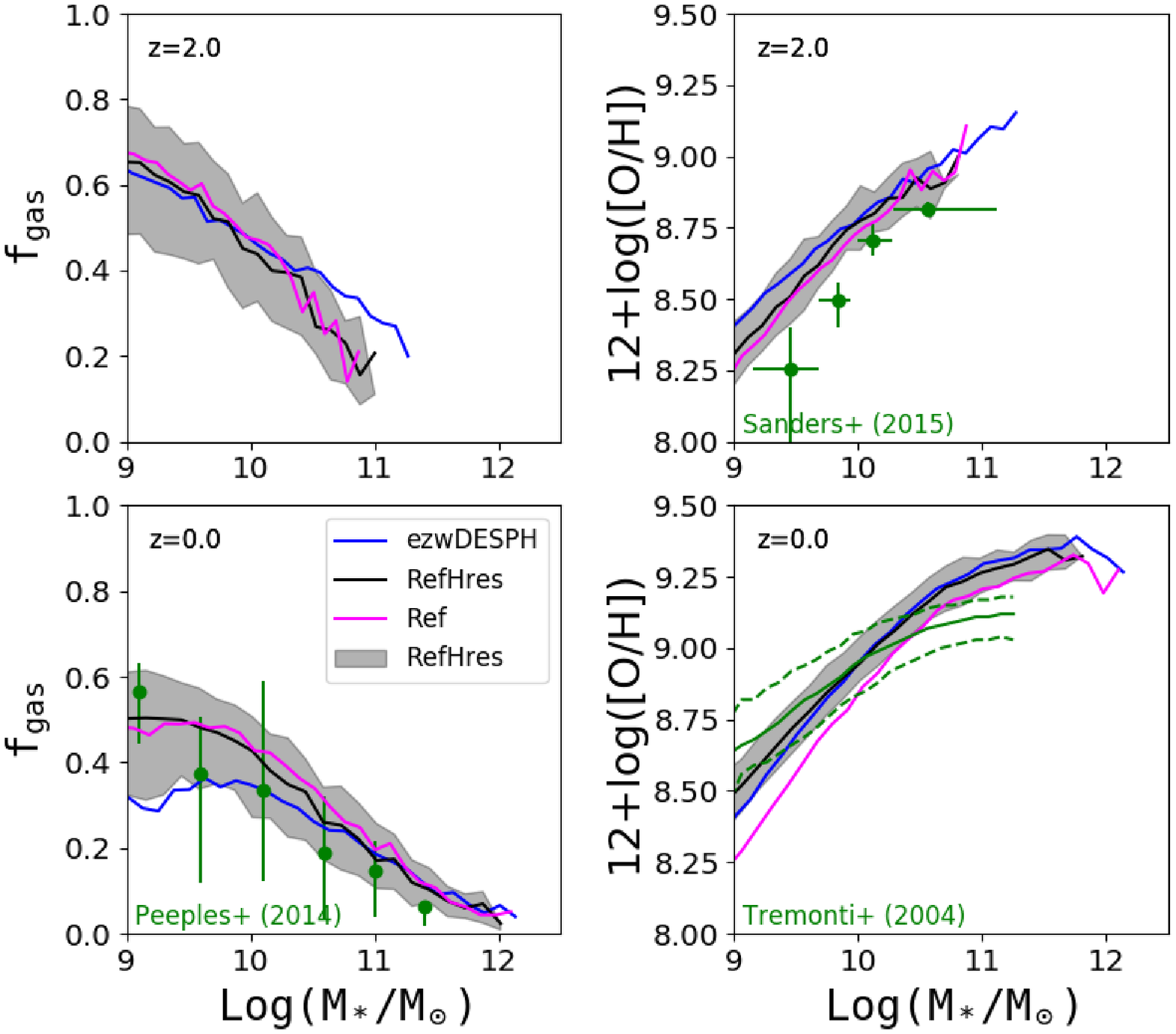}
\caption{\textit{Left panels}: cold gas fractions (defined in the text) as a
function of stellar mass at $z = 2$ (\textit{upper panel}) and $z = 0$
(\textit{lower panel}). The observational data in the lower panel are compiled
by \citet{peeples14}, with error bars denoting the 16\% to 84\% range.
\textit{Right panels}: gas-phase mass-metallicity relations at $z = 2$
(\textit{upper panel}) and $z = 0$ (\textit{lower panel}). The $z = 2$ data are from
\citet{sanders15} and the $z = 0$ data are from \citet{tremonti04}. The shaded area in
each panel shows the 16\% to 84\% range of the results from the fiducial high
resolution RefHres simulation. We also show the medians from the lower resolution Ref as
magenta lines.}
\label{fig:fgas_mzr}
\end{figure*}

\fig{fig:fgas_mzr} shows the cold gas fractions $\fgas$ at $z = 0$ and $z = 2$ in
the left panels. In the simulations, we define $\fgas$ as
\begin{equation}
\label{eqn:f_gas}
\fgas \equiv \frac{\mgas}{\mgas + M_{*}}
\end{equation}
where $M_{*}$ is the total stellar mass of the galaxy, and $\mgas$ is the total
mass of the ISM gas in that galaxy. To determine which SPH particles are
treated as multi-phase ISM gas in our simulations, we assume a physical density
threshold of $n_H > 0.13\ \mathrm{cm}^{-3}$ and a temperature threshold of
$\log T/\mathrm{K} < 4.5$. Any SPH particles within a galaxy that meets these
criteria are included when calculating $\mgas$. These sharp thresholds are
somewhat arbitrary so comparisons to observations should be interpreted with caution
\citep{dave11b}.

At $z = 0$, we add the data from \citet{peeples14}, which are compiled from
various data sets \citep{mcgaugh05, mcgaugh12, leroy08, saintonge11}. The data
points show the averaged atomic + molecular gas fractions in each stellar mass
bin, with error bars indicating the 16th and 84th percentiles, which is the same
range chosen for the simulated data. Our fiducial simulation reproduces the
observed trend very well, though it slightly over-predicts the cold fractions
in massive galaxies.

Compared to the ezwDESPH simulation, small galaxies with $\log(M_*/M_\odot) < 10$
in the fiducial simulation are more gas rich, making the scaling relation at
the faint end agree with the
observations from \citet{peeples14}. Comparing the Ref and RefHres in the
figure shows that this result is resolution independent.
At the massive end, the gas fractions in the fiducial simulation are close to
those from the ezwDESPH simulation. Galaxies in the ezwDESPH
simulation are in general
more massive than their counterparts in the fiducial simulation. Therefore, at
a fixed halo mass galaxies in the fiducial simulation actually contain a higher
gas mass. In future work, we will track gas accretion through cosmic time
in detail to understand the origin of cold gas in galaxies.

At $z = 2$, a detailed comparison with observations is unavailable owing to a lack
of direct measurements of the cold gas content at high redshift. Nevertheless,
it is consistent with the indirect observations of cold gas
\citep[e.g.][]{popping15} that the gas fractions are generally higher than at
$z = 0$ at a fixed stellar mass. 
The differences between the fiducial and the ezwDESPH
simulation are much larger at this redshift, with much lower gas fractions in
massive galaxies than in the fiducial simulation.


In the right panels of \fig{fig:fgas_mzr}, we compare the gas-phase
mass-metallicity relations (MZRs) from our fiducial simulation with observations. We
calculate the gas-phase metallicity for each galaxy by averaging over all
ISM particles within the galaxy, weighted by their SFR. We use oxygen as the
metallicity tracer and adopt a solar value of $[\mathrm{O/H}]_\odot + 12 = 8.69$
\citep{asplund09}. We use the \citet{sanders15} ($z \sim 2.3$) and
\citet{tremonti04} data for comparisons. Since they measure metallicity
using different calibrations, we convert the \citet{sanders15} data to the
\citet{tremonti04} calibration using the fitting formula from \citet{kewley08}.
This increases the overall normalisation of the \citet{sanders15} data by 0.1
to 0.3 dex.

At face value, the comparison in \fig{fig:fgas_mzr} shows a
slight overproduction of
gas phase metallicity with the right overall trend at $z=2$, but a more severe
discrepancy at $z=0$ where the simulations underpredict the metallicity of low mass
galaxies and overpredict the metallicity of high mass galaxies. The caveat is that
calibration and measurement uncertainties have a large impact on the observed
mass metallicity relation \citep{kewley08}. Furthermore, the initial mass function
(IMF) averaged oxygen yield is
uncertain, and it could change with galaxy mass if the IMF itself changes. In
principle, the mass metallicity relation is a strong diagnostic of outflow efficiency
\citep{finlator08}, and it should also be sensitive to the amount of metal recycling
in winds.

\subsection{Intergalactic and Circumgalactic Medium}
Galactic winds are not only important as a feedback mechanism that suppresses
galaxy growth, but are also essential to explain the enrichment of the
intergalactic medium (IGM) and the circumgalactic medium (CGM) as they carry
the metals that are produced inside the galaxy into the outer halo and beyond.
Measurements of the metal content in the IGM/CGM using quasar absorption
spectroscopy (see \citet{tumlinson17} for a review) provide crucial constraints
for cosmological simulations \citep{oppenheimer06, oppenheimer12, ford14,
ford16}. In this section, we show how the new wind algorithm in our fiducial
simulation affects the metal distributions in the IGM/CGM.

To mimic the observational measurements, we create mock quasar absorption
spectra using \textsc{specexbin} as in \citet{huang19}. A more detailed
description of the technique can be found in \citet{oppenheimer06}. In short,
we generate random sightlines covering a redshift range from $z = 0$ to $z =
0.5$ through the simulation volume. On each of these long sightlines, we
calculate the optical depth of multiple ions in redshift space based on the
properties of the surrounding gas, such as the density, temperature, velocity
and metallicity. We use a uniform ultraviolet background \citep{haardt12} to
calculate the ionization level of each ion. We normalise the strength of the
background to match the Lyman $\alpha$ decrement measurements \citep{huang19}.
From these mock spectra, we further
obtain observational quantities such as column densities and equivalent
widths for each ion by fitting their line profiles using the Voigt profile
fitting software \textsc{AUTOVP} \citep{dave97}. In this paper,
we generate 71 sightlines for each of the low resolution simulations 
and 400 sightlines for the RefHres simulation.

\begin{figure} 
\centering
\includegraphics[width=0.96\columnwidth]{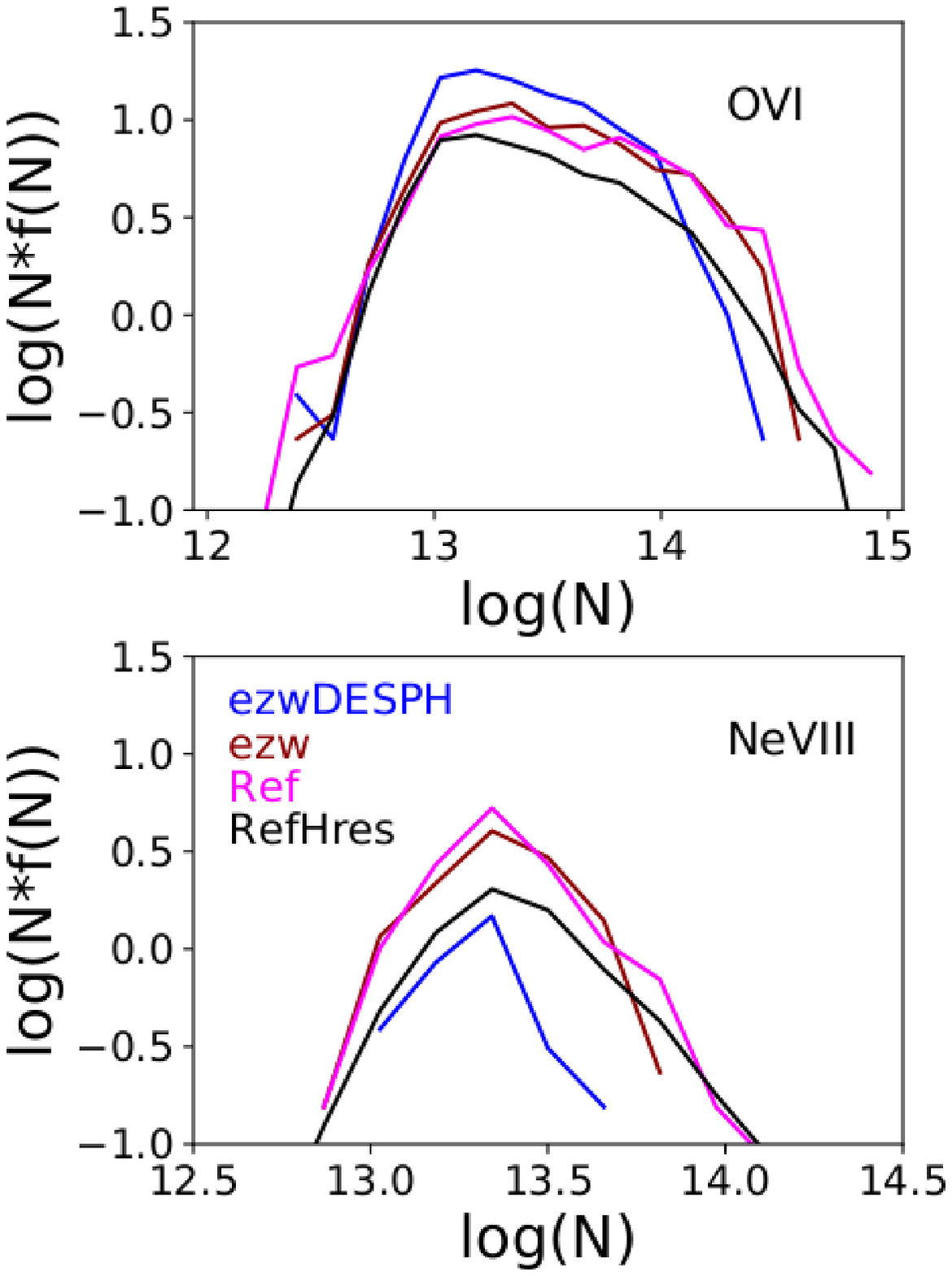}
\caption{The column density distributions of OVI (\textit{upper panel}) 
and NeVIII (\textit{lower panel}). 
We obtain the statistics from random sightlines that span from $z=0.0$ to 
$z=0.5$ as described in the text. Results from the four simulations are colour
coded according to \tab{tab:simulations}}
\label{fig:cddz} 
\end{figure}

\fig{fig:cddz} compares the column density distributions (CDDs) of OVI and
NeVIII from the four simulations. Comparison of ezw to ezwDESPH shows that
numerics have a strong effect on the CDDs
of these ions, as shown by \citet{huang19}. The new wind model
(Ref) slightly increases the number of high column density absorbers
compared to the \ezw wind (ezw) but does not strongly affect the low
column density absorbers. The CDDs are also sensitive to numerical resolution,
as the higher resolution simulation RefHres has fewer absorbers than the lower
resolution simulation.

\begin{figure*} 
\centering
\includegraphics[width=1.9\columnwidth]{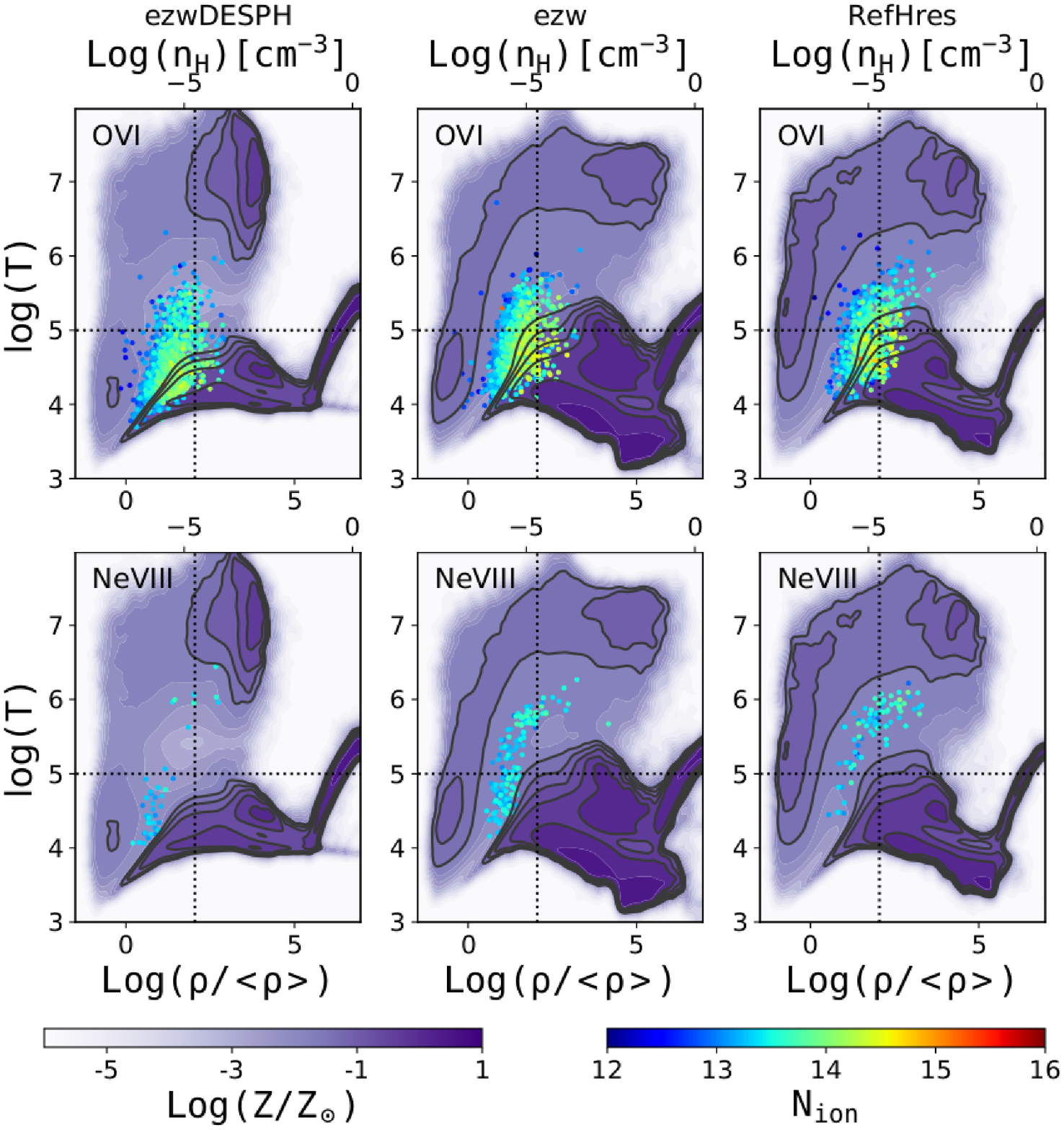}
\caption{The metal distributions at $z = 0$ in phase space from the
ezwDESPH (\textit{left}), ezw (\textit{middle}) and RefHres
(\textit{right}) simulations. The purple background colour scale indicates the
mass weighted average metallicity in each cell. In each panel, we show the OVI
absorbers or the NeVIII absorbers on random sightlines that are
generated using the technique described in the text. The absorbers are colour
coded according to their column densities. The two dotted lines in each panel
divide the phase space into four regions: the warm-hot
intergalactic medium (WHIM, upper left), the diffuse IGM (lower left), hot
halo gas (upper right), and cold dense galactic gas (lower right).
Several contours lines are stressed for better visualisation.}
\label{fig:zrhot_ovi_neviii} 
\end{figure*}

The contours in \fig{fig:zrhot_ovi_neviii} show how metals are distributed in
the temperature-density phase space at $z = 0$ in the three simulations.
Comparing the ezw (middle panels) and the ezwDESPH (left panels) shows the
effects of changing numerics and cooling physics on the metal distributions and
the high-ion absorbers. We have studied those effects in greater detail in
previous work \citep{huang19}. The main effect is that there are more metals
in the warm-hot gas (WHIM; upper-left quadrants) 
owing to better resolved shocks around filaments.

Comparing the RefHres simulation (right panels) to the ezw
simulation (middle panels) shows that the new wind algorithm spreads a
considerable amount of metals into the warm-hot IGM gas and the hot, dense gas
as a result of both the stronger mass loading in low mass galaxies and the
faster wind speed. Since we do not allow metal mixing between the enriched wind
particles and the pristine IGM gas, the enhanced metallicity at below cosmic
mean density comes directly from wind particles that escape into the IGM. The
higher metallicity in the hot gas is likely because of wind particles being
able to remain longer in hot haloes before re-accreting onto the galaxies.


One numerical caveat is that when the original
\textsc{specexbin} calculates the local gas properties such as the temperature
at a given location in a sightline, it averages over all neighbouring particles
close to the sightline without distinguishing wind particles from normal SPH
particles. This potentially leads to errors in a multi-phase gas, such as when
cold, metal-rich wind particles are among hot CGM particles. Therefore, we
modified \textsc{specexbin} to take into account the contribution of each
surrounding particle to the spectra on a particle-by-particle basis. However, we
do not find any significant differences in the results for the high ions from using these two
different methods.

%
%

\section{Summary}
\label{sec:summary}

Galactic winds are crucial to galaxy formation. At present, hydrodynamic
simulations that model cosmological volumes (i.e., many Mpc on a side) lack the
resolution to generate winds from physical processes in the ISM. Such 
simulations, therefore,  employ
sub-grid prescriptions that are designed to capture the phenomenological
behaviours of these processes. In this paper, we revisit a wind implementation
that is based on a numerical algorithm proposed and developed by
\citet{springel05}, \citet{oppenheimer06}, and \citet{dave13}. We take into
account new constraints
from high-resolution zoom-in simulations \citepalias{muratov15} and statistical
properties of galaxies at high redshift, such as their stellar mass functions,
and
make several changes to our wind algorithm. We examine the ability of the new
algorithm to reproduce a wide range of observations and study the sensitivity
of these predictions to variations in model parameters.

The basic design of the wind algorithm is that in star forming galaxies, cold and dense SPH particles are stochastically ejected from their host
galaxies with an initial momentum kick to model large-scale star
formation driven winds. The mass loading factor $\eta$ determines the rate at
which particles are ejected and the wind speed $v_w$ determines the initial
velocity given to the ejected particles. Observations and analytic calculations
have shown that both of these parameters correlate with properties of their
host galaxy or host halo such as the star formation rate and the characteristic
velocity $\sigma$ \citep{rupke05}, but an accurate determination of these
scalings is unknown. Previous wind algorithms often parameterise them as
$\eta \propto \sigma^{-1}$ or $\sigma^{-2}$, and $v_w \propto \sigma$,
following the analytic formulation for momentum-driven or energy-driven winds
\citep{murray05}. 

However, it becomes clear in cosmological simulations that artificial numerical
treatments as well as fine tuning of the model parameters are required to
successfully reproduce key observables, such as the galaxy stellar mass 
function, owing to limitations in the numerical resolution of simulations and
the simplicity of the analytic models. Furthermore, recent zoom-in galaxy 
simulations \citep[e.g.][]{muratov15} suggest different wind scalings than the 
analytic models. Most importantly, simulations necessarily impose these scalings at wind
launch, while they are supposed to hold for gas that has escaped the dense ISM. When we measure 
the resultant wind scalings outside of galaxies, the original scalings no longer
hold. We have therefore altered our wind launch algorithm to reproduce, approximately,
the wind properties measured by \citetalias{muratov15} at 25\% of the halo virial
radius.


Major updates from our previous wind algorithm include: 1. We allow more
freedom when assigning $\eta$ and $v_w$. In particular, we allow a
stronger dependence of $\eta$ on $\sigma$, or equivalently, the halo mass. 2.
We allow newly ejected wind particles to temporarily decouple from their host
galaxies dynamically before they reach a density threshold of $0.1\rho_{SF}$.
The new algorithm may appear to be less deterministic than the original one by
having a few more tunable parameters, but it is an unavoidable compromise to
the uncertainties and limitations of our current knowledge of the nature of galactic winds. The primary focus of this paper is, therefore, not to extensively
search for a set of parameters that best reproduce the observed Universe but
rather to explore and characterise how some of the well-established
observational results on galaxy formation could be affected by a physically
plausible range of wind model parameters. Naturally, we perform this 
exploration within the narrow confines of our wind model.
Differences between the methods used by different simulation groups in the 
literature could be larger.

We find that the faint-end slopes of the GSMFs at $z > 1$ in our simulations
are most sensitive to the power-law index $\beta_\eta$, which determines how
strongly the mass loading factor $\eta$ depends on $\sigma$ in low mass galaxies
(\sect{sec:beta_dependence}).
The energy-driven scaling $\eta \propto \sigma^{-2}$ that was used in our previous simulations
\citep[e.g.][]{dave13} produces a faint-end slope that is too steep compared to
observations. We find that to match the observed flatter slope, we need a
scaling as steep as $\eta \propto \sigma^{-5}$
for $\sigma < 106\kms$ in our fiducial simulation. All of our simulations adopt
$\eta \propto \sigma^{-1}$ at high masses. The need for such 
a strong scaling at low $\sigma$
has also been found in the FIRE simulations, which predict an intermediate
scaling of $\eta \propto \sigma^{-3.3}$ \citepalias{muratov15}, as well as in
semi-analytic works \citep{somerville12, lu14, peeples11} and other cosmological
simulations \citep{pillepich18a} that include kinetic feedback. Even though
$\beta_\eta$ critically affects sub-$L_*$ galaxies at $z = 1$, the different
scalings adopted in our test simulations produce similar faint-end slopes of
the GSMFs at $z = 0$ and also have little effect on the final masses of massive
galaxies. This emphasises that robust statistical properties of dwarf galaxies
at high redshifts are essential to distinguish between different feedback models
and to understand how stellar feedback regulates galaxy growth. Such 
observations will have to await the launch of JWST.

Changing the overall strength of outflows by changing the normalisation factor
$\alpha_\eta$ also has a clear effect on galaxy growth, with a higher mass
loading factor leading to less star formation, especially in dwarf galaxies at
high redshifts (Figure \ref{fig:gsmfs_calib_linfac}). This dependence of $M_*$
on $\alpha_\eta$ can be
qualitatively explained by a simple analytic model that assumes isolated galaxy
growth and negligible wind recycling (Figure \ref{fig:pairs_mstar_linfac}).

The evolution of wind particles in a halo is very sensitive to the initial wind
speed and the gravitational potential near the centre, which is usually
dominated by baryons and is numerically poorly resolved. The winds launched
with our new method have wind velocities that agree with the FIRE simulations
\citepalias{muratov15} at $R_{25}$, while those launched with the original velocity
formula often lose most of their momentum at small radii and even fail to
reach $R_{25}$ in massive haloes (\fig{fig:v25vc}). As a consequence, the initial wind speed has a
strong impact on the growth of massive galaxies. Contrary to some previous
findings that the stellar feedback is only efficient enough to suppress star
formation in sub-$L_*$ galaxies, in some of our simulations, including the
fiducial simulation, the fast winds do significantly reduce star formation in
massive galaxies and bring the massive end of the predicted GSMF at $z = 0$
much closer to observations as long as they are capable of escaping their host
galaxies instead of almost instantly falling back as in our original algorithm. 

Note, however, that the FIRE simulations only explore haloes as massive as
$10^{13}\ \msun$. Below this mass scale, the FIRE simulations are able to reproduce
the stellar mass – halo mass relation without any AGN feedback 
\citep{feldmann17}, supporting our results that stellar feedback alone might be
sufficient to suppress star formation up to this mass scale. However, in our
wind algorithm, we extrapolate the empirical relation between $v_{25}$ and
$v_c$ to even more massive systems by adjusting the initial wind velocities.
Therefore, our results at the massive end of the GSMFs should not be
interpreted as a consequence derived from physical assumptions but rather show
that the wind speed, as well as how winds propagate and stay in the halo, have
strong effects on galaxy evolution. 

We further study how the initial wind speed could affect our simulation
results by comparing our fiducial simulation with the RefSlow simulation,
a simulation with slower wind speeds (Figures \ref{fig:gsmfs_calib_vwind},
\ref{fig:sde_calib} and \ref{fig:smhms_calib}).
Changing the wind speed significantly affects star formation in massive galaxies
but has little effect in low-mass galaxies. In the most massive galaxies of the
two simulations, the average wind speed differs by a factor of $\sim 2$,
and the stellar masses differ by $\sim 0.2 - 0.4$ dex at different redshifts.
This leads to clear differences at the massive end of the GSMFs, where
the statistical variance is large. 

The faster wind speeds
in the fiducial simulation relative to our older \ezw algorithm drive wind particles further from
their host galaxy. It also heats more wind particles to the temperature of the
hot corona, making them have to cool before reaccreting and hence reducing
their reaccretion rate (Figures \ref{fig:sfhistoryz2}, \ref{fig:sfhistory1} and 
\ref{fig:sfhistory2}). Both effects
lengthen the recycling time of the wind particles and make wind recycling less
efficient than simulations with slower wind speeds (Figure \ref{fig:trec}). However, wind recycling
still dominates accretion onto the massive galaxies at low redshifts, fuelling
too much late star formation. Hot accretion is also responsible for 25\% of the
total mass of stars formed in the massive galaxies at $z=0$ and also needs to
be significantly suppressed to have these galaxy stellar masses match observations. Mergers
play only a limited role in the growth of massive galaxies and are nearly
negligible for low-mass and intermediate-mass 
galaxies (Figure \ref{fig:fmergers}). 
However, if one removes
all the late time star formation in massive galaxies required to match 
observations, the merger growth could become much more important.

This sensitivity to the initial launch speed also implies that the simulations
are sensitive to numerical resolution that affects the accuracy of force
calculations near the centre of the haloes and the physical assumptions that
governs the propagation of winds in the haloes. We empirically find that in our
fiducial simulation RefHres, which has twice the spatial resolution and
eight times the mass resolution as
the other simulations, we need to enhance the wind speed by an overall factor
of $\sim 1.14$ to match the constraints at $R_{25}$. After this correction, the
galaxy properties of the fiducial simulation are similar to those of the
corresponding lower resolution simulation. It implies that recalibration of
the initial wind speed at different resolutions is necessary in sub-grid wind
implementations that are similar to ours. Instead of matching observational
constraints such as the stellar mass functions, it is likely sufficient to tune
the parameters to reproduce the same wind speed at a certain radius, after
which wind propagation becomes largely independent of resolution. In this work,
we choose $R_{25}$ because of the constraints from the FIRE simulations \citepalias{muratov15}.

With the new wind model and a fiducial set of wind parameters, 
we run a simulation (RefHres) with higher numerical resolution than these
test simulations. 
This simulation results in GSMFs, SMHMs, and SDEs that are in much better
agreement with observations at all redshifts than the original \ezw wind.
However, it still produces too many stars in massive galaxies at $z=0$. 
The cold gas fractions agree well with observations 
and are not significantly affected by the new wind algorithm. The fiducial
simulation produces slightly more high column density absorbers for high ions
such as OVI and NeVIII, but this result is sensitive to numerical resolution.

Despite many changes in both numerical algorithms and wind implementations, our
new simulations confirm three key conclusions of our previous work: cold accretion
produces most of the gas that forms stars in low mass haloes, hot accretion takes
over from cold accretion in high mass haloes, and wind recycling is an essential
component of galaxy growth at redshifts $z<1$
\citep{keres05, keres09a, keres09b, oppenheimer10}. However, the details of the
wind implementation have a large impact on the amount and mass dependence of
wind recycling. Reproducing the 
observed stellar masses in high mass haloes likely
requires an additional mechanism that suppresses hot gas accretion, and AGN feedback
is a natural candidate for this mechanism \citep{benson03, croton06, bower06}.
However, we should be cautious in drawing lessons about AGN feedback scaling
because the amount of feedback required is sensitive to still uncertain aspects of
galactic winds driven by stellar feedback. In this paper we have focused on the
effects of wind launch algorithms, but our simulations also suffer from under-resolving
the physics of ejected wind gas after it has entered the CGM. This is probably true
of all current cosmological simulations, even zoom-in simulations that attempt to
resolve parsec-level structure on the ISM. Forcing high resolution in the CGM is
one approach to this problem \citep{peeples19, vandevoort19, hummels19}, though even so
it may be difficult to resolve the relevant scales of instabilities and fluid mixing
\citep{scannapieco15, schneider17}. Another approach is to develop an explicit
sub-grid model for evolving wind particles after they leave the galaxy, so that
wind propagation and recycling, which we have shown to critically affect many
simulation results, are controlled by physical parameters instead of unresolved
numerics. We will present such a model in future work.

\section*{Acknowledgements}

We acknowledge support by NSF grant AST-1517503,
NASA ATP grant 80NSSC18K1016, and HST Theory grant HST-AR-14299.
DW acknowledges support of NSF grant AST-1909841.

\bibliography{references}

\begin{thebibliography}{}
\makeatletter
\relax
\def\mn@urlcharsother{\let\do\@makeother \do\$\do\&\do\#\do\^\do\_\do\%\do\~}
\def\mn@doi{\begingroup\mn@urlcharsother \@ifnextchar [ {\mn@doi@}
  {\mn@doi@[]}}
\def\mn@doi@[#1]#2{\def\@tempa{#1}\ifx\@tempa\@empty \href
  {http://dx.doi.org/#2} {doi:#2}\else \href {http://dx.doi.org/#2} {#1}\fi
  \endgroup}
\def\mn@eprint#1#2{\mn@eprint@#1:#2::\@nil}
\def\mn@eprint@arXiv#1{\href {http://arxiv.org/abs/#1} {{\tt arXiv:#1}}}
\def\mn@eprint@dblp#1{\href {http://dblp.uni-trier.de/rec/bibtex/#1.xml}
  {dblp:#1}}
\def\mn@eprint@#1:#2:#3:#4\@nil{\def\@tempa {#1}\def\@tempb {#2}\def\@tempc
  {#3}\ifx \@tempc \@empty \let \@tempc \@tempb \let \@tempb \@tempa \fi \ifx
  \@tempb \@empty \def\@tempb {arXiv}\fi \@ifundefined
  {mn@eprint@\@tempb}{\@tempb:\@tempc}{\expandafter \expandafter \csname
  mn@eprint@\@tempb\endcsname \expandafter{\@tempc}}}

\bibitem[\protect\citeauthoryear{{Agertz}, {Kravtsov}, {Leitner}  \&
  {Gnedin}}{{Agertz} et~al.}{2013}]{agertz13}
{Agertz} O.,  {Kravtsov} A.~V.,  {Leitner} S.~N.,   {Gnedin} N.~Y.,  2013,
  \mn@doi [\apj] {10.1088/0004-637X/770/1/25}, \href
  {https://ui.adsabs.harvard.edu/abs/2013ApJ...770...25A} {770, 25}

\bibitem[\protect\citeauthoryear{{Angl{\'e}s-Alc{\'a}zar},
  {Faucher-Gigu{\`e}re}, {Kere{\v{s}}}, {Hopkins}, {Quataert}  \&
  {Murray}}{{Angl{\'e}s-Alc{\'a}zar} et~al.}{2017}]{angles-alcazar17}
{Angl{\'e}s-Alc{\'a}zar} D.,  {Faucher-Gigu{\`e}re} C.-A.,  {Kere{\v{s}}} D.,
  {Hopkins} P.~F.,  {Quataert} E.,   {Murray} N.,  2017, \mn@doi [\mnras]
  {10.1093/mnras/stx1517}, \href
  {https://ui.adsabs.harvard.edu/abs/2017MNRAS.470.4698A} {470, 4698}

\bibitem[\protect\citeauthoryear{{Asplund}, {Grevesse}, {Sauval}  \&
  {Scott}}{{Asplund} et~al.}{2009}]{asplund09}
{Asplund} M.,  {Grevesse} N.,  {Sauval} A.~J.,   {Scott} P.,  2009, \mn@doi
  [\araa] {10.1146/annurev.astro.46.060407.145222}, \href
  {https://ui.adsabs.harvard.edu/abs/2009ARA&A..47..481A} {47, 481}

\bibitem[\protect\citeauthoryear{{Baldry} et~al.,}{{Baldry}
  et~al.}{2012}]{baldry12}
{Baldry} I.~K.,  et~al., 2012, \mn@doi [\mnras]
  {10.1111/j.1365-2966.2012.20340.x}, \href
  {https://ui.adsabs.harvard.edu/abs/2012MNRAS.421..621B} {421, 621}

\bibitem[\protect\citeauthoryear{{Behroozi}, {Wechsler}  \&
  {Conroy}}{{Behroozi} et~al.}{2013}]{behroozi13}
{Behroozi} P.~S.,  {Wechsler} R.~H.,   {Conroy} C.,  2013, \mn@doi [\apj]
  {10.1088/0004-637X/770/1/57}, \href
  {https://ui.adsabs.harvard.edu/abs/2013ApJ...770...57B} {770, 57}

\bibitem[\protect\citeauthoryear{{Benson}, {Bower}, {Frenk}, {Lacey}, {Baugh}
  \& {Cole}}{{Benson} et~al.}{2003}]{benson03}
{Benson} A.~J.,  {Bower} R.~G.,  {Frenk} C.~S.,  {Lacey} C.~G.,  {Baugh} C.~M.,
    {Cole} S.,  2003, \mn@doi [\apj] {10.1086/379160}, \href
  {https://ui.adsabs.harvard.edu/abs/2003ApJ...599...38B} {599, 38}

\bibitem[\protect\citeauthoryear{{Bernardi}, {Meert}, {Sheth}, {Vikram},
  {Huertas-Company}, {Mei}  \& {Shankar}}{{Bernardi} et~al.}{2013}]{bernardi13}
{Bernardi} M.,  {Meert} A.,  {Sheth} R.~K.,  {Vikram} V.,  {Huertas-Company}
  M.,  {Mei} S.,   {Shankar} F.,  2013, \mn@doi [\mnras]
  {10.1093/mnras/stt1607}, \href
  {https://ui.adsabs.harvard.edu/abs/2013MNRAS.436..697B} {436, 697}

\bibitem[\protect\citeauthoryear{{Bower}, {Benson}, {Malbon}, {Helly}, {Frenk},
  {Baugh}, {Cole}  \& {Lacey}}{{Bower} et~al.}{2006}]{bower06}
{Bower} R.~G.,  {Benson} A.~J.,  {Malbon} R.,  {Helly} J.~C.,  {Frenk} C.~S.,
  {Baugh} C.~M.,  {Cole} S.,   {Lacey} C.~G.,  2006, \mn@doi [\mnras]
  {10.1111/j.1365-2966.2006.10519.x}, \href
  {http://adsabs.harvard.edu/abs/2006MNRAS.370..645B} {370, 645}

\bibitem[\protect\citeauthoryear{{Chabrier}}{{Chabrier}}{2003}]{chabrier03}
{Chabrier} G.,  2003, \mn@doi [\pasp] {10.1086/376392}, \href
  {http://adsabs.harvard.edu/abs/2003PASP..115..763C} {115, 763}

\bibitem[\protect\citeauthoryear{{Christensen}, {Dav{\'e}}, {Governato},
  {Pontzen}, {Brooks}, {Munshi}, {Quinn}  \& {Wadsley}}{{Christensen}
  et~al.}{2016}]{christensen16}
{Christensen} C.~R.,  {Dav{\'e}} R.,  {Governato} F.,  {Pontzen} A.,  {Brooks}
  A.,  {Munshi} F.,  {Quinn} T.,   {Wadsley} J.,  2016, \mn@doi [\apj]
  {10.3847/0004-637X/824/1/57}, \href
  {https://ui.adsabs.harvard.edu/abs/2016ApJ...824...57C} {824, 57}

\bibitem[\protect\citeauthoryear{{Conroy}, {Gunn}  \& {White}}{{Conroy}
  et~al.}{2009}]{conroy09}
{Conroy} C.,  {Gunn} J.~E.,   {White} M.,  2009, \mn@doi [\apj]
  {10.1088/0004-637X/699/1/486}, \href
  {https://ui.adsabs.harvard.edu/abs/2009ApJ...699..486C} {699, 486}

\bibitem[\protect\citeauthoryear{{Crain} et~al.,}{{Crain}
  et~al.}{2015}]{crain15}
{Crain} R.~A.,  et~al., 2015, \mn@doi [\mnras] {10.1093/mnras/stv725}, \href
  {https://ui.adsabs.harvard.edu/abs/2015MNRAS.450.1937C} {450, 1937}

\bibitem[\protect\citeauthoryear{{Croton} et~al.,}{{Croton}
  et~al.}{2006}]{croton06}
{Croton} D.~J.,  et~al., 2006, \mn@doi [\mnras]
  {10.1111/j.1365-2966.2005.09675.x}, \href
  {http://adsabs.harvard.edu/abs/2006MNRAS.365...11C} {365, 11}

\bibitem[\protect\citeauthoryear{{Cullen} \& {Dehnen}}{{Cullen} \&
  {Dehnen}}{2010}]{cd10}
{Cullen} L.,  {Dehnen} W.,  2010, \mn@doi [\mnras]
  {10.1111/j.1365-2966.2010.17158.x}, \href
  {http://adsabs.harvard.edu/abs/2010MNRAS.408..669C} {408, 669}

\bibitem[\protect\citeauthoryear{{Dalla Vecchia} \& {Schaye}}{{Dalla Vecchia}
  \& {Schaye}}{2008}]{dallavecchia08}
{Dalla Vecchia} C.,  {Schaye} J.,  2008, \mn@doi [\mnras]
  {10.1111/j.1365-2966.2008.13322.x}, \href
  {http://adsabs.harvard.edu/abs/2008MNRAS.387.1431D} {387, 1431}

\bibitem[\protect\citeauthoryear{{Dav{\'e}}, {Hernquist}, {Weinberg}  \&
  {Katz}}{{Dav{\'e}} et~al.}{1997}]{dave97}
{Dav{\'e}} R.,  {Hernquist} L.,  {Weinberg} D.~H.,   {Katz} N.,  1997, \mn@doi
  [\apj] {10.1086/303712}, \href
  {http://adsabs.harvard.edu/abs/1997ApJ...477...21D} {477, 21}

\bibitem[\protect\citeauthoryear{{Dav{\'e}}, {Oppenheimer}, {Katz}, {Kollmeier}
   \& {Weinberg}}{{Dav{\'e}} et~al.}{2010}]{dave10}
{Dav{\'e}} R.,  {Oppenheimer} B.~D.,  {Katz} N.,  {Kollmeier} J.~A.,
  {Weinberg} D.~H.,  2010, \mn@doi [\mnras] {10.1111/j.1365-2966.2010.17279.x},
  \href {http://adsabs.harvard.edu/abs/2010MNRAS.408.2051D} {408, 2051}

\bibitem[\protect\citeauthoryear{{Dav{\'e}}, {Oppenheimer}  \&
  {Finlator}}{{Dav{\'e}} et~al.}{2011a}]{dave11a}
{Dav{\'e}} R.,  {Oppenheimer} B.~D.,   {Finlator} K.,  2011a, \mn@doi [\mnras]
  {10.1111/j.1365-2966.2011.18680.x}, \href
  {http://adsabs.harvard.edu/abs/2011MNRAS.415...11D} {415, 11}

\bibitem[\protect\citeauthoryear{{Dav{\'e}}, {Finlator}  \&
  {Oppenheimer}}{{Dav{\'e}} et~al.}{2011b}]{dave11b}
{Dav{\'e}} R.,  {Finlator} K.,   {Oppenheimer} B.~D.,  2011b, \mn@doi [\mnras]
  {10.1111/j.1365-2966.2011.19132.x}, \href
  {http://adsabs.harvard.edu/abs/2011MNRAS.416.1354D} {416, 1354}

\bibitem[\protect\citeauthoryear{{Dav{\'e}}, {Katz}, {Oppenheimer}, {Kollmeier}
   \& {Weinberg}}{{Dav{\'e}} et~al.}{2013}]{dave13}
{Dav{\'e}} R.,  {Katz} N.,  {Oppenheimer} B.~D.,  {Kollmeier} J.~A.,
  {Weinberg} D.~H.,  2013, \mn@doi [\mnras] {10.1093/mnras/stt1274}, \href
  {http://adsabs.harvard.edu/abs/2013MNRAS.434.2645D} {434, 2645}

\bibitem[\protect\citeauthoryear{{Dav{\'e}}, {Thompson}  \&
  {Hopkins}}{{Dav{\'e}} et~al.}{2016}]{mufasa}
{Dav{\'e}} R.,  {Thompson} R.,   {Hopkins} P.~F.,  2016, \mn@doi [\mnras]
  {10.1093/mnras/stw1862}, \href
  {http://adsabs.harvard.edu/abs/2016MNRAS.462.3265D} {462, 3265}

\bibitem[\protect\citeauthoryear{{Feldmann}, {Quataert}, {Hopkins},
  {Faucher-Gigu{\`e}re}  \& {Kere{\v{s}}}}{{Feldmann}
  et~al.}{2017}]{feldmann17}
{Feldmann} R.,  {Quataert} E.,  {Hopkins} P.~F.,  {Faucher-Gigu{\`e}re} C.-A.,
   {Kere{\v{s}}} D.,  2017, \mn@doi [\mnras] {10.1093/mnras/stx1120}, \href
  {https://ui.adsabs.harvard.edu/abs/2017MNRAS.470.1050F} {470, 1050}

\bibitem[\protect\citeauthoryear{{Finlator} \& {Dav{\'e}}}{{Finlator} \&
  {Dav{\'e}}}{2008}]{finlator08}
{Finlator} K.,  {Dav{\'e}} R.,  2008, \mn@doi [\mnras]
  {10.1111/j.1365-2966.2008.12991.x}, \href
  {http://adsabs.harvard.edu/abs/2008MNRAS.385.2181F} {385, 2181}

\bibitem[\protect\citeauthoryear{{Ford}, {Oppenheimer}, {Dav{\'e}}, {Katz},
  {Kollmeier}  \& {Weinberg}}{{Ford} et~al.}{2013}]{ford13}
{Ford} A.~B.,  {Oppenheimer} B.~D.,  {Dav{\'e}} R.,  {Katz} N.,  {Kollmeier}
  J.~A.,   {Weinberg} D.~H.,  2013, \mn@doi [\mnras] {10.1093/mnras/stt393},
  \href {http://adsabs.harvard.edu/abs/2013MNRAS.432...89F} {432, 89}

\bibitem[\protect\citeauthoryear{{Ford}, {Dav{\'e}}, {Oppenheimer}, {Katz},
  {Kollmeier}, {Thompson}  \& {Weinberg}}{{Ford} et~al.}{2014}]{ford14}
{Ford} A.~B.,  {Dav{\'e}} R.,  {Oppenheimer} B.~D.,  {Katz} N.,  {Kollmeier}
  J.~A.,  {Thompson} R.,   {Weinberg} D.~H.,  2014, \mn@doi [\mnras]
  {10.1093/mnras/stu1418}, \href
  {http://adsabs.harvard.edu/abs/2014MNRAS.444.1260F} {444, 1260}

\bibitem[\protect\citeauthoryear{{Ford} et~al.,}{{Ford} et~al.}{2016}]{ford16}
{Ford} A.~B.,  et~al., 2016, \mn@doi [\mnras] {10.1093/mnras/stw595}, \href
  {http://adsabs.harvard.edu/abs/2016MNRAS.459.1745F} {459, 1745}

\bibitem[\protect\citeauthoryear{{Furlong} et~al.,}{{Furlong}
  et~al.}{2015}]{furlong15}
{Furlong} M.,  et~al., 2015, \mn@doi [\mnras] {10.1093/mnras/stv852}, \href
  {https://ui.adsabs.harvard.edu/abs/2015MNRAS.450.4486F} {450, 4486}

\bibitem[\protect\citeauthoryear{{Gonz{\'a}lez}, {Labb{\'e}}, {Bouwens},
  {Illingworth}, {Franx}  \& {Kriek}}{{Gonz{\'a}lez} et~al.}{2011}]{gonzalez11}
{Gonz{\'a}lez} V.,  {Labb{\'e}} I.,  {Bouwens} R.~J.,  {Illingworth} G.,
  {Franx} M.,   {Kriek} M.,  2011, \mn@doi [\apjl]
  {10.1088/2041-8205/735/2/L34}, \href
  {https://ui.adsabs.harvard.edu/abs/2011ApJ...735L..34G} {735, L34}

\bibitem[\protect\citeauthoryear{{Haardt} \& {Madau}}{{Haardt} \&
  {Madau}}{2012}]{haardt12}
{Haardt} F.,  {Madau} P.,  2012, \mn@doi [\apj] {10.1088/0004-637X/746/2/125},
  \href {http://adsabs.harvard.edu/abs/2012ApJ...746..125H} {746, 125}

\bibitem[\protect\citeauthoryear{{Heckman} \& {Borthakur}}{{Heckman} \&
  {Borthakur}}{2016}]{heckman16}
{Heckman} T.~M.,  {Borthakur} S.,  2016, \mn@doi [\apj]
  {10.3847/0004-637X/822/1/9}, \href
  {https://ui.adsabs.harvard.edu/abs/2016ApJ...822....9H} {822, 9}

\bibitem[\protect\citeauthoryear{{Hopkins}}{{Hopkins}}{2013}]{hopkins13}
{Hopkins} P.~F.,  2013, \mn@doi [\mnras] {10.1093/mnras/sts210}, \href
  {http://adsabs.harvard.edu/abs/2013MNRAS.428.2840H} {428, 2840}

\bibitem[\protect\citeauthoryear{{Hopkins}, {Quataert}  \& {Murray}}{{Hopkins}
  et~al.}{2012}]{hopkins12}
{Hopkins} P.~F.,  {Quataert} E.,   {Murray} N.,  2012, \mn@doi [\mnras]
  {10.1111/j.1365-2966.2012.20593.x}, \href
  {http://adsabs.harvard.edu/abs/2012MNRAS.421.3522H} {421, 3522}

\bibitem[\protect\citeauthoryear{{Hopkins} et~al.,}{{Hopkins}
  et~al.}{2018}]{fire2}
{Hopkins} P.~F.,  et~al., 2018, \mn@doi [\mnras] {10.1093/mnras/sty1690}, \href
  {http://adsabs.harvard.edu/abs/2018MNRAS.480..800H} {480, 800}

\bibitem[\protect\citeauthoryear{{Huang} et~al.,}{{Huang}
  et~al.}{2019}]{huang19}
{Huang} S.,  et~al., 2019, \mn@doi [\mnras] {10.1093/mnras/stz057}, \href
  {https://ui.adsabs.harvard.edu/abs/2019MNRAS.484.2021H} {484, 2021}

\bibitem[\protect\citeauthoryear{Hummels et~al.,}{Hummels
  et~al.}{2019}]{hummels19}
Hummels C.~B.,  et~al., 2019, \mn@doi [\apj] {10.3847/1538-4357/ab378f}, 882,
  156

\bibitem[\protect\citeauthoryear{{Ilbert} et~al.,}{{Ilbert}
  et~al.}{2013}]{ilbert13}
{Ilbert} O.,  et~al., 2013, \mn@doi [\aap] {10.1051/0004-6361/201321100}, \href
  {https://ui.adsabs.harvard.edu/abs/2013A&A...556A..55I} {556, A55}

\bibitem[\protect\citeauthoryear{{Kere{\v s}}, {Katz}, {Weinberg}  \&
  {Dav{\'e}}}{{Kere{\v s}} et~al.}{2005}]{keres05}
{Kere{\v s}} D.,  {Katz} N.,  {Weinberg} D.~H.,   {Dav{\'e}} R.,  2005, \mn@doi
  [\mnras] {10.1111/j.1365-2966.2005.09451.x}, \href
  {http://adsabs.harvard.edu/abs/2005MNRAS.363....2K} {363, 2}

\bibitem[\protect\citeauthoryear{{Kere{\v s}}, {Katz}, {Fardal}, {Dav{\'e}}  \&
  {Weinberg}}{{Kere{\v s}} et~al.}{2009a}]{keres09a}
{Kere{\v s}} D.,  {Katz} N.,  {Fardal} M.,  {Dav{\'e}} R.,   {Weinberg} D.~H.,
  2009a, \mn@doi [\mnras] {10.1111/j.1365-2966.2009.14541.x}, \href
  {http://adsabs.harvard.edu/abs/2009MNRAS.395..160K} {395, 160}

\bibitem[\protect\citeauthoryear{{Kere{\v s}}, {Katz}, {Dav{\'e}}, {Fardal}  \&
  {Weinberg}}{{Kere{\v s}} et~al.}{2009b}]{keres09b}
{Kere{\v s}} D.,  {Katz} N.,  {Dav{\'e}} R.,  {Fardal} M.,   {Weinberg} D.~H.,
  2009b, \mn@doi [\mnras] {10.1111/j.1365-2966.2009.14924.x}, \href
  {http://adsabs.harvard.edu/abs/2009MNRAS.396.2332K} {396, 2332}

\bibitem[\protect\citeauthoryear{{Kewley} \& {Ellison}}{{Kewley} \&
  {Ellison}}{2008}]{kewley08}
{Kewley} L.~J.,  {Ellison} S.~L.,  2008, \mn@doi [\apj] {10.1086/587500}, \href
  {https://ui.adsabs.harvard.edu/abs/2008ApJ...681.1183K} {681, 1183}

\bibitem[\protect\citeauthoryear{{Leroy}, {Walter}, {Brinks}, {Bigiel}, {de
  Blok}, {Madore}  \& {Thornley}}{{Leroy} et~al.}{2008}]{leroy08}
{Leroy} A.~K.,  {Walter} F.,  {Brinks} E.,  {Bigiel} F.,  {de Blok} W.~J.~G.,
  {Madore} B.,   {Thornley} M.~D.,  2008, \mn@doi [\aj]
  {10.1088/0004-6256/136/6/2782}, \href
  {http://adsabs.harvard.edu/abs/2008AJ....136.2782L} {136, 2782}

\bibitem[\protect\citeauthoryear{{Li} \& {White}}{{Li} \& {White}}{2009}]{li09}
{Li} C.,  {White} S. D.~M.,  2009, \mn@doi [\mnras]
  {10.1111/j.1365-2966.2009.15268.x}, \href
  {https://ui.adsabs.harvard.edu/abs/2009MNRAS.398.2177L} {398, 2177}

\bibitem[\protect\citeauthoryear{{Lu} et~al.,}{{Lu} et~al.}{2014}]{lu14}
{Lu} Y.,  et~al., 2014, \mn@doi [\apj] {10.1088/0004-637X/795/2/123}, \href
  {https://ui.adsabs.harvard.edu/abs/2014ApJ...795..123L} {795, 123}

\bibitem[\protect\citeauthoryear{{Martin}}{{Martin}}{2005}]{martin05}
{Martin} C.~L.,  2005, \mn@doi [\apj] {10.1086/427277}, \href
  {https://ui.adsabs.harvard.edu/abs/2005ApJ...621..227M} {621, 227}

\bibitem[\protect\citeauthoryear{{McGaugh}}{{McGaugh}}{2005}]{mcgaugh05}
{McGaugh} S.~S.,  2005, \mn@doi [\apj] {10.1086/432968}, \href
  {https://ui.adsabs.harvard.edu/abs/2005ApJ...632..859M} {632, 859}

\bibitem[\protect\citeauthoryear{{McGaugh}}{{McGaugh}}{2012}]{mcgaugh12}
{McGaugh} S.~S.,  2012, \mn@doi [\aj] {10.1088/0004-6256/143/2/40}, \href
  {https://ui.adsabs.harvard.edu/abs/2012AJ....143...40M} {143, 40}

\bibitem[\protect\citeauthoryear{{Mitchell}, {Lacey}, {Baugh}  \&
  {Cole}}{{Mitchell} et~al.}{2013}]{mitchell13}
{Mitchell} P.~D.,  {Lacey} C.~G.,  {Baugh} C.~M.,   {Cole} S.,  2013, \mn@doi
  [\mnras] {10.1093/mnras/stt1280}, \href
  {https://ui.adsabs.harvard.edu/abs/2013MNRAS.435...87M} {435, 87}

\bibitem[\protect\citeauthoryear{{Moster}, {Naab}  \& {White}}{{Moster}
  et~al.}{2018}]{moster18}
{Moster} B.~P.,  {Naab} T.,   {White} S. D.~M.,  2018, \mn@doi [\mnras]
  {10.1093/mnras/sty655}, \href
  {https://ui.adsabs.harvard.edu/abs/2018MNRAS.477.1822M} {477, 1822}

\bibitem[\protect\citeauthoryear{{Moustakas} et~al.,}{{Moustakas}
  et~al.}{2013}]{moustakas13}
{Moustakas} J.,  et~al., 2013, \mn@doi [\apj] {10.1088/0004-637X/767/1/50},
  \href {https://ui.adsabs.harvard.edu/abs/2013ApJ...767...50M} {767, 50}

\bibitem[\protect\citeauthoryear{{Muratov}, {Kere{\v{s}}},
  {Faucher-Gigu{\`e}re}, {Hopkins}, {Quataert}  \& {Murray}}{{Muratov}
  et~al.}{2015}]{muratov15}
{Muratov} A.~L.,  {Kere{\v{s}}} D.,  {Faucher-Gigu{\`e}re} C.-A.,  {Hopkins}
  P.~F.,  {Quataert} E.,   {Murray} N.,  2015, \mn@doi [\mnras]
  {10.1093/mnras/stv2126}, \href
  {https://ui.adsabs.harvard.edu/abs/2015MNRAS.454.2691M} {454, 2691}

\bibitem[\protect\citeauthoryear{{Murray}, {Quataert}  \& {Thompson}}{{Murray}
  et~al.}{2005}]{murray05}
{Murray} N.,  {Quataert} E.,   {Thompson} T.~A.,  2005, \mn@doi [\apj]
  {10.1086/426067}, \href {http://adsabs.harvard.edu/abs/2005ApJ...618..569M}
  {618, 569}

\bibitem[\protect\citeauthoryear{{Murray}, {Quataert}  \& {Thompson}}{{Murray}
  et~al.}{2010}]{murray10}
{Murray} N.,  {Quataert} E.,   {Thompson} T.~A.,  2010, \mn@doi [\apj]
  {10.1088/0004-637X/709/1/191}, \href
  {http://adsabs.harvard.edu/abs/2010ApJ...709..191M} {709, 191}

\bibitem[\protect\citeauthoryear{{Murray}, {M{\'e}nard}  \&
  {Thompson}}{{Murray} et~al.}{2011}]{murray11}
{Murray} N.,  {M{\'e}nard} B.,   {Thompson} T.~A.,  2011, \mn@doi [\apj]
  {10.1088/0004-637X/735/1/66}, \href
  {https://ui.adsabs.harvard.edu/abs/2011ApJ...735...66M} {735, 66}

\bibitem[\protect\citeauthoryear{{Muzzin} et~al.,}{{Muzzin}
  et~al.}{2013}]{muzzin13}
{Muzzin} A.,  et~al., 2013, \mn@doi [\apj] {10.1088/0004-637X/777/1/18}, \href
  {https://ui.adsabs.harvard.edu/abs/2013ApJ...777...18M} {777, 18}

\bibitem[\protect\citeauthoryear{{Oppenheimer} \& {Dav{\'e}}}{{Oppenheimer} \&
  {Dav{\'e}}}{2006}]{oppenheimer06}
{Oppenheimer} B.~D.,  {Dav{\'e}} R.,  2006, \mn@doi [\mnras]
  {10.1111/j.1365-2966.2006.10989.x}, \href
  {http://adsabs.harvard.edu/abs/2006MNRAS.373.1265O} {373, 1265}

\bibitem[\protect\citeauthoryear{{Oppenheimer} \& {Dav{\'e}}}{{Oppenheimer} \&
  {Dav{\'e}}}{2008}]{oppenheimer08}
{Oppenheimer} B.~D.,  {Dav{\'e}} R.,  2008, \mn@doi [\mnras]
  {10.1111/j.1365-2966.2008.13280.x}, \href
  {http://adsabs.harvard.edu/abs/2008MNRAS.387..577O} {387, 577}

\bibitem[\protect\citeauthoryear{{Oppenheimer}, {Dav{\'e}}, {Kere{\v s}},
  {Fardal}, {Katz}, {Kollmeier}  \& {Weinberg}}{{Oppenheimer}
  et~al.}{2010}]{oppenheimer10}
{Oppenheimer} B.~D.,  {Dav{\'e}} R.,  {Kere{\v s}} D.,  {Fardal} M.,  {Katz}
  N.,  {Kollmeier} J.~A.,   {Weinberg} D.~H.,  2010, \mn@doi [\mnras]
  {10.1111/j.1365-2966.2010.16872.x}, \href
  {http://adsabs.harvard.edu/abs/2010MNRAS.406.2325O} {406, 2325}

\bibitem[\protect\citeauthoryear{{Oppenheimer}, {Dav{\'e}}, {Katz}, {Kollmeier}
   \& {Weinberg}}{{Oppenheimer} et~al.}{2012}]{oppenheimer12}
{Oppenheimer} B.~D.,  {Dav{\'e}} R.,  {Katz} N.,  {Kollmeier} J.~A.,
  {Weinberg} D.~H.,  2012, \mn@doi [\mnras] {10.1111/j.1365-2966.2011.20096.x},
  \href {http://adsabs.harvard.edu/abs/2012MNRAS.420..829O} {420, 829}

\bibitem[\protect\citeauthoryear{Peeples \& Shankar}{Peeples \&
  Shankar}{2011}]{peeples11}
Peeples M.~S.,  Shankar F.,  2011, \mn@doi [\mnras]
  {10.1111/j.1365-2966.2011.19456.x}, 417, 2962

\bibitem[\protect\citeauthoryear{{Peeples}, {Werk}, {Tumlinson}, {Oppenheimer},
  {Prochaska}, {Katz}  \& {Weinberg}}{{Peeples} et~al.}{2014}]{peeples14}
{Peeples} M.~S.,  {Werk} J.~K.,  {Tumlinson} J.,  {Oppenheimer} B.~D.,
  {Prochaska} J.~X.,  {Katz} N.,   {Weinberg} D.~H.,  2014, \mn@doi [\apj]
  {10.1088/0004-637X/786/1/54}, \href
  {https://ui.adsabs.harvard.edu/abs/2014ApJ...786...54P} {786, 54}

\bibitem[\protect\citeauthoryear{Peeples et~al.,}{Peeples
  et~al.}{2019}]{peeples19}
Peeples M.~S.,  et~al., 2019, \mn@doi [\apj] {10.3847/1538-4357/ab0654}, 873,
  129

\bibitem[\protect\citeauthoryear{{Pillepich} et~al.,}{{Pillepich}
  et~al.}{2018a}]{pillepich18a}
{Pillepich} A.,  et~al., 2018a, \mn@doi [\mnras] {10.1093/mnras/stx2656}, \href
  {https://ui.adsabs.harvard.edu/abs/2018MNRAS.473.4077P} {473, 4077}

\bibitem[\protect\citeauthoryear{{Pillepich} et~al.,}{{Pillepich}
  et~al.}{2018b}]{pillepich18b}
{Pillepich} A.,  et~al., 2018b, \mn@doi [\mnras] {10.1093/mnras/stx3112}, \href
  {https://ui.adsabs.harvard.edu/abs/2018MNRAS.475..648P} {475, 648}

\bibitem[\protect\citeauthoryear{{Popping} et~al.,}{{Popping}
  et~al.}{2015}]{popping15}
{Popping} G.,  et~al., 2015, \mn@doi [\mnras] {10.1093/mnras/stv2136}, \href
  {https://ui.adsabs.harvard.edu/abs/2015MNRAS.454.2258P} {454, 2258}

\bibitem[\protect\citeauthoryear{{Read} \& {Hayfield}}{{Read} \&
  {Hayfield}}{2012}]{read12}
{Read} J.~I.,  {Hayfield} T.,  2012, \mn@doi [\mnras]
  {10.1111/j.1365-2966.2012.20819.x}, \href
  {http://adsabs.harvard.edu/abs/2012MNRAS.422.3037R} {422, 3037}

\bibitem[\protect\citeauthoryear{{Rupke}, {Veilleux}  \& {Sanders}}{{Rupke}
  et~al.}{2005}]{rupke05}
{Rupke} D.~S.,  {Veilleux} S.,   {Sanders} D.~B.,  2005, \mn@doi [\apjs]
  {10.1086/432889}, \href {http://adsabs.harvard.edu/abs/2005ApJS..160..115R}
  {160, 115}

\bibitem[\protect\citeauthoryear{Sadoun, Shlosman, Choi  \&
  Romano-D{\'{\i}}az}{Sadoun et~al.}{2016}]{sadoun16}
Sadoun R.,  Shlosman I.,  Choi J.-H.,   Romano-D{\'{\i}}az E.,  2016, \mn@doi
  [\apj] {10.3847/0004-637x/829/2/71}, 829, 71

\bibitem[\protect\citeauthoryear{{Saintonge} et~al.,}{{Saintonge}
  et~al.}{2011}]{saintonge11}
{Saintonge} A.,  et~al., 2011, \mn@doi [\mnras]
  {10.1111/j.1365-2966.2011.18677.x}, \href
  {https://ui.adsabs.harvard.edu/abs/2011MNRAS.415...32S} {415, 32}

\bibitem[\protect\citeauthoryear{{Sanders} et~al.,}{{Sanders}
  et~al.}{2015}]{sanders15}
{Sanders} R.~L.,  et~al., 2015, \mn@doi [\apj] {10.1088/0004-637X/799/2/138},
  \href {https://ui.adsabs.harvard.edu/abs/2015ApJ...799..138S} {799, 138}

\bibitem[\protect\citeauthoryear{{Scannapieco} \& {Br{\"u}ggen}}{{Scannapieco}
  \& {Br{\"u}ggen}}{2015}]{scannapieco15}
{Scannapieco} E.,  {Br{\"u}ggen} M.,  2015, \mn@doi [\apj]
  {10.1088/0004-637X/805/2/158}, \href
  {http://adsabs.harvard.edu/abs/2015ApJ...805..158S} {805, 158}

\bibitem[\protect\citeauthoryear{{Scannapieco} et~al.,}{{Scannapieco}
  et~al.}{2012}]{scannapieco12}
{Scannapieco} C.,  et~al., 2012, \mn@doi [\mnras]
  {10.1111/j.1365-2966.2012.20993.x}, \href
  {http://adsabs.harvard.edu/abs/2012MNRAS.423.1726S} {423, 1726}

\bibitem[\protect\citeauthoryear{{Schaye} et~al.,}{{Schaye}
  et~al.}{2015}]{schaye15}
{Schaye} J.,  et~al., 2015, \mn@doi [\mnras] {10.1093/mnras/stu2058}, \href
  {http://adsabs.harvard.edu/abs/2015MNRAS.446..521S} {446, 521}

\bibitem[\protect\citeauthoryear{{Schneider} \& {Robertson}}{{Schneider} \&
  {Robertson}}{2017}]{schneider17}
{Schneider} E.~E.,  {Robertson} B.~E.,  2017, \mn@doi [\apj]
  {10.3847/1538-4357/834/2/144}, \href
  {https://ui.adsabs.harvard.edu/abs/2017ApJ...834..144S} {834, 144}

\bibitem[\protect\citeauthoryear{{Sembolini} et~al.,}{{Sembolini}
  et~al.}{2016a}]{sembolini16a}
{Sembolini} F.,  et~al., 2016a, \mn@doi [\mnras] {10.1093/mnras/stw250}, \href
  {http://adsabs.harvard.edu/abs/2016MNRAS.457.4063S} {457, 4063}

\bibitem[\protect\citeauthoryear{{Sembolini} et~al.,}{{Sembolini}
  et~al.}{2016b}]{sembolini16b}
{Sembolini} F.,  et~al., 2016b, \mn@doi [\mnras] {10.1093/mnras/stw800}, \href
  {http://adsabs.harvard.edu/abs/2016MNRAS.459.2973S} {459, 2973}

\bibitem[\protect\citeauthoryear{{Smit} et~al.,}{{Smit} et~al.}{2014}]{smit14}
{Smit} R.,  et~al., 2014, \mn@doi [\apj] {10.1088/0004-637X/784/1/58}, \href
  {https://ui.adsabs.harvard.edu/abs/2014ApJ...784...58S} {784, 58}

\bibitem[\protect\citeauthoryear{{Somerville}, {Gilmore}, {Primack}  \&
  {Dom{\'\i}nguez}}{{Somerville} et~al.}{2012}]{somerville12}
{Somerville} R.~S.,  {Gilmore} R.~C.,  {Primack} J.~R.,   {Dom{\'\i}nguez} A.,
  2012, \mn@doi [\mnras] {10.1111/j.1365-2966.2012.20490.x}, \href
  {https://ui.adsabs.harvard.edu/abs/2012MNRAS.423.1992S} {423, 1992}

\bibitem[\protect\citeauthoryear{{Song} et~al.,}{{Song} et~al.}{2016}]{song16}
{Song} M.,  et~al., 2016, \mn@doi [\apj] {10.3847/0004-637X/825/1/5}, \href
  {https://ui.adsabs.harvard.edu/abs/2016ApJ...825....5S} {825, 5}

\bibitem[\protect\citeauthoryear{{Springel}}{{Springel}}{2005}]{springel05}
{Springel} V.,  2005, \mn@doi [\mnras] {10.1111/j.1365-2966.2005.09655.x},
  \href {http://adsabs.harvard.edu/abs/2005MNRAS.364.1105S} {364, 1105}

\bibitem[\protect\citeauthoryear{{Springel} \& {Hernquist}}{{Springel} \&
  {Hernquist}}{2003}]{springel03}
{Springel} V.,  {Hernquist} L.,  2003, \mn@doi [\mnras]
  {10.1046/j.1365-8711.2003.06207.x}, \href
  {http://adsabs.harvard.edu/abs/2003MNRAS.339..312S} {339, 312}

\bibitem[\protect\citeauthoryear{{Stinson}, {Seth}, {Katz}, {Wadsley},
  {Governato}  \& {Quinn}}{{Stinson} et~al.}{2006}]{stinson06}
{Stinson} G.,  {Seth} A.,  {Katz} N.,  {Wadsley} J.,  {Governato} F.,   {Quinn}
  T.,  2006, \mn@doi [\mnras] {10.1111/j.1365-2966.2006.11097.x}, \href
  {https://ui.adsabs.harvard.edu/abs/2006MNRAS.373.1074S} {373, 1074}

\bibitem[\protect\citeauthoryear{{Tomczak} et~al.,}{{Tomczak}
  et~al.}{2014}]{tomczak14}
{Tomczak} A.~R.,  et~al., 2014, \mn@doi [\apj] {10.1088/0004-637X/783/2/85},
  \href {https://ui.adsabs.harvard.edu/abs/2014ApJ...783...85T} {783, 85}

\bibitem[\protect\citeauthoryear{{Tremonti} et~al.,}{{Tremonti}
  et~al.}{2004}]{tremonti04}
{Tremonti} C.~A.,  et~al., 2004, \mn@doi [\apj] {10.1086/423264}, \href
  {https://ui.adsabs.harvard.edu/abs/2004ApJ...613..898T} {613, 898}

\bibitem[\protect\citeauthoryear{{Tumlinson}, {Peeples}  \& {Werk}}{{Tumlinson}
  et~al.}{2017}]{tumlinson17}
{Tumlinson} J.,  {Peeples} M.~S.,   {Werk} J.~K.,  2017, \mn@doi [\araa]
  {10.1146/annurev-astro-091916-055240}, \href
  {https://ui.adsabs.harvard.edu/abs/2017ARA&A..55..389T} {55, 389}

\bibitem[\protect\citeauthoryear{Valentini, Murante, Borgani, Monaco, Bressan
  \& Beck}{Valentini et~al.}{2017}]{valentini17}
Valentini M.,  Murante G.,  Borgani S.,  Monaco P.,  Bressan A.,   Beck A.~M.,
  2017, \mn@doi [\mnras] {10.1093/mnras/stx1352}, 470, 3167

\bibitem[\protect\citeauthoryear{Weinberger et~al.,}{Weinberger
  et~al.}{2016}]{weinberger18}
Weinberger R.,  et~al., 2016, \mn@doi [\mnras] {10.1093/mnras/stw2944}, 465,
  3291

\bibitem[\protect\citeauthoryear{{Wiersma}, {Schaye}  \& {Smith}}{{Wiersma}
  et~al.}{2009}]{wiersma09}
{Wiersma} R.~P.~C.,  {Schaye} J.,   {Smith} B.~D.,  2009, \mn@doi [\mnras]
  {10.1111/j.1365-2966.2008.14191.x}, \href
  {http://adsabs.harvard.edu/abs/2009MNRAS.393...99W} {393, 99}

\bibitem[\protect\citeauthoryear{{Zhang}}{{Zhang}}{2018}]{zhang18}
{Zhang} D.,  2018, \mn@doi [Galaxies] {10.3390/galaxies6040114}, \href
  {https://ui.adsabs.harvard.edu/abs/2018Galax...6..114Z} {6, 114}

\bibitem[\protect\citeauthoryear{{van de Voort}, {Springel}, {Mandelker}, {van
  den Bosch}  \& {Pakmor}}{{van de Voort} et~al.}{2019}]{vandevoort19}
{van de Voort} F.,  {Springel} V.,  {Mandelker} N.,  {van den Bosch} F.~C.,
  {Pakmor} R.,  2019, \mn@doi [\mnras] {10.1093/mnrasl/sly190}, \href
  {https://ui.adsabs.harvard.edu/abs/2019MNRAS.482L..85V} {482, L85}

\makeatother
\end{thebibliography}

\appendix

\end{document}